\documentclass[
    aps,
    prl,
    reprint,
    preprintnumbers,
    superscriptaddress,
    amsmath,
    amssymb,
    nofootinbib,
    nobibnotes,
    longbibliography
]{revtex4-2}

\usepackage{graphicx}   
\usepackage{dcolumn}    
\usepackage{bm}         
\usepackage{mathtools}  
\usepackage{xcolor}    
\usepackage{microtype}  
\usepackage{orcidlink}
\usepackage{booktabs}
\usepackage{hyperref}
\hypersetup{
    hidelinks
}
\usepackage{amsmath}
\usepackage{mathrsfs}
\usepackage{float}
\usepackage{tikz} 

\newcommand{\Tr}{\operatorname{Tr}}

\begin{document}


\title{\boldmath From $Z$ to \textbf{all}: cyclic sum rules for thermal CFT partition functions}

\preprint{TIFR/TH/26-34}

\author{Arnav Advant\orcidlink{0009-0001-1678-5573}}
\email{23b1824@iitb.ac.in}
\affiliation{Department of Physics, Indian Institute of Technology Bombay, Powai, Mumbai 400076, India}

\author{Harsh Anand\orcidlink{0009-0000-2281-3609}}
\email{harsh.anand@tifr.res.in}
\affiliation{Department of Theoretical Physics, Tata Institute of Fundamental Research, Homi Bhabha Rd, Mumbai 400005, India}

\author{Nathan Benjamin\,\orcidlink{0000-0003-3661-6563}}
\email{nathanbe@usc.edu}
\affiliation{Department of Physics and Astronomy, University of Southern California, Los Angeles, CA 90089, USA}

\author{Abhijit Gadde\orcidlink{0000-0003-0784-9616}}
\email{abhijit@theory.tifr.res.in}
\affiliation{Department of Theoretical Physics, Tata Institute of Fundamental Research, Homi Bhabha Rd, Mumbai 400005, India}

\author{Vipul Kumar\,\orcidlink{0009-0000-4317-608X}}
\email{vipul.kumar@tifr.res.in}
\affiliation{Department of Theoretical Physics, Tata Institute of Fundamental Research, Homi Bhabha Rd, Mumbai 400005, India}

\author{Shiraz Minwalla\,\orcidlink{0000-0002-6789-6978}}
\email{minwalla.theory@tifr.res.in}
\affiliation{Department of Theoretical Physics, Tata Institute of Fundamental Research, Homi Bhabha Rd, Mumbai 400005, India}

\author{Jyotirmoy Mukherjee\,\orcidlink{0000-0002-1822-2264}}
\email{jyotirmoy.mukherjee\_119@tifr.res.in}
\affiliation{Department of Theoretical Physics, Tata Institute of Fundamental Research, Homi Bhabha Rd, Mumbai 400005, India}

\author{Sridip Pal\,\orcidlink{0000-0002-3813-9513}}
\email{sridip@theory.tifr.res.in}
\affiliation{Department of Theoretical Physics, Tata Institute of Fundamental Research, Homi Bhabha Rd, Mumbai 400005, India}

\author{Asikur Rahaman\,\orcidlink{0009-0008-7814-3059}}
\email{asikur.rahaman@tifr.res.in}
\affiliation{Department of Theoretical Physics, Tata Institute of Fundamental Research, Homi Bhabha Rd, Mumbai 400005, India}

\author{Pabitra Ray\,\orcidlink{0000-0002-5553-7003}}
\email{raypabitra96@gmail.com}
\affiliation{Department of Theoretical Physics, Tata Institute of Fundamental Research, Homi Bhabha Rd, Mumbai 400005, India}

\date{\today}

\begin{abstract}

The high-temperature expansion of twisted $S^{d-1}\times S^1$ partition functions of CFT$_d$,  $d = 2\kappa$, has recently been shown to obey an infinite tower of universal constraints. We determine these constraints explicitly and demonstrate that they can be packaged into a single equation involving a  large conformal diffeomorphism isomorphic to $\mathbb{Z}_{\kappa+1}$, which cyclically permutes the thermal circle and the rotation angles. Our equation asserts that the $\mathbb{Z}_{\kappa+1}$-invariant part of $\ln Z/\beta$  is a constant fixed by the a-anomaly. Our derivation assumes CP-invariance and locality of the thermal effective action, whose structure we  constrain using  an algebraic formalism starting from embedding space.

\end{abstract}

\maketitle

\subsection{Introduction}

By the state-operator correspondence, the thermal partition function of a CFT$_d$ on a twisted $S^{d-1} \times S^1$
\begin{equation}\label{pfsd}
{Z}(\beta;\omega_i) :={\rm Tr}_{S^{d-1}} \ e^{ -\beta(D -\omega_1 J_1- \cdots -\omega_{\kappa } J_{\kappa} )}~,\quad  \kappa:=\left\lfloor \frac d2 \right\rfloor
\end{equation}
fully encodes the scaling dimensions $\Delta$, spins $J_i$ and degeneracies of all local operators. While the detailed structure of $Z(\beta; \omega_i)$ depends sensitively on the particular CFT, it was shown almost two decades ago \cite{Bhattacharyya_2008} that its high-temperature behavior always takes the hydrodynamic form \begin{equation}\label{pfhtu}
\ln Z= \frac{c'}{\beta^{d-1}\displaystyle \prod_{i=1}^{\kappa}( 1-\omega_i^2)}
+ {\cal O}\left(\beta^{3-d }\right)~,~~~ \beta\to 0^+~, ~~~ \omega_i\ \rm{fixed}~.
\end{equation}
This remarkably universal result was subsequently rederived using the formalism of  `equilibrium partition functions' \cite{Banerjee:2012iz, Jensen:2012jh, Shaghoulian:2015lcn}, an approach that was further developed and named {\it thermal effective theory} in \cite{Benjamin:2023qsc}.

A systematic analysis of subleading corrections to Eq.~\eqref{pfhtu} was presented in the recent paper \cite{Advant:2026cqt}. The analysis of \cite{Advant:2026cqt} was largely performed in $d=4$; the generalization to CP-preserving\footnote{For reasons spelled out in \cite{Advant:2026cqt}, the results of this paper apply to all theories in $d=4, 8, 12, \ldots$ dimensions,  but only to CP-preserving theories in $d=6, 10, 14,  \ldots$ dimensions.} CFTs in arbitrary even dimensions $d=2\kappa$  is presented below. One finds    
\begin{equation}\label{expallord}\begin{split}
&\ln Z=\ln Z^{\rm rot}+\ln Z^{\rm sh}~,\\
&\ln Z^{\rm rot}=-\frac{1}{\beta^{d-1} x_1 x_2\ldots x_{\kappa}}
\left(\sum_{q=0}^\infty \beta^{2q}  S_q(x_i) \right)~,\\
&\frac{\ln Z^{\rm sh}}{\beta}=\kappa!a_{2\kappa}
\left[
(-1)^\kappa +
\frac{1}{2^{2\kappa-1}}
\sum_{i=1}^{\kappa}
\int_0^1 ds\,
\prod_{\substack{j=1\\ j\neq i}}^{\kappa}
\frac{
	\left(sx_i-2x_j\right)^2
}{
	x_j\left(x_i-x_j\right)
}
\right]~. \\ 
\end{split}
\end{equation}
$\ln Z^{\rm sh}$, the  universal  `anomalous shift' of the partition function, is proportional to the $a$ anomaly $a_{2 \kappa}$.\footnote{Strictly speaking, the expression for $\ln Z^{\rm sh}$ above is a conjecture, whose  correctness we have checked in $d=4, 6, \ldots, 20$ (see  \hyperref[sec:supplemental]{\texttt{SM}}: throughout this paper \hyperref[sec:supplemental]{\texttt{SM}} denotes the Supplemental Material section). In particular in $d=4$, this expression reduces to $\ln Z^{\rm sh}=\beta\frac{a_4}{12}\left(\frac{x_1}{x_2}+\frac{x_2}{x_1}+7\right)$ in agreement with Eq. (A.3) in \cite{Advant:2026cqt}, and Eq. (3.7) and (3.18) of \cite{Benjamin:2023qsc}.}  The `non-anomalous' part, $\ln Z^{\rm rot}$, is non-universal. It is 
characterized by the coefficient functions $S_q$, which turn out to be (not necessarily homogeneous) symmetric degree-$q$ polynomials\footnote{This  had previously been expected by the authors of \cite{Benjamin:2023qsc}, 
based on extrapolations from the results of computations performed at low orders in $\beta^2$.\label{fn:analytic-structure}} in the variables $\omega_i^2$, or equivalently in 
\begin{equation}\label{xot}
x_i:=1-\omega_i^2~, ~~~(i=1\ldots \kappa)~.
\end{equation}

While the polynomials $S_q$ are theory dependent, their coefficients are constrained,  by the locality of the thermal effective action, to obey 
several additional sharp universal constraints \cite{Advant:2026cqt}. These constraints force $S_q$ to lie in a $d_q$ codimensional subspace \cite{Advant:2026cqt} (of the space of arbitrary symmetric degree $q$ polynomials), where 
$d_q$ is defined by\footnote{$d=2$ is an exception because not every $S_q$ with $I_q^{(\kappa)} = 0$ (see Eq. \eqref{eq:Iq}) is produced by a local integrand. There are no local integrands except the cosmological constant.} 
\begin{equation}\label{constcount}
\frac{t^{\kappa}}{\prod_{i=1}^{\kappa+1}(1-t^i)}= \sum_{q=0}^\infty d_q t^q~; 
\end{equation}
at large $q$, $d_q \sim \frac{q^\kappa}{\kappa! (\kappa+1)!}$.
While the precise form of
the constraints on $S_q$ was worked out  in \cite{Advant:2026cqt} at the first nontrivial value, $q=\kappa$, the detailed structure of the infinite tower of constraints at $q\geq \kappa+1$ was left undetermined. A central result of this paper is the following. The entire infinite tower of constraints is encapsulated, at every $\kappa\geq 2$, by a strikingly simple universal equation involving a
$\mathbb{Z}_{\kappa+1}$ action, generated by $\Omega$, which cyclically permutes the thermal circle and angular directions on $S^{2\kappa-1}\times S^1$, via 
\begin{equation}\label{permb}
\begin{split}
\mathrm{\Omega}:~~&(b_0, b_1,\ldots,b_\kappa)
	\to
	 ( b_1,\ldots,b_{\kappa},b_0)~,\\
{\rm where}~~~&(b_0, b_1,\ldots,b_\kappa):= 
(\beta, \beta \omega_1, \ldots,\beta \omega_\kappa)~.\\
\end{split}
\end{equation}
Our central equation  asserts the vanishing of the $\mathbb{Z}_{\kappa+1}$-invariant part of $\frac{\ln Z^{\rm rot}}{\beta}$,
\begin{equation}\label{omegapartzed}
\begin{split}
&\Big(1+\Omega+\Omega^2+\ldots+\Omega^{\kappa}\Big)\circ \left[\frac{\ln Z^{\rm rot}}{\beta} \right] 
 = 0~.
\end{split}
\end{equation}
Eq. \eqref{omegapartzed},  together with our (conjectured) expression for $\ln {Z}^{\rm sh}$,  then implies \begin{equation}\label{omegapartzedfull}
\begin{split}
&\Big(1+\Omega+\Omega^2+\ldots+\Omega^{\kappa}\Big)\circ \left[\frac{\ln Z}{\beta} \right] 
 = (-1)^{\kappa}\kappa!a_{2\kappa}~.    
\end{split}
\end{equation}

Our demonstration of Eq.~\eqref{omegapartzedfull} applies only in an order-by-order sense in $\beta^2$. Eq.~\eqref{omegapartzedfull} is thus best viewed as a generating function for an infinite tower of constraints on coefficient polynomials (see Eq.~\eqref{expallord})
\begin{equation}\label{tqdef}
T_q:=\beta^{2q} S_q~,
\end{equation}
that are {\it homogeneous} of degree $q$ in $\big(b_0^2, b_1^2, \ldots, b_\kappa^2\big)$. While $T_q$
are symmetric under the interchange $b_i^2 \leftrightarrow b_j^2$ for $i,j\neq 0$, they are not necessarily invariant under permutations involving $b_0^2$, and so can be rewritten (see Appendix \ref{app:poly}) as
\begin{equation}\label{formof}
T_q= I_q^{(0)}+ b_0^2 I_q^{(1)} +b_0^4 I_q^{(2)}+\ldots+b_{0}^{2\kappa}I^{(\kappa)}_{q}
\end{equation}
where $I^{(r)}_q$ is a {\it completely symmetric} polynomial of degree $(q-r)$ in $\big(b_0^2, b_1^2, \ldots,b_\kappa^2\big)$ and absent for $r>q$. Eq.~\eqref{omegapartzed}  is equivalent to
\begin{equation}
\label{eq:Iq}
I_q^{(\kappa)}=0~,~~\forall q \geq \kappa~,
\end{equation}
effectively imposing 
$d_q$ constraints on $S_q$, in agreement with the expectations described around Eq.~\eqref{constcount}. We demonstrate the constraining power below explicitly for $\kappa=2$ and delegate the derivation of \eqref{eq:Iq} for $\kappa>2$ to Appendix \ref{app:poly} and the \hyperref[sec:supplemental]{\texttt{SM}}, and the counting of constraints to Appendix \ref{appcounting}.

As a highly nontrivial check on the results of this paper, we verify that the partition function of Kerr-AdS$_{2 \kappa+1}$ black holes\footnote{These are dual, via AdS/CFT,  to e.g. strongly coupled ${\cal N}=4$ Yang-Mills when $\kappa=2$, or the $(2,0)$ theory of the $M_5$ brane worldvolume when $\kappa=3$.} obeys the central equation   Eq.~\eqref{omegapartzedfull} as an exact non-perturbative identity between functions of $\big(b_0^2, b_1^2, \ldots, b_{\kappa}^2\big)$. 

In the rest of this paper, we first present details of our derivations for the simplest nontrivial dimension $d=4$, and then sketch generalizations to arbitrary dimensions.

\subsection{Derivation of the central equation}\label{deriv}

\subsubsection{Simplifying the measure}

As explained in \cite{Advant:2026cqt}, the most general local thermal  effective action for a $d=2\kappa$ dimensional CFT takes the form   Eq.~\eqref{expallord}, with 
\begin{equation}\label{stpt}
\begin{split}
\ln Z^{\rm rot}
&= \int d^dx \frac{ \sqrt{g_d}}{(v^2)^{\kappa}} {\mathcal I}~. \\
\end{split}
\end{equation}
Specializing to $d=4$, ${\mathcal I}$ is a local Weyl invariant scalar constructed out of both the metric of $S^3 \times R$
\begin{equation}\label{metform}
\begin{split}
ds^2&=   d\tau^2 +d\theta^2 + \sin^2\theta d\phi_1^2 + \cos^2\theta d\phi_2^2~, \\
\end{split}
\end{equation}
as well as the `identification vector'
\begin{equation}\label{vedef}
\begin{split}
v&= \beta \left( \partial_\tau -i\omega_1 \partial_{\phi_1} -i \omega_2 \partial_{\phi_2} \right)\\ &= b_0 \partial_\tau + b_1 (-i \partial_{\phi_1}) + b_2 (-i \partial_{\phi_2})
\end{split}
\end{equation}
which determines the boundary conditions of all dynamical fields $\chi$ via \begin{equation}\label{ident}
\chi(\tau, \phi_1, \phi_2, \theta)= e^{{\mathcal L}_v} \chi(\tau, \phi_1, \phi_2, \theta)~. 
\end{equation} 

Translational invariance  allows the integrals over $\tau$, $\phi_1$ and $\phi_2$ to be evaluated. Changing integration variables from $\theta$ to\footnote{$\gamma^2$ below is the time dilation factor of the rotating fluid, see \cite{Anand:2025mfh}.} 
\begin{equation}\label{hatzed}
\begin{split}
&~~z:=\beta^2\left(1-x_1 x_2\gamma^2 \right)~,\\\mathrm{where}\quad&\gamma^2:= \frac{1}{1-\omega_2^2+\sin^2 \theta (\omega_2^2-\omega_1^2) }~,    
\end{split} 
\end{equation}
we find 
(see around \eqref{hg} in \hyperref[sec:supplemental]{\texttt{SM}})
\begin{equation}\label{pdef}
\begin{split}
&\frac{\sqrt{g_4} }{(v^2)^2}  d \theta \int_0^\beta d\tau \int_0^{2 \pi} d \phi_1 \int_0^{2 \pi}  d\phi_2 =\beta P dz~,\\\mathrm{where}~~&
P:=\frac{2 \pi^2}{(b^2_0-b^2_1) (b^2_1 -b^2_2)(b_2^2-b_0^2) }~. 
\end{split}
\end{equation}
  Eq.~\eqref{stpt} 
can thus be rewritten as 
\begin{equation}\label{lnzchin}
\frac{\ln Z^{\rm rot}}{\beta} =P 
\int_{b^2_1}^{b^2_2} d  z  ~{\mathcal I}_{b_0, b_1, b_2} ( z )~. 
\end{equation}
Notice that while $P$ is antisymmetric under interchanges of $b_i^2$ and $b_j^2$, it is symmetric under $\mathbb{Z}_{3}$ cyclical permutations. 
In   Eq.~\eqref{lnzchin} we have renamed  ${\mathcal I}={\mathcal I}_{b_0, b_1, b_2} $ to emphasize the dependence of this function on the `compactification data'  $(b_0, b_1, b_2)$.

\subsubsection{The $\mathbb{Z}_{3}$ invariance of ${\mathcal I}_{b_0, b_1, b_2}$}

As explained in \cite{Advant:2026cqt}, Weyl invariance and locality together ensure that ${\mathcal I}$ can be expressed as a total derivative plus a power series in the three local Weyl invariant scalars $ {\rm tr}(\omega^2)$, $v^2 \mathcal{S}$  and  $v^2 \mathcal{B}_\mu (\omega^2)^{\mu\nu} \mathcal{B}_\nu$ (see \cite{Advant:2026cqt} for definitions). The coefficients of this expansion are theory dependent but  independent of the `compactification data'. As a consequence,  ${\mathcal I}$ depends on $(b_0, b_1, b_2)$ only through the local Weyl invariant scalars.  A straightforward computation reveals that 
\begin{equation}\label{explicitf}
\begin{split}
& \frac{ {\rm tr}(\omega^2)}{8} =  z~, \qquad v^2 \mathcal{S} = b_0^2+b_1^2+b_2^2-3 z~, \\& v^2 \mathcal{B}_\mu (\omega^2)^{\mu\nu} \mathcal{B}_\nu =- 4( z -b_0^2)( z -b_1^2)( z -b^2_2)~.
\end{split}
\end{equation}
Note that each of the quantities in Eq.~\eqref{explicitf} is a polynomial both in $z$ as well as in $b_i^2$ (see the discussion below Eq. \eqref{vabintro} for a structural explanation of this feature). 

Each of the scalars listed in   Eq.~\eqref{explicitf}, and so ${\mathcal I}_{b_0, b_1, b_2}$, is manifestly invariant  under $S_3$ permutations of $(b_0, b_1, b_2)$, so that
\begin{equation}\label{identi}
{\mathcal I}_{b_0, b_1, b_2}({z})={\mathcal I}_{b_1, b_2, b_0}({z})= {\mathcal I}_{b_2, b_0, b_1}(z) = {\mathcal I}({z})~.
\end{equation}

\subsubsection{Derivation using Cauchy's theorem}

When $d=4$, our central equation   Eq.~\eqref{omegapartzedfull}
abbreviates to  
\begin{equation}\label{centab}
3 \mathfrak{S} \circ \left[  \frac{\ln Z}{\beta} \right] = 2a_4~, 
\end{equation}
where $\mathfrak{S}$ is the $\mathbb{Z}_{3}$ symmetrization operator 
\begin{equation}\label{pdefg}
\begin{split}
\mathfrak{S}&:= \frac{(1+\Omega+ \Omega^2)}{3}~. 
\end{split}
\end{equation}
Since $P$ is $\mathbb{Z}_{3}$ invariant,  
$\mathfrak{S}\circ P=P$. Using  Eq.~\eqref{lnzchin}
\begin{equation}\label{finarg}
\begin{split}
&\frac{3}{P} \times  \mathfrak{S} \circ \left[  \frac{\ln Z^{\rm rot}}{\beta} \right] = 3\mathfrak{S} \circ \left[  \frac{\ln Z^{\rm rot}}{\beta} \times \frac{1}{P}  \right] \\
& =
\int_{b^2_1}^{b^2_2} d  z  ~{\mathcal I}_{b_0, b_1, b_2}(z)  + \int_{b^2_2}^{b^2_0} d  z  ~{\mathcal I}_{b_1, b_2, b_0}(z)\\
& + \int_{b^2_0}^{b^2_1} d  z  ~{\mathcal I}_{b_2, b_0, b_1}(z) \\
&=\oint d  z  ~{\mathcal I}(z)=0~.\\
& \therefore \mathfrak{S} \circ \left[  \frac{\ln Z^{\rm rot}}{\beta} \right]=0~.
\end{split}
\end{equation}
The equality in the fourth line above follows order by order 
in $\beta^2$; since $\mathcal{I}$ is a polynomial in $z$ with coefficients that are polynomials in $b_i^2$, the integral of each monomial over the closed contour $[b_1^2,b_2^2]\cup[b_2^2,b_0^2]\cup[b_0^2,b_1^2]$ vanishes trivially.
A simple computation also shows that 
\begin{equation}\label{omegarem}
3 \mathfrak{S} \circ \frac{\ln Z^{\rm sh}}{\beta} = 3 \mathfrak{S} \circ \left[ \frac{a_4}{12}
\left( \frac{x_1}{x_2} + \frac{x_2}{x_1} + 7\right)  \right]=2a_4~.
\end{equation}
Plugging  Eq.~\eqref{finarg} and  Eq.~\eqref{omegarem} into the first of  Eq.~\eqref{expallord}, we recover  Eq.~\eqref{centab}. In Appendix \ref{simplification} we demonstrate that  Eq.~\eqref{centab} simplifies at the  special angular velocities
$\big(\omega^2_1,\omega^2_2\big)=\left(e^{\frac{2 \pi i}{3}}, e^{\frac{4 \pi i}{3}}\right)$, 
where it exactly reproduces  Eq.~(7) and Eq.~(22) of  \cite{Advant:2026cqt}.

 In Appendix \ref{lcm} we present a physically intriguing rederivation of  Eq.~\eqref{identi}
 that does not invoke the computational result Eq.~\eqref{explicitf}. Briefly, we construct a $\mathbb{Z}_{3}$ group of diffeomorphisms, which cyclically permute $(\tau, i \phi_1, i \phi_2)$, but leave the coordinate $z$ invariant. These conformally map  (see Fig. \ref{sads} in \hyperref[sec:supplemental]{\texttt{SM}})
\begin{equation}\label{conformalcycle}
S^3 \times R 
\xrightarrow{\Omega} AdS_3 \times S^1 \xrightarrow{\Omega}  ~AdS_3' \times S^{\prime 1} \xrightarrow{\Omega}
S^3 \times R
\end{equation}
where $AdS_3 \times S^1$
and $AdS_3' \times S^{\prime 1}$ are distinct. Remarkably,  an analytic continuation in $z$ turns each of the $AdS\times S^1$ spacetimes into $S^3 \times R$, but with a cyclically permuted $v$ vector.  Eq.~\eqref{identi} then follows from the Weyl invariance of ${\mathcal I}$.

\subsection{
The vanishing of $I^{(2)}_q$
in $d=4$}\label{itwo}

We will now show that the small $\beta^2$ expansion of the constraint Eq.~\eqref{omegapartzed} on $\ln Z^{\rm rot}$ yields Eq. \eqref{eq:Iq} in $d=4$.

The second line of   Eq.~\eqref{expallord} can be rewritten as
\footnote{Indeed,   Eq.~\eqref{expallord} can be derived as follows.  As the integral on the RHS of   Eq.~\eqref{lnzchin} vanishes when $b_1=b_2$, it follows that the RHS of   Eq.~\eqref{lnzchin} equals $P$ times $\left(b^2_2-b^2_1\right)$ times a power series expansion in  $(b_0, b_1, b_2)$.} 
\begin{equation}\label{zrotnew}
\begin{split}
&\frac{\ln Z^{\rm rot}}{\beta} 
=-\frac{P}{2\pi^2} \sum_{q=0}^\infty \left(b^2_2-b^2_1\right) T_q\\
&=-\frac{1}{2\pi^2}\sum_{q=0}^\infty  \Big[ P  I^{(0)}_q \left(b^2_2-b^2_1\right)+ P I^{(1)}_q \left(b^2_2 b^2_0-b^2_1b^2_0\right)\\&\qquad\qquad + P I^{(2)}_q \left(b^2_2b_0^4-b^2_1 b_0^4\right)\Big]~, 
\end{split}
\end{equation}
where we have used   Eq.~\eqref{formof}. 
 As $P$, $I_q^{(0)}$, $I_q^{(1)}$ and $I_q^{(2)}$ are all invariant under $\mathbb{Z}_{3}$ cyclical shifts,
\begin{equation}\label{symop}
\begin{split}
&3\mathfrak{S}\circ \left[ \frac{\ln Z^{\rm rot}}{\beta} \right]=\frac{3}{2\pi^2} 
\sum_{q=0}^\infty \Big[ \big\{\mathfrak{S}\circ\left(b^2_1-b^2_2 \right)\!\big\} P I^{(0)}_q\\& + \big\{\mathfrak{S}\circ\left(b^2_1b^2_0-b^2_2b^2_0 \right)\!\big\} P I^{(1)}_q +\big\{\mathfrak{S}\circ\left(b^2_1b^4_0-b_2^2b^4_0 \right)\!\big\} P I^{(2)}_q \Big] \\&=-\sum_{q=0}^\infty I^{(2)}_q~.
\end{split}
\end{equation}
(The vanishing of the first two terms and the simplification of the last term follow from the identities Eq.~\eqref{sbot} in the \hyperref[sec:supplemental]{\texttt{SM}}). Since $I_q^{(2)}$ is homogeneous of degree $(q-2)$, we conclude that it must vanish for each $q$. 

See around \eqref{actgonfin} in \hyperref[sec:supplemental]{\texttt{SM}} for a 
generalization of this derivation to $d=2\kappa$, and Appendix \ref{appcounting} for a counting of these constraints.

\subsection{Generalization to $d=2\kappa$ dimensions for arbitrary $\kappa$}\label{gen}

The generalization of our discussion to arbitrary even dimensions (see  
\hyperref[sec:supplemental]{\texttt{SM}}, 
between \eqref{wfom} and \eqref{hyperplane} for all details) proceeds in two steps.

We first express all local EFT terms using a common set of scalar variables. In the embedding space $R^{2\kappa,2}$, boundary spacetime points  are null rays $N$, while the CKV $v$ corresponds to an adjoint 
$SO(2\kappa,2)$ matrix $A$. Using the geometrically natural coordinates
\begin{equation}\label{gim}
m^{(a)}=\frac{N\cdot B^{a}N}{N\cdot B N}~,~~~~ B=-A^2~,
\end{equation}
we find that Eq. \eqref{stpt} becomes
\begin{equation}\label{stptnew}
\begin{split}
\left(\frac{\ln Z^{\rm rot}}{\beta}\right)&=
\frac{(-1)^\kappa 2\pi^\kappa}{{\mathcal V}} \int_{{\mathcal D}_0} 
\Upsilon \\
\Upsilon&=dm_2 \wedge dm_3 \ldots \wedge d m_\kappa ~{\mathcal I}\left(E_r,m^{(a)} \right)\\
{\mathcal V}&=\prod_{0\leq i<j\leq \kappa}\big(b_j^2-b_i^2\big)
\end{split}
\end{equation}
where $E_r$ are defined in \eqref{invar}, and the integration range ${\mathcal D}_0$ is a simplex built out of the $\kappa$ vectors 
\begin{equation}\label{vabintro}
\left(p_{0b}\right)_j=\frac{b_0^{2j}-b_b^{2j}}{b^2_0-b^2_b}, ~~~(b=1\ldots \kappa).
\end{equation}

In the Weyl frame defined by the condition $N\cdot B N=-1$, the variables $m^{(a)}$ are polynomial in $N$ and $B$. In this Weyl frame, the  projector $\Pi$ onto the  CFT submanifold of $R^{2 \kappa, 2}$ is easily computed, and turns out to also be polynomial in these variables (see \eqref{proj} in \hyperref[sec:supplemental]{\texttt{SM}}). Using that covariant derivatives on a submanifold are computed by projecting ordinary derivatives in the flat embedding space onto the submanifold, we argue that this polynomial nature is preserved under the operation of taking covariant derivatives, and conclude that
${\mathcal I}(E_r,m^{(a)} )$ is a polynomial in its arguments. 
Upon integrating, it follows from Eq.~\eqref{stptnew} that $\ln Z$ takes the analytic form Eq.~\eqref{expallord}. 

Next, we act on \eqref{stptnew} with the $\mathbb{Z}_{\kappa+1}$ symmetrization operator (see \eqref{omegapartzed}).
$E_r$ and $m_i$ are unchanged by permutations, and $\Omega^a\circ {\mathcal V}=(-1)^{\kappa a}{\mathcal V}$. It follows that $\Omega^a$ effectively acts only on the integration range ${\cal D}_0$, turning it into the new simplex ${\cal D}_a$ constructed out of the vectors 
\begin{equation}\label{vabintrapab}
\left(p_{ab} \right)_j=\frac{b_a^{2j}-b_b^{2j}}{b^2_a-b^2_b}~, ~~~(b=0\ldots \kappa, ~b\neq a)~,
\end{equation}
with orientation and signs described around Eq.~\eqref{vorder} in \hyperref[sec:supplemental]{\texttt{SM}}.
\begin{figure}[!ht]
\centering
\includegraphics[width=0.66\linewidth]{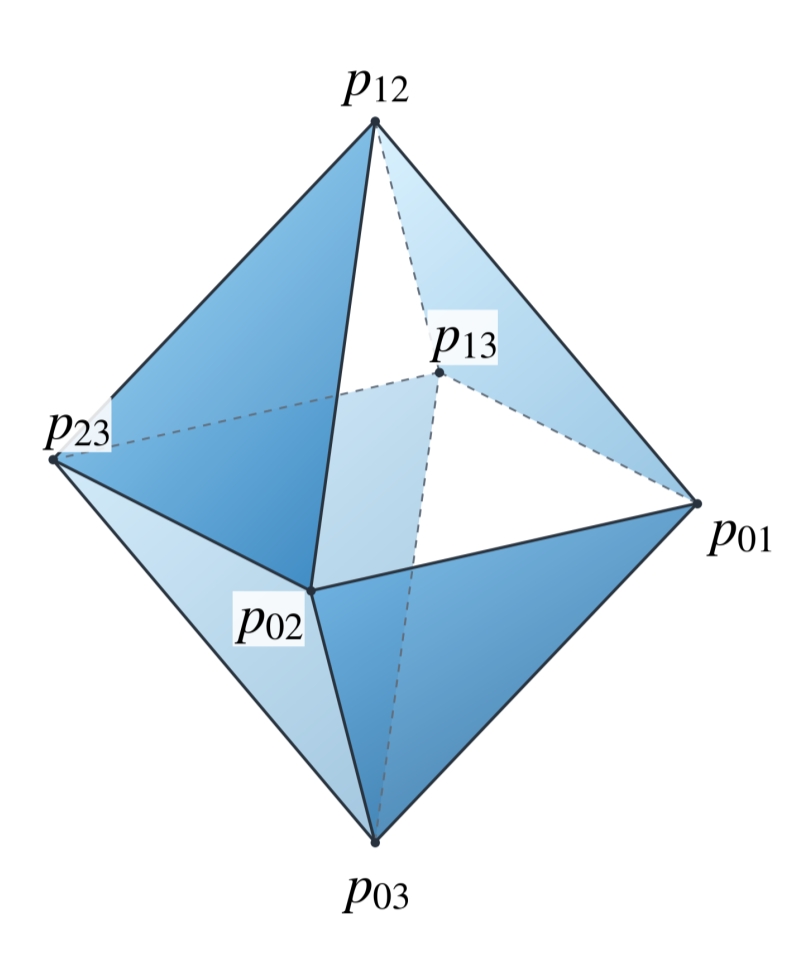}
\caption{The $d=6$ polytope $\Delta(2,4)$.  Its six corners are labeled by the six unordered pairs $(a,b)$.  The four blue shaded triangular faces constitute the domain of integration, while the other four triangular faces are the `holes' ${\cal H}_a$.}
\label{fig:genalg-octahedron}
\end{figure}

The $(\kappa+1)$ distinct simplices above each share exactly one corner with every other simplex. The topological manner in which these simplices are sewn together can  be visualized as follows. Consider the intersection of a codimension one plane with the unit cube in a Euclidean $R^{\kappa+1}$ space defined by 
\begin{equation}\label{twk}
  \Delta(2,\kappa+1)
=\left\{(t_0,\ldots,t_\kappa):0\leq t_a\leq1,\ \sum_a t_a=2\right\}~, 
\end{equation}
(see Fig.~\ref{fig:polytope-kappa2} in \hyperref[sec:supplemental]{\texttt{SM}} for a visualization when $\kappa=2$). For $\kappa \geq 3$, this $\kappa$ dimensional solid shape has $2(\kappa+1)$ distinct boundary surfaces. We identify the $(\kappa+1)$  surfaces $t_a=1$ ($a=0\ldots \kappa$) with our integration simplices ${\mathcal D}_a$. We identify the remaining  $(\kappa+1)$ surfaces $t_a=0$ ($a=0\ldots \kappa$) with ${\mathcal H}_a$, the gaps between integration regions. In the case $d=6$, the  relative arrangement of the ${\mathcal D}_a$ and the holes ${\mathcal H}_a$ is depicted in Fig. \ref{fig:genalg-octahedron}.

As ${\mathcal I}(E_r,m^{(a)} )$ is polynomial 
in $m^{(a)}$, it follows from Stokes' theorem and the exactness of the polynomial top form  $\Upsilon$, that the integral of $\Upsilon$ over the closed oriented surface consisting of all ${\mathcal D}_a$ plus ${\mathcal H}_a$ vanishes. However,  we demonstrate around 
\eqref{hyperplane} of 
\hyperref[sec:supplemental]{\texttt{SM}}
that in our physical set up, the `holes'  ${\mathcal H}_a$ can be chosen to lie in a $(\kappa-2)$ dimensional hyperplane, and so are
of one lower dimension than the integration regions ${\mathcal D}_a$. As a consequence, the integral over each of the  ${\mathcal H}_a$ vanishes, so the same is true of the sum of integrals over  appropriately oriented ${\mathcal D}_a$, implying Eq.~\eqref{omegapartzed}.

\subsection{Checks}

\subsubsection{Exact verification from the Kerr-AdS$_{2\kappa+1}$ black hole}\label{bh}

Consider Kerr--AdS in $(2\kappa+1)$ bulk dimension with horizon radius $r_+$ and independent rotation parameters $a_i$. We choose units so that the AdS radius is unity. The thermodynamics of Kerr--AdS black holes was worked out in \cite{Gibbons:2004uw,Gibbons_2005}. In $d=2\kappa$, we define the black hole variables $(q_0,q_1,\cdots,q_\kappa)$ such that 
\begin{equation}
q_0:=\frac{1}{r_+}~,\quad q_i:=\frac{a_i}{r_+}~,\quad\mathrm{for}~~i=1,\cdots,\kappa~.    
\end{equation}
The black hole thermodynamic potentials $(b_0,b_1,\cdots,b_\kappa)$ (defined in Eq.~(\ref{permb})) turn out to be given in terms of $q_i$ by
\begin{equation}\label{betainv}
\begin{split}
b_i
&=
\frac{
2\pi q_i\prod_{\substack{j=0\\ j\neq i}}^{\kappa}\left(1+q^2_j\right)
}{\mathrm{\Xi}
}~,\quad\mathrm{for}~~i=0,1,\cdots,\kappa
\end{split}    
\end{equation}
where $\mathrm{\Xi}$ is given by
\begin{equation}\label{Delta}
\begin{split}
\Xi:=\sum_{i=0}^{\kappa+1}(\kappa-i)\,E_i\!\left(q_0^2,\dots,q_\kappa^2\right)~.
\end{split}  
\end{equation}
Here $E_i$'s are defined in Eq.~\eqref{invar}. 

The Kerr-AdS$_{2\kappa+1}$ black hole partition function equals  
\begin{equation}
\ln Z_{\rm BH}=-I_{2\kappa+1}^{\mathrm{th}}
    =
    \frac{\beta\pi^{\kappa-1}}
{8(\kappa-1)!G_{2\kappa+1}}
(1-q_0^2)
\prod_{i=1}^{\kappa}
\frac{1+q_i^2}{q_0^2-q_i^2}~.
    \label{eq:general-thermal-action}
\end{equation}
Now, consider a  $\mathbb{Z}_{\kappa+1}$ transformation that permutes $(q_0,q_1,\cdots,q_\kappa)$ as (compare  Eq.~\eqref{permb})
\begin{equation}\label{zthreebh}
\begin{split}
\mathrm{\Omega}_{\rm bh}:&\quad(q_0, q_1,\ldots,q_\kappa)
	\to
	( q_1,\ldots,q_{\kappa},q_0)~.
\end{split}
\end{equation}
It follows immediately from  Eq.~\eqref{betainv} and  Eq.~\eqref{Delta} (which is $\mathbb{Z}_{\kappa+1}$ invariant) that  Eq.~\eqref{zthreebh} induces the action  Eq.~\eqref{permb} on $(b_0, b_1,\cdots,b_\kappa)$ so that the $\mathbb{Z}_{\kappa+1}$ generator $\Omega_{\rm bh}$ reproduces the action of $\Omega$.  

Using  Eq.~\eqref{zthreebh} we have checked (see around \eqref{bhident} in \hyperref[sec:supplemental]{\texttt{SM}}) that 
\begin{equation}\label{bhidentmain}
\begin{split}
\Big(1+\mathrm{\Omega}+\cdots+\mathrm{\Omega}^\kappa\Big)\circ\left[\frac{\ln Z_{\rm BH}}{\beta}\right]&=(-1)^\kappa\frac{\pi^{\kappa-1}}{4(\kappa-1)!G_{2\kappa+1}}\\&=(-1)^\kappa\kappa! a_{2\kappa}    
\end{split} 
\end{equation}
in perfect agreement\footnote{Note that  $a_{2\kappa}=\frac{\pi^{\kappa-1}}{4\kappa!(\kappa-1)!G_{2\kappa+1}}$ is the standard answer for the holographic Weyl anomaly.} with our general prediction   Eq.~\eqref{omegapartzedfull}. Eq.~\eqref{bhidentmain} holds as an exact (non-perturbative)  identity, and is true at all values of the parameters.

\subsubsection{Direct checks at low orders in the small $x_i$ expansion}\label{explicit}

While this paper has focused on the study of $\ln Z$ in the high temperature expansion, it was argued in  \cite{Anand:2025mfh} (see also \cite{Komargodski:2026ain, Lee:2026azy, Mukherjee:2026dfu, Miscioscia:2026bvl}) that $\ln Z$ also admits a closely related second `semi-universal' small $x_i$ expansion. Order-by-order in this expansion,  constraints on $\ln Z$ take the form of differential equations on temperature dependent coefficient functions 
\cite{Advant:2026cqt}. In \hyperref[sec:supplemental]{\texttt{SM}} (see around \eqref{supp_tlist10}) we have used the methods of \cite{Advant:2026cqt} to work out the explicit form of the constraints
on $\frac{\ln Z^{\rm rot}}{\beta}$ up to seventh subleading order in small $x_i$, in $d=4$. We have  then independently reevaluated these constraints, by inserting the small $x_i$ semi-universal expansion (see Eq.~(11) of \cite{Advant:2026cqt}) into \eqref{omegapartzed}, Taylor expanding up to the same order in $x_i$, and equating coefficients. The two different methods yield identical constraint differential equations. In particular, at lowest nontrivial order in $x_i$, both methods yield the single constraint Eq. \eqref{firstconst} in  \hyperref[sec:supplemental]{\texttt{SM}}, in perfect agreement with Eq. (16) of \cite{Advant:2026cqt} at $a=0$.

\subsection{Discussion}

The authors of \cite{Advant:2026cqt} had predicted the existence of an infinite tower of constraints on coefficients of the high temperature expansion, but had no expectation for the precise form of these equations. In this paper we have demonstrated that this entire tower of relations 
is equivalent to the single, beautiful constraint equation, Eq.~\eqref{omegapartzed}. The order-by-order expansion of this equation in $\beta^2$ yields the simple constraints on polynomial coefficients  Eq.~\eqref{eq:Iq}. 

It would be good to understand the precise conditions on CFTs that guarantee that 
$\ln Z$ is accurately described by a local thermal effective action as assumed in this paper (see \cite{Advant:2026cqt, Anand:2025mfh} for relevant discussion). It would also be useful to generalize our analysis to CP violating  theories in  dimensions $d=6, 10, \ldots$ and to other scenarios \cite{Buric:2024kxo,Buric:2025uqt,Buric:2026pes,Parmentier:2026aqh,Kusuki:2025pgx,Banihashemi:2025qqi,David:2025tqn} where thermal EFT has been useful.

Our striking central equation \eqref{omegapartzed}, which asserts a relationship between thermodynamics in the time direction and in the rotation planes, calls for a direct physical interpretation, perhaps along the lines of  Appendix \ref{lcm}. 
However, we emphasize that Eq. \eqref{omegapartzed} has only been derived order-by-order in the high temperature expansion (i.e. in an expansion around $(b_0, b_1, \ldots, b_\kappa)=(0, 0, \ldots, 0)$). Since $\ln Z$ in  Eq.~\eqref{pfsd} converges only when $|{\rm Re} (b_i)| < b_0$ $\forall i \neq 0$ (see e.g. \cite{Kim:2023sig, Bajaj:2024utv, Choi:2024xnv, Choi:2025lck, Anand:2025mfh}), and as $\Omega$ maps allowed (convergent) values of $b_i$ to disallowed (divergent) $b_i$ (see  Eq.~\eqref{permb}), any non-perturbative 
generalization of Eq.~\eqref{omegapartzed} would seem to require an analytic continuation of $\ln Z$ to divergent values. It is not clear whether such a continuation is actually meaningful (see e.g. \cite{Felder_2000, Basar:2014mha,McGady:2017rzv,McGady:2018rmo} for relevant discussion). 

Relatedly, it was explained in \cite{Benjamin:2024kdg} that all `bulk' terms (see \cite{Benjamin:2024kdg} for terminology) in the all-orders expansion of $\ln Z$ around the rational point
$\left(b_0, b_1 \ldots b_\kappa\right)=\left(0, \frac{2 \pi i r_1}{s_1}, \frac{2 \pi i r_2}{s_2}, \ldots \frac{2 \pi i r_\kappa}{s_\kappa}\right)$ are formally given by $\frac{1}{S}$ times the high temperature partition function of this paper, evaluated as a function of $S \delta b_i$. Here 
$S$ is the LCM of $s_1, s_2 ,\ldots ,s_{\kappa}$, and 
$\delta b_i$ are the small deviations of $b_i$ away from this rational point. Thus the analysis of this paper determines not just the all-orders expansion of $\ln Z$ in the high temperature limit, but also a large part of the expansion about each of the singular rational points above. One may wonder whether the nonperturbative partition function can be reconstructed from a suitably defined sum over these `saddle points'; a proposal with echoes of modular invariance (see \cite{Lei:2024oij} for a potentially related study of free theories). 

Finally, it is possible that an analysis similar to that 
of \cite{Advant:2026cqt} and the current paper can also be 
carried out for the superconformal indices \cite{Kinney:2005ej, Bhattacharya:2008zy} in the so-called Cardy limit (see e.g. \cite{DiPietro:2014bca,ArabiArdehali:2014otj,ArabiArdehali:2014cey, ArabiArdehali:2015iow,ArabiArdehali:2015ybk,Choi:2018hmj, Kim:2019yrz, Cabo-Bizet:2019osg, Gadde:2020bov, Goldstein:2020yvj, Jejjala:2021hlt, Cassani:2021fyv,Ardehali:2021irq, Ohmori:2021dzb,Jejjala:2022lrm}). Such a study would require the generalization to partition functions with additional conserved $R$ charges (and possibly critical chemical potentials \cite{Bhattacharyya:2007sa, Choi:2024xnv}).

\subsection{Acknowledgments}
We thank E. D'Hoker, T. Dumitrescu, H. T. Lam, O. Parrikar, P. Rath, V. Schomerus, S. Trivedi, and Y. Wang for very useful discussions. We would also like to thank K. Bajaj,  J. Choi, K. Jensen, Z. Komargodski, A. Miscioscia, C. Patel, F. Popov, S. Van Leueven, and S. Wadia for comments on a preliminary version of this manuscript.  N.B. is supported by the U.S. Department of Energy, Office of Science, under grant Contract Number DE-SC0026324. The work of A.A.,  H.A., A.G.,  V.K., S.M., J.M., S.P., A.R. and P.R. was supported by the Department of Atomic Energy, Government of India, under Project Identification Number RTI-4012 and the Infosys Endowment for the study of the Quantum Structure of Spacetime. A.A., H.A., A.G., V.K., S.M., J.M., S.P., A.R. and P.R. would also like to acknowledge their debt to the
people of India for their steady support of the study of basic science.

\subsection{AI Disclosure Statement}
 The formula in the last line of Eq.~\eqref{expallord} was developed in discussions with Claude Fable 5.1; the study of $\kappa \geq 3$ (the material between Eq. \eqref{wfom} and \eqref{hyperplane} in \hyperref[sec:supplemental]{\texttt{SM}}) was developed in discussions with ChatGPT Astra 6.1 and independently with Claude Fable 5.1; and the material presented 
between Eq. \eqref{ql} and \eqref{sumsum} in \hyperref[sec:supplemental]{\texttt{SM}} were developed in discussions with ChatGPT Sol 6.1. In addition, the authors used Claude Opus 5.5 to assist with checks of all calculations and preparation of the figures. Every part of this manuscript (including subsections whose material was developed in discussion with AI tools) is written entirely by the authors. The authors take full responsibility for the scientific content of the paper. 
 
\bibliography{biblio}

\clearpage

\appendix

\refstepcounter{section}
\label{app:poly}
\section{Appendix A: Constraining polynomials}

Every symmetric polynomial of $(y_0, y_1,\cdots, y_\kappa)$  can famously be recast as a polynomial of the free generators  
\begin{equation}\label{invar}
\begin{split}
&E_0(y_0, y_1,\cdots, y_\kappa):=1~,\\&
E_{r}(y_0, y_1,\cdots, y_\kappa)\\&\qquad:=\sum_{0\leq i_1<\cdots<i_r\leq \kappa} y_{i_1} \cdots y_{i_r}  ~,~~\mathrm{for}~~r=1,\cdots,\kappa+1~.    
\end{split}
\end{equation}
Note that each of $(y_0, y_1 \ldots, y_\kappa)$ clearly obeys the equation
\begin{equation}\label{chareq}
w^{\kappa+1}-E_1 w^\kappa\cdots+(-1)^r E_r w^{\kappa+1-r}\cdots +(-1)^{\kappa+1}E_{\kappa+1}=0~.
\end{equation}

Now consider a  polynomial of the above variables that is completely symmetric in $y_1 \ldots y_\kappa$, but has no particular symmetry under interchanges involving $y_0$. Any such polynomial can be written as a polynomial in $E_1,E_2,\cdots, E_{\kappa+1}$ and $y_0$. As $y_0$ obeys  Eq.~\eqref{chareq}, however, powers of $y_0$ that are $\geq \kappa+1$ can all be traded for lower powers of $y_0$ multiplying arbitrary polynomials of $E_r$. It 
thus follows that the most general polynomial symmetric in $y_1 \ldots y_\kappa$ can be written in the form  Eq.~\eqref{formof}, with $y_i=b_i^2$.

In \hyperref[sec:supplemental]{\texttt{SM}} (see around \eqref{actgonfin}) we insert \eqref{formof} into Eq.~\eqref{expallord}, and then insert this expansion into  Eq.~\eqref{omegapartzed}, and equate powers of $\beta^2$. While this yields no constraints on the polynomials  $I^{(0)}_q, I^{(1)}_q,\ldots,I^{(\kappa-1)}_{q}$, it forces
\begin{equation}\label{itwoo}
I_q^{(\kappa)}=0~, ~~\forall ~q \geq \kappa~.
\end{equation} 
Eq.~\eqref{itwoo} expresses a simple universal constraint on the coefficient functions of the high temperature expansion. 

\refstepcounter{section}
\label{simplification}
\section{Appendix B: Simplification at cyclotomic points}

Throughout this Appendix we work  at the special values 
\begin{equation}\label{specialvalues}
\big(\omega^2_1,\omega^2_2 \ldots \omega_\kappa^2\big)=\left(e^{\frac{2 \pi i}{\kappa+1}}, e^{\frac{4 \pi i}{\kappa+1}}, \ldots, e^{\frac{2 \pi \kappa i}{\kappa+1}} \right)
\end{equation}
which implies 
\begin{equation}\label{yall}
\begin{split}
&\big(b^2_0, b^2_1, \ldots  b^2_\kappa\big)= \beta^2 \left(1,e^{\frac{2 \pi i}{\kappa+1}}, \ldots, e^{\frac{2 \pi \kappa i}{\kappa+1}} \right) \\
\implies &\Omega\circ \left(b^2_0, b^2_1, \ldots,b^2_\kappa \right)
=\beta^2\left(e^{\frac{2 \pi i}{\kappa+1}},e^{\frac{4 \pi i}{\kappa+1}} ,\ldots, e^{\frac{2 \pi \kappa i}{\kappa+1}},1 \right)
\\&\qquad\qquad\qquad\qquad= e^{\frac{2 \pi i}{\kappa+1}} \left(b_0^2,b_1^2,b^2_2, \ldots,b_\kappa^2  \right)~.
\end{split}
\end{equation}
Consider any degree $h$ homogeneous function $f$ of the $b_\mu^2$, $\mu=0\ldots \kappa$, where $h$ is an integer. It follows immediately from \eqref{yall} that the action of  
the cyclic symmetrization operator
\begin{equation}\label{symmet}
{\mathfrak S}= \frac{1+\Omega+\Omega^2 + \ldots +\Omega^\kappa}{\kappa+1}
\end{equation}
on $f$ yields $f$ when $h$ is an integral multiple of $\kappa+1$, but equals zero otherwise. As the homogeneity of $\frac{\ln Z}{\beta}$ in $b_\mu^2$ is the same as its homogeneity in $\beta^2$, it follows that the action of ${\mathfrak S}$ on this function (at the special values of $\omega_i^2$) is simply a projector onto the terms in its power series expansion proportional to $(\beta^2)^{q(\kappa+1)}$ for all integer $q$. At these special values of $\omega_i^2$ it thus follows that the LHS of \eqref{omegapartzedfull} equals 
\begin{equation}\label{LHSh}
(\kappa+1) {\mathfrak S} \circ \left(  \frac{\ln Z}{\beta} \right) 
=(\kappa+1) \sum_{q=0}^\infty (\beta^2)^{(\kappa+1)q} \left[ \frac{\ln Z}{\beta} \right]\bigg|_{(\beta^2)^{(\kappa+1)q}}~,
\end{equation} 
where $|_{\beta^n}$ denotes the coefficient of $\beta^n$. 
Inserting \eqref{LHSh} into \eqref{omegapartzedfull} and comparing powers of $\beta^2$ we conclude that 
\begin{equation}\label{spcval}
(\kappa+1) ~\times \left[ \frac{\ln Z}{\beta} \right]\bigg|_{(\beta^2)^{(\kappa+1)q}}= (-1)^\kappa \kappa !a_{2\kappa} \delta_{q,0}~,~~\forall ~q~\in \mathbb{Z}_{\geq 0}
\end{equation}
in perfect agreement with Eq.~(7) and Eq.~(22) of \cite{Advant:2026cqt}.

\refstepcounter{section}
\label{lcm}
\section{Appendix C: Large Conformal Maps}

In this Appendix, we construct three large conformal maps that cycle through $S^3\times R$, $AdS_3\times S^1$, and a second copy of $AdS_3\times S^1$ (see \eqref{conformalcycle}) closing into the $\mathbb{Z}_3$ action on the $b_i$.

Consider the metric $g_{\mu\nu}$ 
\begin{equation}
\begin{split}\label{ftmetric}
ds^2&=   d\tau^2 +d\theta^2 + \sin^2\theta d\phi_1^2 + \cos^2\theta d\phi_2^2 \\
&= d\tau^2 + \frac{dl^2}{4(1-l)l} + l d\phi_1^2 + \left(1-l\right) d\phi_2^2, \\
\end{split}
\end{equation}
where $l = \sin^2 \theta$, $l \in (0, 1)$.

\subsection{Map to $AdS_3 \times S^1$} \label{mapto}

Pulling out an overall factor of $l$, we conclude \eqref{ftmetric} $\sim$ 
\begin{equation}\label{mettransf}
\begin{split} 
ds^2&=  \frac{d \tau^2}{l}  + d \phi_1^2+ \frac{1-l}{l} d \phi_2^2 + \frac{dl^2}{4 l^2 (1-l)}  \\
\end{split}
\end{equation}
where $\sim$ denotes Weyl equivalence. Note that $\phi_1$ is non-contractible in  Eq.~\eqref{mettransf}. Changing coordinates,
\begin{equation}\label{coordchange}
\begin{split}
l&= \frac{1}{1-{\tilde l}}, ~~~~{\tilde l} \in (-\infty, 0),\\ 
ds^2&=(1-\tilde l)\,d\tau^2
+d\phi_1^2
-\tilde l\,d\phi_2^2
-\frac{d\tilde l^2}{4\tilde l(1-\tilde l)}
\end{split}
\end{equation}

The further redefinition ${\tilde l}= -\sinh^2 \rho$, $\rho\in (0, \infty)$ gives 
\begin{equation} \label{adsmet}
\begin{split}
ds^2&= d\phi_1^2+d\rho^2
+\cosh^2\rho\,d\tau^2
+\sinh^2\rho\,d\phi_2^2,  
\end{split}   
\end{equation}
the metric of (Euclidean) $AdS_3 \times S^1$. The 
codimension one boundary $ \rho \to \infty$ of the $AdS_3$ metric is a `blow up' of the codimension two surface $l=0$  on which the $\phi_1$ circle shrinks 
in the original metric  Eq.~\eqref{ftmetric}.

The CFT path integral on  $S^3 \times R$ is thus Weyl equivalent to the path integral on  $H_3 \times S^1$. The identification Eq.~\eqref{ident}
carries over to the $H_3\times S^1$ path integral, which thus computes 
\begin{equation}\label{hamint}
{\rm Tr}_{H_3 \times S^1} e^{-\beta H + \beta \omega_2 J +\beta \omega_1 Q}
\end{equation}
where $H$ is the usual global Hamiltonian of $AdS_3$, 
$J$ generates  angular momentum (translation on the boundary $S^1$) of $AdS_3$ and $Q$ is the charge on the non-contractible $S^1$ parametrized by the angle $\phi_1$. We thus conclude that the trace  Eq.~\eqref{hamint} is identical to the $S^3 \times S^1$ trace studied in this paper. We emphasize that the equality Eq.~\eqref{hamint} is a nonperturbative statement; the partition function on $S^3\times\mathbb{R}$ and on $AdS_3\times S^1$, with the identification Eq.~\eqref{ident}, are literally the same function. In \hyperref[sec:supplemental]{\texttt{SM}}, we check this equivalence for a free conformally coupled scalar. 

\subsection{Analytic continuation to $S^3\times R$}\label{ancont}

The complex variable change   
\begin{equation}\label{confdiff}
(\tau, \phi_1, \phi_2)
=\left(i\tilde{\phi}_2, -i\tilde{\tau}, \tilde{\phi}_1\right)
\end{equation}
turns the $AdS_3 \times S^1$ metric in \eqref{adsmet} to  
\begin{equation}\label{tildemetnew}
ds^2=- \left(d {\tilde \tau}^2+\frac{d\tilde l^2}{4\tilde l(1-\tilde l)}+\tilde l\,d{\tilde \phi}_1^2+(1-\tilde l)\,d{\tilde \phi}_2^2
\right)~. 
\end{equation}

Up to a Weyl transformation that removes the overall minus sign,   Eq.~\eqref{tildemetnew} has the same form in tilde coordinates, as the metric  Eq.~\eqref{ftmetric} had in the old coordinates, and so yields $S^3 \times R$. We conclude that the analytic continuation of  ${\tilde l} $ from the `physical range' listed in  Eq.~\eqref{coordchange} to the `unphysical' range ${\tilde l} \in (0, 1)$ turns 
$AdS_3\times S^1$ into a {\it second} $S^3\times R$ 
spacetime. 

Since 
\begin{equation}\label{rotvec}
\left(\partial_\tau, - i \partial_{\phi_1}, -i \partial_{\phi_2} \right) 
= \left(-i \partial_{ {\tilde \phi}_2}, \partial_{\tilde{\tau}},-i\partial_{\tilde{\phi}_1}\right),  
\end{equation}
it follows that 
\begin{equation}
\begin{split}
 v&= b_0\partial_\tau + b_1 (-i\partial_{\phi_1}) +  b_2 (-i \partial_{\phi_2}) \\&= {\tilde b}_0 \partial_{\tilde{\tau}} +{\tilde b}_1 (-i\partial_{\tilde{\phi}_1})  + {\tilde b}_2 (-i\partial_{\tilde{\phi}_2}) \\
 {\rm where }~~~ &({\tilde b}_0, {\tilde b}_1, {\tilde b_2})= (b_1, b_2, b_0)~.
\end{split}
\end{equation}

The variable $z$ defined in  Eq.~\eqref{hatzed} is related to $l$ via 
\begin{equation}\label{hzdef}
\begin{split}
{z}&= \frac{
(1-l)b^2_0b^2_1-b^2_1b^2_2+l b^2_2b^2_0
}{
b^2_0-lb^2_1+(l-1)b^2_2
}~,\\
\tilde{{z}} &= \frac{
(1-\tilde{l})\tilde{b}^2_0\tilde{b}^2_1-\tilde{b}^2_1\tilde{b}^2_2+\tilde{l} \tilde{b}^2_2\tilde{b}^2_0
}{
\tilde{b}^2_0-\tilde{l}\tilde{b}^2_1+(\tilde{l}-1)\tilde{b}_2^2
}~.\\
\end{split}
\end{equation}
Inserting  Eq.~\eqref{coordchange} into  Eq.~\eqref{hzdef}, it is easily verified that ${\tilde {z}} ={z}$. 

Now consider the quantity ${\mathcal I}$ evaluated in the original geometry. Following ${\mathcal I}$ through the coordinate transforms performed above, and using its Weyl invariance, we find
\begin{equation}\label{mathcain}
{\mathcal I}_{b_0, b_1, b_2}\left( {z}\right) ={\mathcal I}_{b_1, b_2, b_0}(z) 
\end{equation}
yielding a structural explanation for   Eq.~\eqref{identi}. 

We can now rerun this analysis with the role of $\phi_1$ and $\phi_2$ exchanged. We pull out the factor $(1-l)$ (rather than $l$) from Eq.~\eqref{ftmetric}, to obtain a second $AdS_3\times S^1$. The analogous analytic continuation  then produces a third 
$S^3 \times R$, explaining the full $\mathbb{Z}_{3}$ action. 

The workout of this Appendix generalizes in a straightforward manner to $d=2\kappa$ for arbitrary $\kappa \geq 3$. 

\refstepcounter{section}
\label{appcounting}\section{Appendix D: Counting constraints    Eq.~\eqref{itwoo}}

Completely Bose symmetric degree-$q$ homogeneous polynomials of $\big(b_0^2, b_1^2 \ldots  b_\kappa^2\big)$ are freely generated by the degree $1, 2, \ldots \kappa+1$ generators,  $E_1$, $E_2 \ldots , E_{\kappa+1}$ (see   Eq.~\eqref{invar}). Such polynomials thus constitute an  $n_q^{\rm sym}$ dimensional vector space where 
\begin{equation}\label{symcount}
\frac{1}{\prod_{i=1}^{\kappa+1} (1-t^i)} =\sum_{q=0}^\infty n_q^{\rm sym} t^q~.
\end{equation}

It follows from  Eq.~\eqref{formof} that polynomials symmetrical only under permutations of $\big(b_1^2, b_2^2, \ldots, b_\kappa^2\big)$ lie in an $n_q^{\rm b}$ dimensional vector space, with 
\begin{equation}\label{firstcount}
\frac{\sum_{i=0}^\kappa t^i} {\prod_{i=1}^{\kappa+1} (1-t^i)} =\frac{1}{(1-t)^2\prod_{i=2}^{\kappa} (1-t^i)}=\sum_{q=0}^\infty n_q^{\rm b} t^q
\end{equation}
in agreement with the `naive count',   Eq.~(C.3) of \cite{Advant:2026cqt}. 

Similarly, degree 
$q$ polynomials that take the form  Eq.~\eqref{formof},  but with $I_q^{(\kappa)}=0$, lie in an $n_q^{\rm m}$ dimensional space, with 
\begin{equation}\label{secondcount}
\begin{split}
\frac{\sum_{i=0}^{\kappa-1} t^i} {\prod_{i=1}^{\kappa+1} (1-t^i)} &=\frac{1}{(1-t)^2(1-t^{\kappa+1})\prod_{i=2}^{\kappa-1} (1-t^i)}\\&=\sum_{q=0}^\infty n_q^{\rm m} t^q
\end{split}
\end{equation}
in agreement with the local count Eq. C.4 of \cite{Advant:2026cqt}. 

By subtracting  Eq.~\eqref{secondcount} from  Eq.~\eqref{firstcount}, we see by comparison with  Eq.~\eqref{constcount} that the condition that $I^{(\kappa)}_q=0$ imposes $n_q^{\rm b}- n_q^{\rm m} =d_q$  constraints on the polynomial $T_q$ in perfect agreement with the predictions of \cite{Advant:2026cqt}. See around  Eq.~\eqref{pfandco} in  \hyperref[sec:supplemental]{\texttt{SM}}
for an $S_3$ representation take on all these formulae.

\clearpage
\onecolumngrid

\setcounter{equation}{0}
\renewcommand{\theequation}{S\arabic{equation}}

\section*{Supplemental Material}
\label{sec:supplemental}

\refstepcounter{section}
\label{app:measure}
\section{Computation of the measure factor}\label{measure}

In the coordinates Eq.~\eqref{metform}, the measure factor in  Eq.~\eqref{stpt} equals
\begin{equation}\label{hg}
\begin{split} 
d^4x\frac{\sqrt{g_4}}
{(v^2)^2}&=\frac{\cos \theta \sin \theta d \phi_1 d \phi_2 d \tau d\theta }{\beta^4(1-\omega_1^2 \sin^2\theta -\omega_2^2 \cos^2\theta)^2}\\
&= \frac{1}{2\beta^4} \frac{d \phi_1 d \phi_2 d\tau  d ( \sin^2 \theta)}{\left[1-\omega_2^2+\sin^2 \theta (\omega_2^2-\omega_1^2) \right]^2}\\
&= \frac{d \phi_1 d \phi_2 d\tau }{2\beta^4(\omega_1^2-\omega_2^2)} d \left( \frac{1}{1-\omega_2^2+\sin^2 \theta (\omega_2^2-\omega_1^2)} \right) \\
&= \frac{d \gamma^2 d \phi_1 d \phi_2 d\tau}{2\beta^4(\omega_1^2-\omega_2^2)}  \\
&= -\frac{d z d \phi_1 d \phi_2 d\tau }{2(b^2_1-b^2_2)(b^2_0-b^2_1)(b^2_0-b^2_2)} 
\end{split}
\end{equation}
where $\gamma^2$ was defined in  Eq.~\eqref{hatzed} and $z$ was defined in  Eq.~\eqref{hatzed}. 
\footnote{The paper \cite{Advant:2026cqt} used a third variable \begin{equation}
\begin{split}
    &\chi=\frac{x_1\ell}{x_2+(x_1-x_2)\ell}, ~~~(\ell=\sin^2\theta)\\
   &d\chi=\frac{x_1x_2}{(x_2+(x_1-x_2)\ell)^2}d\ell\\
   &=\frac{x_1x_2}{(x_2-x_1)}d \gamma^2
    \end{split}
\end{equation}
so that 
\begin{equation}
    \gamma^2=\frac{(1-\chi)x_1+x_2\chi}{x_1 x_2}
\end{equation} }
Consequently,  Eq.~\eqref{stpt} reduces to  Eq.~\eqref{lnzchin}.

\section{Explicit constraints at low orders}

In the main text of this paper, we have
verified that the expansion of Eq.~\eqref{omegapartzed}, at  order $\beta^{2q}$, yields $d_q$ constraints (see around   Eq.~\eqref{constcount} for a definition of $d_q$), in perfect agreement with the constraints derived and counted in \cite{Advant:2026cqt} using a different method. Recall that the  
constraints derived in \cite{Advant:2026cqt} follow from the fact that ${\mathcal I}$ (the integrand of the thermal effective action) admits a power series expansion in the 
scalars listed in Eq. \eqref{explicitf}, up to total derivatives that do not affect the final partition function. Since (as we have demonstrated around Eq. \eqref{finarg}), Eq. \eqref{omegapartzed} also follows from exactly the same 
assumptions, it follows immediately that the partition functions built out of such ${\mathcal I}$ must be a subset of the space of partition functions that obey Eq. \eqref{omegapartzed}. As the space of partition functions that obey Eq.~\eqref{omegapartzed} has exactly the same dimension as the space of partition functions with a local  ${\mathcal I}$ built out of 
the components in Eq. \eqref{explicitf}, it follows that 
Eq.~\eqref{omegapartzed} imposes no additional constraints on 
${\mathcal I}$. Consequently the space of partition functions carved out by the methods of this paper must be identical to the space of partition functions described by the paper \cite{Advant:2026cqt}.

As a sanity check, in this section we provide some direct calculational evidence for the conclusion of the previous paragraph.   We work in the simplest dimension $d=4$, and present our evidence in an expansion around the `semi-universal' limit 
\begin{equation}
	x_1,x_2 \rightarrow 0, \quad \left( \frac{x_1}{x_2} ~~{\rm and}~~\beta\right)~  \text{fixed}
\end{equation}
(see  \cite{Anand:2025mfh, Advant:2026cqt}), rather than the `high temperature' expansion in small $\beta^2$, employed in most of the rest of this paper. We make this choice for the following reason. In the high temperature expansion  Eq.~\eqref{expallord}, the  coefficient of $\beta^{2q}$ is a polynomial of degree $q$
 in the variables $x_i$. As degree $q$ polynomials are characterized by a finite number of coefficients, we find only a finite number, $d_q$ constraints,  at any given value of $q$. In the small $x_i$ `semi-universal' expansion, on the other hand,  coefficients of $x_1^m x_2^n$ are functions $h_{m, n}(\beta^2)$. \footnote{These functions admit a power series expansion in $\beta^2$. This expansion starts at order $\beta^{2m+2n}$, and generically includes terms of all higher order in $\beta^2$ (see Eq. (11) of  \cite{Advant:2026cqt}).} At every order $x_i^q$, we obtain $\left\lfloor \frac{q}{2} \right\rfloor - \left\lfloor \frac{q}{3} \right\rfloor$ new {\it functional equations} on these constraint functions at order $x_i^q$ (see \cite{Advant:2026cqt}). \footnote{These equations  relate linear combinations of $h_{m,n}(\beta^2)$ with $m+n=q$ to successively higher derivatives (w.r.t. $\beta^2$) of
$h_{m,n}(\beta^2)$ with $m+n<q$.}  Power series expanding each of these differential relations in $\beta^2$ yields an {\it infinite} number of constraints on coefficients. At a practical level, therefore, the small $x_i$ expansion is much more efficient than the small $\beta^2$ expansion; working up to order $q$ in small $x_i$ gives each of the high temperature constraints up to order $\beta^{2q}$, plus infinitely more. 

In order to extract the constraints that follow from  Eq.~\eqref{omegapartzed} in a small $x_i$ expansion, we start with the equation  Eq.~\eqref{steq} 
\begin{equation}\label{steq}
    \Big(1 + \Omega + \Omega^2\Big)\circ\left[\frac{\ln{Z}^{\rm rot} (b_0^2,b_1^2,b_2^2)}{\beta}\right] =0
\end{equation}
where the $\mathbb{Z}_{3}$ generator $\Omega$, and $(b_0, b_1, b_2)$ were defined in    Eq.~\eqref{permb}.
It is useful to define the function $H(b_0^2, b_1^2, b_2^2)$ via 
\begin{equation}\label{Hdef}
\frac{\ln Z^{\rm rot}}{\beta} = - \frac{H}{(b^2_0-b^2_1)(b^2_0-b^2_2)}~.
\end{equation} 
\footnote{We have explained in the main text that the function $H$ admits a polynomial expansion $\sum_{n=0}^\infty T_n(b_i^2)$. Since we are interested in the semi-universal rather than the high-temperature limit, we will not employ  this expansion, but instead use  Eq.~\eqref{Hexp} in what follows.} Plugging  Eq.~\eqref{Hdef} into  Eq.~\eqref{steq} yields the equation 
\begin{equation}\label{cyc}
\begin{split}
&(b^2_1-b^2_2) ~H\left( b^2_0, b^2_1, b^2_2 \right) + (b^2_2-b^2_0) ~H\left( b^2_1, b^2_2, b^2_0 \right) + (b^2_0-b^2_1) ~H\left( b^2_2, b^2_0, b^2_1\right)=0~.
\end{split}
\end{equation}
We now choose to view $H$ as a function of $(\beta^2, x_1, x_2)$ rather than $(b^2_0, b^2_1, b^2_2)$. With this presentation,  Eq.~\eqref{cyc} becomes 
\begin{equation}\label{H_consteaint}
(x_2 - x_1)H(\beta^2,x_1,x_2) - x_2H(\beta^2(1-x_1),\Omega x_1,\Omega x_2) + x_1H(\beta^2(1-x_2),\Omega^2 x_1,\Omega^2 x_2) = 0~, 
\end{equation}
where $\Omega x_i$ denotes the action of $\Omega$ on $x_i$, and takes the explicit form 
\begin{equation}\label{rrce}
    \begin{split}
        &\Omega (x_1,x_2) = \left(\frac{x_2 - x_1}{1 - x_1},\frac{x_1}{x_1-1}\right)~,
        \\&\Omega^2 (x_1,x_2) = \left(\frac{x_2}{x_2-1},\frac{x_1 - x_2}{1 - x_2}\right)~.
    \end{split}
\end{equation}
In the second and third terms in  Eq.~\eqref{H_consteaint}, the first argument $\beta^2$ has been replaced by $\beta^2(1-x_1)$ and $\beta^2(1-x_2)$ respectively. Consequently the functional relation  Eq.~\eqref{H_consteaint} is effectively `non-local' in $\beta^2$ (i.e. the three different terms in that equation are evaluated at different values of $\beta^2$). In the small $x_i$ expansion, however, the difference in arguments is small, and can be Taylor expanded. For any function $F(\beta^2)$, 
\begin{equation}\label{thetadef} 
    F(\beta^2 (1 - x_i)) = (1 - x_i)^{\theta}F(\beta^2)~, \quad \theta = \beta^2\frac{d}{d\beta^2}
\end{equation}
(one way to verify this equation is to expand $F$ in a Taylor series and explicitly work out the RHS). Consequently,  Eq.~\eqref{H_consteaint} can be rewritten in the form 
\begin{equation}\label{almofineq}
	(x_2 - x_1)H(\beta^2,x_1,x_2) - x_2 (1-x_1)^\theta H(\beta^2,\Omega x_1,\Omega x_2) + x_1 (1-x_2)^\theta  H(\beta^2,\Omega^2 x_1,\Omega^2 x_2) = 0 
\end{equation}
where
\begin{equation}\label{formalx}
(1-x_i)^\theta= 1- \theta x_i  +\frac{\theta(\theta-1)}{2!} x_i^2	
- \frac{\theta(\theta-1)(\theta -2)}{3!} x_i^3 + \ldots
\end{equation} 
and $\theta$ is the differential operator given in  Eq.~\eqref{thetadef}.

We now follow \cite{Advant:2026cqt} (See Eq. (11) of \cite{Advant:2026cqt}) to expand 
\begin{equation}\label{Hexp}
H= \sum_{m, n=0}^{\infty} h_{m, n}(\beta^2)x_1^m x_2^n
\end{equation} 
with $h_{m,n}(\beta^2)=h_{n,m}(\beta^2)$. 
Plugging  Eq.~\eqref{Hexp} into  Eq.~\eqref{almofineq}, Taylor expanding $\Omega x_i$ and $\Omega^2 x_i$ in $x_i$, collecting terms and requiring the vanishing of the coefficients of every independent monomial $x_1^m x_2^n$
yields several constraint equations. The equation corresponding to the coefficient of monomials with $m+n = p$ takes the form 
\begin{equation}\label{formofconsti}
\sum_{\substack{a,b\ge0, ~a \geq b\\a+b\le p}} D_{a, b} \left[  h_{a, b}(\beta^2) \right]=0~.
\end{equation} 	
Here $D_{a, b}$ is a polynomial in $\theta$ that is generically of order $p-(a+b)$, and so is a differential operator in $\beta$ that is generically of the same order.

We have implemented this procedure for all monomials with $p=m+n \leq 7$.  \footnote{This procedure is easy to implement in Mathematica as $\theta$ can be treated as a commuting variable in all computations.} The operators $D_{a, b}$ that result from this process are listed in  Eq.~\eqref{supp_tlist10}; this table can be read as follows. Each row in  Eq.~\eqref{supp_tlist10} corresponds to a constraint, at the value of $p$ denoted in the first column of that table. The entries in each row list the values of $D_{a, b}$ for $a+b \leq p$ in the following `dictionary' order:
\begin{equation}\label{orderlist}
(b, a)=(0,0), ~(0, 1), \ldots (0, p), ~(1,1), ~(1, 2) \ldots (1,p-1), (2,2)  \ldots 
\end{equation}  
In other words we first list $D_{a,b}$ with $b= 0$  and $a$ ranging from $0$ to $p$. Then we do the same for $b=1$ and $a$ ranging from $1$ to $p-1$ and so on. The total number of
distinct values of $D_{a,b}$ equals 
\begin{equation}\label{noterms}
(p+1)+(p-1)+(p-3) + \ldots \left(p+1-2 \left[\frac{p+1}{2}\right]\right)=\left\lfloor
\frac{(p+2)^2}{4}
\right\rfloor.
\end{equation}
So we have $4$ terms at $p=2$, $9$ terms at $p=4$, $12$ terms at $p=5$, 16 terms at $p=6$, 20 terms at $p=7$. These terms are all listed in the corresponding rows of  Eq.~\eqref{supp_tlist10}.

As an example of the use of the table  Eq.~\eqref{supp_tlist10}, we use the $p=2$ line of this table to read off the form of the constraint at $p=2$, which then takes the form 
\begin{equation}\label{firstconst}
\begin{split}
&\binom{\theta}{2}h_{0,0}+3\binom{\theta-1}{1}h_{1,0}+3\left(h_{1,1}+h_{2,0}\right)=0\\\Rightarrow\quad&\left(\beta^2\frac{d}{d\beta^2} - 1\right)\left(\beta^2\frac{d}{d\beta^2}\right)h_{0,0} + 6\left(\beta^2\frac{d}{d\beta^2}-1\right)h_{1,0} + 6(h_{2,0} + h_{1,1}) = 0  
\end{split}
\end{equation}
\eqref{firstconst} is in perfect agreement with Eq.~(16) of  \cite{Advant:2026cqt} after setting $a=0$ in that equation (this is appropriate since we are studying 
$\ln Z^{\rm rot}$ rather than $\ln Z$ in this section). 
\begin{equation}\label{supp_tlist10} 
  \begin{tabular}{||c|c||}
  \hline
  $p^{\text{th}}$ sub-leading order & Constraint \\
  \hline 
  $p = 0$ & NA  \\ 
  \hline
  $p = 1$ & NA \\ 
  \hline
  $p = 2$ & $\left( \binom{\theta}{2},3\binom{\theta-1}{1},3,3\right)$ \\
  \hline
  $p = 3$ & NA \\  
  \hline
  $p = 4$ & $\left( \binom{\theta}{4}, 2\binom{\theta-1}{3}, 2\binom{\theta-2}{2}, 3\binom{\theta-3}{1}, 5, \binom{\theta-2}{2}, 2\binom{\theta-3}{1}, 4, 1 \right)$\\ 
  \hline
  $p = 5$ & $\left( \binom{\theta}{5}, 2\binom{\theta-1}{4}, 2\binom{\theta-2}{3}, 2\binom{\theta-3}{2}, \binom{\theta-4}{1}, -2, \binom{\theta-2}{3}, 2\binom{\theta-3}{2}, 2\binom{\theta-4}{1},0, \binom{\theta-4}{1}, 2 \right)$\\ 
  \hline
  $p = 6$ & $\begin{aligned}
  \left( \binom{\theta}{6}, 2\binom{\theta-1}{5}, 2\binom{\theta-2}{4}, 2\binom{\theta-3}{3}, 2\binom{\theta-4}{2}, 3\binom{\theta-5}{1}, 7, \binom{\theta-2}{4}, \right. \\
  \left. 2\binom{\theta-3}{3}, 2\binom{\theta-4}{2}, 2\binom{\theta-5}{1}, 4, \binom{\theta-4}{2}, 2\binom{\theta-5}{1}, 2, 1 \right)
  \end{aligned}$\\ 
  \hline
  $p = 7$ & $\begin{aligned}
  \left( \binom{\theta}{7}, 2\binom{\theta-1}{6}, 2\binom{\theta-2}{5}, 2\binom{\theta-3}{4}, 2\binom{\theta-4}{3}, 2\binom{\theta-5}{2}, \binom{\theta-6}{1}, -4, \binom{\theta-2}{5}, 2\binom{\theta-3}{4}, \right. \\
  \left. 2\binom{\theta-4}{3}, 2\binom{\theta-5}{2}, 2\binom{\theta-6}{1},0, \binom{\theta-4}{3}, 2\binom{\theta-5}{2}, 2\binom{\theta-6}{1}, 2, \binom{\theta-6}{1}, 2 \right)
  \end{aligned}$\\ 
  \hline
  \end{tabular}  
\end{equation}

As we have explained above, the table  Eq.~\eqref{supp_tlist10} lists the form of the semi-universal constraints (for $p \leq 7)$ obtained from the Taylor expansion of  Eq.~\eqref{almofineq}. We now compare these constraints with those obtained using the methods of \cite{Advant:2026cqt}, i.e. by proceeding as follows. We simply plug the expansion Eq. (S.13)
of \cite{Advant:2026cqt} into  Eq.~\eqref{lnzchin}, and then plug
\eqref{explicitf} into the resultant equation, and perform the integral over $z$. As the integrand is a polynomial in $z$, the integral is very easy to perform, and yields an $H$ function of the form  Eq.~\eqref{Hexp}. While the number of distinct functions $h_{m, n}(\beta^2)$  with $m+n=p$ equals $\left\lfloor \frac{p}{2}+1\right\rfloor $, the number of local terms (in Eq. (S.13) of \cite{Advant:2026cqt}) that contribute to the partition function at this order equals 
the smaller number $\left\lfloor \frac{p}{3}+1\right\rfloor $. Summing these numbers, we see that while the number of distinct $h_{m, n}(\beta^2)$  with $m+n \leq p$  equals 
$\left \lfloor\frac{(p+2)^2}{4} \right\rfloor$, the number of local structures (i.e. terms listed in Eq. (S.13) of \cite{Advant:2026cqt} with $m+3n \leq 
p$) equals $\left\lfloor \frac{(p+2)(p+3)}{6}\right\rfloor$. 

Consequently, the integral described above can be thought of as generating a rectangular matrix with $\left\lfloor \frac{(p+2)^2}{4}\right\rfloor $ rows and $\left\lfloor \frac{(p+2)(p+3)}{6}\right\rfloor $ columns. The matrix in question acts on the $\left\lfloor \frac{(p+2)(p+3)}{6}\right\rfloor $ length column of free local functions that appear in Eq. (S.13) of \cite{Advant:2026cqt}, and creates a $\left\lfloor \frac{(p+2)^2}{4}\right\rfloor$ length column of functions $h_{m,n}(\beta^2)$ with $m+n \leq p$. 
\footnote{The matrix in question is built out of the differential operators $\theta$, but since all terms in the matrix are independent of $\beta$, the matrix can be manipulated as if $\theta$ were a number.} 

Now any matrix with $r$ rows and $c$ columns has a `left null space' of dimension at least $r-c$, i.e. an $r-c$ dimensional  subspace of rows which, when acting on the matrix from the left simply gives zero. In the situation described above 
$r-c=\left\lfloor \frac{(p+2)^2}{4}\right\rfloor-\left\lfloor \frac{(p+2)(p+3)}{6}\right\rfloor $ which $>0$ for $p\geq 2$. Consequently, the constraints generated by this method also take the form  Eq.~\eqref{formofconsti}, with $D_{a, b}$ given by (any arbitrary) basis vectors for this null space. It is easy to convince oneself that if $w$ is a null vector of the matrix 
described above at some $p=p'$, it remains a null vector at every $p \geq p'$. We have used Mathematica to perform the 
integrals necessary to evaluate the matrices above up to $p=7$, and have determined the explicit form of all left null vectors. Each increase in $p$ up to $p=7$ increases the dimension of the null space by no more than unity, and we have verified that the row vectors listed in table  Eq.~\eqref{supp_tlist10} are, indeed, the row vectors that we need to add (to the row vectors listed at every lower level) to correctly generate the null space of the matrix described above at the given value of $p$. 

We have thus explicitly verified that the semi-universal constraints that follow from  Eq.~\eqref{steq} are identical 
to those obtained using the methods of \cite{Advant:2026cqt}, at least for $p\leq 7$.

\section{More about $\ln Z^{\rm sh}$}

In Eq. \eqref{expallord} we have presented a conjecture for the form of the quantity $\ln Z^{\rm sh}$. In this section we explain how we arrived at this conjecture. 

The explicit form of the trace of the stress tensor in terms of the various Weyl anomalies is, of course, very well known. By integrating this expression, it is, in principle, not difficult to evaluate $\ln Z^{\rm sh}$ in any given dimension. However these computations grow increasingly complicated as the dimension is increased. In order to circumvent this complication, we proceeded as follows. We first used Claude Opus 5.5 to compute $\frac{\ln Z^{\text{sh}}}\beta$ for $d$ up to $16$ exactly. From this it proposed the general form Eq. \eqref{expallord}, which reproduced all 65 exact coefficients for $d \leq 16$. Finally, we checked this general $d$ prediction for $d = 18, 20$ numerically to very high precision, again with assistance from Claude Opus 5.5.

In low dimensions, we have also compared the expression in Eq. \eqref{expallord} against results in the published literature. In $d=4, 6, 8$, the $s$-integral in the expression for $\ln Z^{\rm sh}$ in Eq. \eqref{expallord} is easily evaluated, yielding the following more explicit versions of our conjectured expression
\begin{equation}
\begin{split}
 \frac{\ln Z^{\rm sh}}{\beta}\Bigg|_{d=4}&=
\frac{a_4}{12}
\left[
\frac{x_1}{x_2}
+
\frac{x_2}{x_1}
+
7
\right]~,\\  \frac{\ln Z^{\rm sh}}{\beta}\Bigg|_{d=6}&=\frac{3a_6}{80}
\left[
\frac{x_1^2}{x_2x_3}
+\frac{x_2^2}{x_1x_3}
+\frac{x_3^2}{x_1x_2}
-4\left(
\frac{x_1}{x_2}
+\frac{x_1}{x_3}
+\frac{x_2}{x_1}
+\frac{x_2}{x_3}
+\frac{x_3}{x_1}
+\frac{x_3}{x_2}
\right)
-29
\right]~,\\ \frac{\ln Z^{\rm sh}}{\beta}\Bigg|_{d=8}&=\frac{a_8}{560}\Bigg[
15\Bigg(
\frac{x_1^3}{x_2x_3x_4}
+\frac{x_2^3}{x_1x_3x_4}
+\frac{x_3^3}{x_1x_2x_4}
+\frac{x_4^3}{x_1x_2x_3}
\Bigg)-55\Bigg(
\frac{x_1^2}{x_2x_3}
+\frac{x_1^2}{x_2x_4}
+\frac{x_1^2}{x_3x_4}
\\
&\qquad
+\frac{x_2^2}{x_1x_3}
+\frac{x_2^2}{x_1x_4}
+\frac{x_2^2}{x_3x_4}
+\frac{x_3^2}{x_1x_2}
+\frac{x_3^2}{x_1x_4}
+\frac{x_3^2}{x_2x_4}
+\frac{x_4^2}{x_1x_2}
+\frac{x_4^2}{x_1x_3}
+\frac{x_4^2}{x_2x_3}
\Bigg)
\\
&\qquad\quad+29\Bigg(
\frac{x_1x_2}{x_3x_4}
+\frac{x_1x_3}{x_2x_4}
+\frac{x_1x_4}{x_2x_3}
+\frac{x_2x_3}{x_1x_4}
+\frac{x_2x_4}{x_1x_3}
+\frac{x_3x_4}{x_1x_2}
\Bigg)
+211\Bigg(
\frac{x_1}{x_2}
+\frac{x_1}{x_3}
+\frac{x_1}{x_4}
\\
&\qquad\qquad
+\frac{x_2}{x_1}
+\frac{x_2}{x_3}
+\frac{x_2}{x_4}
+\frac{x_3}{x_1}
+\frac{x_3}{x_2}
+\frac{x_3}{x_4}
+\frac{x_4}{x_1}
+\frac{x_4}{x_2}
+\frac{x_4}{x_3}
\Bigg)+1569
\Bigg]~.
\end{split}    
\end{equation}
The above three expressions for $d=4,6,8$ match exactly with the equations (44), (45) and (49)\footnote{There is a typographical inconsistency in the form
$+55\,\Xi_d^3\Xi_c$ of equation (49), whereas permutation symmetry requires the same
$-55$ coefficient as for the other terms of the form $x_i^3x_j$. After correcting that sign, our expression matches exactly. } of the paper \cite{Olavarria:2013xqa}. It should be noted that $\frac{\ln Z^{\rm sh}}{\beta}\Big|_{d=4,6,8}=E^{(d=4,6,8)}_{\rm vac}$ in the convention of \cite{Olavarria:2013xqa} and $a_4=\frac{\pi}{8G_5},~a_6=\frac{\pi^2}{48G_7},~a_8=\frac{\pi^3}{576 G_9}$ as explained in Eq.~\eqref{bhident}.

\section{The $\mathbb{Z}_{\kappa+1}$ orbit sum over $(\ln Z^{\rm sh})/\beta$}

In this section we evaluate the orbit sum of $\frac{\ln Z^{\rm sh}}{\beta}$ for a general $\kappa$. In terms of the variables $b_i$ for $i=0,1,\ldots,\kappa$, the expression becomes 
\begin{equation}
	\frac{\ln Z^{\rm sh}}{\beta}= \kappa! a_{2\kappa}
	\left[
	(-1)^\kappa+
	\frac{1}{2^{2\kappa-1}}
	\sum_{i=1}^{\kappa}
	\int_0^1 ds\,
	\prod_{\substack{j=1\\ j\neq i}}^{\kappa}
	\frac{
		\left[(s-2)b_0^2+2b_j^2-sb_i^2\right]^2
	}{
		\left(b_0^2-b_j^2\right)
		\left(b_j^2-b_i^2\right)
	}
	\right]~.
\end{equation}   
Applying $\Omega^p$ on the above equation, we get
\begin{equation}\label{orbitcasimir}
	\Omega^p\circ\left[\frac{\ln Z^{\rm sh}}{\beta}\right]= \kappa! a_{2\kappa}
	\left[
	(-1)^\kappa+
	\frac{1}{2^{2\kappa-1}}
	\sum_{\substack{i=0\\i\neq p}}^{\kappa}
	\int_0^1 ds\,
	\prod_{\substack{j=0\\j\neq i,p}}^{\kappa}
	\frac{
		\left[(s-2)b_p^2+2b_j^2-sb_i^2\right]^2
	}{
		\left(b_p^2-b_j^2\right)
		\left(b_j^2-b_i^2\right)
	}
	\right]~,\quad\mathrm{for}~~p=0,1,\cdots,\kappa~.
\end{equation} 
Before going into the actual proof, we will define notation. Set
\begin{equation}\label{ql}
Q(z):=\prod_{i=0}^{\kappa}\big(z-b^2_i\big)\,,\qquad
L_i(z):=\prod_{\substack{ j=0\\j\neq i}}^\kappa\frac{z-b_j^2}{b_i^2-b_j^2}=\frac{Q(z)}{(z-b_i^2)\,Q'(b^2_i)}\,,
\end{equation}
so that $L_i(z)$ is a basis (Lagrange basis) polynomial, with $L_i(b^2_j)=\delta_{ij}$. We will apply the residue theorem to the following function
\begin{equation}
f(\zeta)=\frac{1}{(z-\zeta)\,Q(\zeta)}\,.
\end{equation}
Its residues are $\dfrac{1}{(z-b_i^2)Q'(b^2_i)}$ at $\zeta=b^2_i$ and $-\dfrac{1}{Q(z)}$ at $\zeta=z\neq b_i^2$ for all $i=0,1,\cdots,\kappa$.
The residue at infinity vanishes because $f(\zeta)\sim-\zeta^{-\kappa-2}$. Since the residues sum to zero,
\begin{equation}\label{eq:pou}
\sum_{i=0}^{\kappa}L_i(z)=1~.
\end{equation}
Now consider the distinguished variable in slot $p$ (for $\frac{\ln Z^{\rm sh}}{\beta}$ itself, $p=0$), take $i\neq p$, and define
\begin{equation}
z=b^2_p+\tfrac{s}{2}\big(b^2_i-b^2_p\big)~.
\end{equation}
The integration range of $s$ from $0$ to $1$, becomes the range of $z$ from $b_p^2$ to $\tfrac12(b^2_p+b^2_i)$. Then
\begin{align}
(s-2)b^2_p+2b^2_j-s\,b^2_i&=2\big(b^2_j-z\big)\,,\\
\prod_{j\neq p,i}(b^2_j-z)^2&=\frac{Q(z)^2}{(z-b^2_p)^2(z-b^2_i)^2}\,,\\
\prod_{j\neq p,i}\big(b^2_p-b^2_j\big)\big(b^2_j-b^2_i\big)&=(-1)^\kappa\,\frac{Q'(b^2_p)\,Q'(b^2_i)}{\big(b^2_i-b^2_p\big)^2}\,.
\end{align}
The integrand of \eqref{orbitcasimir} therefore becomes
\begin{equation}
\frac{1}{2^{2\kappa-1}}\prod_{j\neq p,i}\frac{
		\left[(s-2)b_p^2+2b_j^2-sb_i^2\right]^2
	}{
		\left(b_p^2-b_j^2\right)
		\left(b_j^2-b_i^2\right)
	}=\frac{(-1)^\kappa}{2}\big(b^2_i-b^2_p\big)^2\frac{L_p(z)L_i(z)}{\big(z-b^2_p\big)\big(z-b^2_i\big)}\,.
\end{equation}
Changing variables with $ds=2\,dz/\big(b^2_i-b^2_p\big)$, the prefactor $\frac{1}{2}$ cancels
completely, and using partial fraction decomposition of
$\dfrac{b^2_i-b^2_p}{\big(z-b^2_p\big)\big(z-b^2_i\big)}$, we get
\begin{equation}
\begin{split}
\frac{1}{2^{2\kappa-1}}
	\int_0^1 ds\,
	\prod_{j\neq i,p}
	\frac{
		\left[(s-2)b_p^2+2b_j^2-sb_i^2\right]^2
	}{
		\left(b_p^2-b_j^2\right)
		\left(b_j^2-b_i^2\right)
	}
=(-1)^\kappa\int_{b^2_p}^{\tfrac12(b^2_p+b^2_i)}dz\,L_p(z)L_i(z)\left[\frac{1}{z-b^2_i}-\frac{1}{z-b^2_p}\right]\,.
\end{split}
\end{equation}
We will use the following identity
\begin{equation}
\frac{L'_i(z)}{L_i(z)}=\frac{Q'(z)}{Q(z)}-\frac{1}{z-b^2_i}~,\quad\mathrm{for}~~i=0,1,\cdots,\kappa~. 
\end{equation}
So we can write
\begin{equation}
\frac{1}{2^{2\kappa-1}}
	\int_0^1 ds\,
	\prod_{j\neq i,p}
	\frac{
		\left[(s-2)b_p^2+2b_j^2-sb_i^2\right]^2
	}{
		\left(b_p^2-b_j^2\right)
		\left(b_j^2-b_i^2\right)
	}
= (-1)^\kappa\int_{b^2_p}^{\tfrac12(b^2_p+b^2_i)}dz\,\Big[L'_p(z)L_i(z)-L_p(z)L_i'(z)\Big]~.   
\end{equation}
We will sum the indices $i,p$ in the above equation and get
\begin{equation}
\begin{split}
&\sum_{\substack{i,p\\i\neq p}}\frac{1}{2^{2\kappa-1}}
	\int_0^1 ds\,
	\prod_{j\neq i,p}
	\frac{
		\left[(s-2)b_p^2+2b_j^2-sb_i^2\right]^2
	}{
		\left(b_p^2-b_j^2\right)
		\left(b_j^2-b_i^2\right)
	}
\\&= \frac{(-1)^\kappa}{2}\sum_{\substack{i,p\\i\neq p}}\left[\int_{b^2_p}^{\tfrac12(b^2_p+b^2_i)}dz\,\Big[L'_p(z)L_i(z)-L_p(z)L_i'(z)\Big]+ \int_{b^2_i}^{\tfrac12(b^2_p+b^2_i)}dz\,\Big[L'_i(z)L_p(z)-L_i(z)L_p'(z)\Big]\right]\\&= \frac{(-1)^\kappa}{2}\sum_{\substack{i,p\\i\neq p}}\int_{b^2_p}^{b_i^2}dz\,\Big[L'_p(z)L_i(z)-L_p(z)L_i'(z)\Big]\\&=(-1)^\kappa\sum_{i,p=0}^{\kappa}\int_{b^2_p}^{b_i^2}dz\,L'_p(z)L_i(z)  
\end{split}    
\end{equation}
where in the second step of the manipulation, we have symmetrized the index $(i,p)$ by introducing a half factor in front and in the third step, we have eliminated the upper limit $\tfrac12(b^2_p+b^2_i)$ from the two integrals. The constraint $i\neq p$ goes away because the summand for $i=p$ in the third step vanishes. Consider the summand integral and split it at $z=c\neq b_i^2$ such that
\begin{equation}
\int_{b^2_p}^{b_i^2}dz\,L'_p(z)L_i(z) =\int_{c}^{b_i^2}dz\,L'_p(z)L_i(z)-\int_{c}^{b_p^2}dz\,L'_p(z)L_i(z)~.
\end{equation}
Hence the summation can be written as
\begin{equation} \label{sumsum}
\begin{split}
\sum_{i,p=0}^{\kappa}\int_{b^2_p}^{b_i^2}dz\,L'_p(z)L_i(z)&=\sum_{i=0}^{\kappa}\int_{c}^{b_i^2}dz\,\left(\sum_{p=0}^\kappa L_p(z)\right)'L_i(z)-\sum_{p=0}^{\kappa}\int_{c}^{b_p^2}dz\,L'_p(z)\left(\sum_{i=0}^\kappa L_i(z)\right)\\&=0-\sum_{p=0}^\kappa\Big[L_p(b_p^2)-L_p(c)\Big]=-\kappa~.     
\end{split}   
\end{equation}
Summing over $p$ in Eq.~\eqref{orbitcasimir}, we have
\begin{equation}
\Big(1+\mathrm{\Omega}+\cdots+\mathrm{\Omega}^\kappa\Big)\circ\left[\frac{\ln Z^{\rm sh}}{\beta}\right]=(-1)^\kappa \kappa!a_{2\kappa}\big(\kappa+1-\kappa\big)=(-1)^\kappa \kappa!a_{2\kappa}~,
\end{equation}
which is our cyclic identity for the anomalous part.

\section{Simple symmetrization identities}

In the last step in  Eq.~\eqref{symop} we have used  
\begin{equation}\label{sbot}
\begin{split}
&3\mathfrak{S}\circ\left(b^2_1-b^2_2\right)
=(b^2_1-b^2_2)+(b^2_2-b^2_0) +(b^2_0-b^2_1)=0~,\\
&3\mathfrak{S}\circ\left(b^2_1b^2_0-b^2_2b^2_0\right)= (b^2_1b^2_0 -b^2_2 b^2_0) + (b^2_2b^2_1-b^2_0b^2_1) +(b^2_0b^2_2-b^2_1b^2_2)
=0~,\\
&3\mathfrak{S}\circ\left(b^2_1b^4_0-b^2_2b^4_0\right)= (b^2_1b^4_0 -b^2_2 b^4_0) + (b^2_2b^4_1-b^2_0b^4_1) +(b^2_0b^4_2-b^2_1b^4_2)=-(b^2_0-b^2_1)(b^2_1-b^2_2)(b^2_2-b^2_0)= -\frac{2 \pi^2}{P}~.
\end{split}
\end{equation}
Equating the coefficients of distinct powers of $\beta^{2q}$ in the last of  Eq.~\eqref{symop}
yields  Eq.~\eqref{itwoo}.

\subsection{The vanishing of $I_q^{(\kappa)}$ in $2\kappa$ dimensions for $\kappa\geq 2$}

In this Appendix we will demonstrate that the first of 
Eq.~\eqref{omegapartzed} implies \eqref{itwoo} is true in arbitrary even dimensions, i.e. at arbitrary integer values of $\kappa$. 

From \eqref{expallord} it follows that 
\begin{equation}\label{expahd}
\frac{\ln Z^{\rm rot}}{\beta} =-\frac{1}{\prod_{i=1}^\kappa (b_0^2-b_i^2)} \left( \sum_{q=0}^\infty  T_q \right)~.  
\end{equation}
The denominator in \eqref{expahd} does not transform in a particularly simple manner under either $\mathbb{Z}_{\kappa +1}$ or the full $S_{\kappa +1}$ permutation group. We can fix this by multiplying the numerator and denominator of this expression by 
\begin{equation}\label{deltanotdef}
\mathcal{V}_0 =\det
\begin{pmatrix}
1 & 1 & \cdots & 1 \\
b_1^2 & b_2^2 & \cdots & b_\kappa^2 \\
\vdots & \vdots & \ddots & \vdots \\
b_1^{2(\kappa-1)} & b_2^{2(\kappa-1)} & \cdots & b_\kappa^{2(\kappa-1)}
\end{pmatrix}
=
\prod_{1\leq i<j\leq\kappa}
\left(b_j^2-b_i^2\right)~. 
\end{equation}
On doing this, \eqref{expahd} can be rewritten as 
\begin{equation}\label{expahdpo}
\frac{\ln Z^{\rm rot}}{\beta} =\frac{(-1)^{\kappa+1}}{\mathcal{V}} \left( \sum_{q=0}^\infty \mathcal{V}_0  T_q \right)~,  
\end{equation}
where the denominator is now the Vandermonde
\begin{equation}\label{deltadef}
\mathcal{V} =\det
\begin{pmatrix}
1 & 1 & \cdots & 1 \\
b_0^2 & b_1^2 & \cdots & b_\kappa^2 \\
\vdots & \vdots & \ddots & \vdots \\
b_0^{2\kappa} & b_1^{2\kappa} & \cdots & b_\kappa^{2\kappa}
\end{pmatrix}
=
\prod_{0\leq i<j\leq\kappa}
\left(b_j^2-b_i^2\right)~,
\end{equation}
which lies in the completely antisymmetric representation of $S_{\kappa +1}$. $\Omega$, the $\mathbb{Z}_{\kappa +1}$ generator, can be thought of as a product of $\kappa$ interchanges, and so its action on the completely antisymmetric object $\mathcal{V}$ is given by 
\begin{equation}\label{omegadelta}
\Omega\circ \mathcal{V} = (-1)^{\kappa} \mathcal{V}~. 
\end{equation}
For later use we also define 
\begin{equation}
    \mathcal{V}_a = \prod_{\substack{0\leq i<j\leq\kappa\\ i,j\neq a}}
\big(b_j^2 - b_i^2\big)~.
\end{equation}
Inserting \eqref{formof} into \eqref{expahdpo} we see that 
$\frac{\ln Z^{\rm rot}}{\beta}$ can be written as 
\begin{equation}\label{expahdpon}
\frac{\ln Z^{\rm rot}}{\beta} =\frac{(-1)^{\kappa+1}}{\mathcal{V}} \left( \sum_{q=0}^\infty \mathcal{V}_0  \left(\sum_{m=0}^\kappa b_0^{2m} I_q^{(m)} \right) \right)~.  
\end{equation}
We now insert \eqref{expahdpo} into \eqref{omegapartzed}. 
Using \eqref{omegadelta}, the permutation invariance of $I_q^{(m)}$, we obtain 
\begin{equation}\label{actong}
\Omega^a\circ \left[\frac{\mathcal{V}_0}{\mathcal{V}}\right]=(-1)^a \frac{\mathcal{V}_a}{\mathcal{V}}~,
\end{equation}
\begin{equation} \label{actgon}
\mathfrak{S}\circ \left[\frac{\ln Z^{\rm rot}}{\beta}\right] 
= \frac{(-1)^{\kappa+1}}{\mathcal{V}(\kappa+1)} \sum_{q=0}^\infty \sum_{m=0}^\kappa\sum_{a=0}^{\kappa} \left[ \left(\mathcal{V}_a (-1)^a I_q^{(m)} b_a^{2m}  \right) \right]~. 
\end{equation} 
Where 
\begin{equation}\label{symdef}
{\mathfrak S}=\frac{1}{\kappa+1} \sum_{a=0}^\kappa \Omega^a~. 
\end{equation}
Now recall that any $n\times n$ matrix $A_{ij}$ obeys the identity 
\begin{equation}\label{idenmat}
\sum_{j=1}^n (-1)^{j+n} A_{mj} {\hat A}_{nj} = \delta_{m n} \det (A)
\end{equation}
where ${\hat A}_{ij}$ is the minor of the element $A_{ij}$. Applying the identity \eqref{idenmat} to the matrix listed in \eqref{deltadef} with $m=0, 1, 2 \ldots, \kappa-1$ and $n=\kappa$ we obtain 
\begin{equation}\label{eqonvand}
\sum_{a=0}^{\kappa} b_a^{2n} \mathcal{V}_a (-1)^a =   0 
~~~~(n=0,1, 2, \ldots \kappa-1)~.
\end{equation}
Applying the same identity with $m=\kappa$ gives 
\begin{equation}\label{eqonvandspc}
\sum_{a=0}^{\kappa} b_a^{2\kappa} (-1)^a\mathcal{V}_a   
= (-1)^\kappa\mathcal{V}~.
\end{equation}
Inserting \eqref{eqonvand} and \eqref{eqonvandspc} into 
\eqref{actgon}, we see that the coefficients of $I_q^{(0)},$
$I_q^{(1)}$, $\ldots $ $I_q^{(\kappa-1)}$ all vanish, and 
\eqref{actgon} simplifies to 
\begin{equation} \label{actgonfin}
\mathfrak{S}\circ \left[\frac{\ln Z^{\rm rot}}{\beta}\right] 
= - \frac{1}{\kappa+1} \sum_{q=0}^\infty I_q^{(\kappa)}~. 
\end{equation}
Expanding \eqref{actgonfin} term by term, (and using that $I_q^{(\kappa)}$ is homogeneous of degree $(q-\kappa)$ in $b_\mu^2$, $\mu=0\ldots \kappa$), we arrive at 
\eqref{itwoo}.

\section{Relation to $S_3$ representation theory}\label{app:rep}

We have demonstrated in the main text that if the $\ln Z^{\rm rot}$ of a CFT$_4$ is well described by a local thermal effective action, then it is constrained to obey  Eq.~\eqref{finarg}, which asserts that 
\begin{equation}\label{finaragain}
\mathfrak{S} \circ \left[  \frac{\ln Z^{\rm rot}}{\beta} \right]=0,
\end{equation} 
where $\mathfrak{S}$
is defined in  Eq.~\eqref{pdefg}. In this section we will view this condition (and resultant equations) from the lens of $S_3$ representation theory. 

The permutation group $S_3$ acts on any  function of $(b^2_0, b^2_1, b^2_2)$ in the obvious way (by permuting the $b^2_i$). $S_3$ has three irreducible representations, namely the one dimensional ${\bf 1}_S$ totally symmetric and the ${\bf 1}_A$ totally antisymmetric representations, together with the two dimensional mixed representation 
${\bf 2}_M$. \footnote{See e.g. \cite{Choudhury:2018iwf} for a review of aspects of $S_3$ representation theory relevant to this section.}
A little thought will convince the reader that the quantities 
\begin{equation}\label{projector}
\begin{split}
P_{\mathbf{1}_{S}}
&=
\frac{1}{6}
\left[
1+(01)+(02)+(12)+(012)+(021)
\right]~,\\
P_{\mathbf{1}_{A}}
&=
\frac{1}{6}
\left[
1-(01)-(02)-(12)+(012)+(021)
\right]~,\\
P_{\mathbf{2}_M}
&=
1-P_{\mathbf{1}_S}-P_{\mathbf{1}_A}
=
\frac{1}{3}
\left[
2-(012)-(021)
\right]~.
\end{split}
\end{equation}
act as projectors onto the relevant representation spaces. In  Eq.~\eqref{projector} we have used the following notation: $1$ is the identity permutation, $(ij)$ is the permutation that interchanges $i$ and $j$, and $(ijk)$ is the permutation that implements  $\left(i\rightarrow j\rightarrow k \rightarrow i\right)$. The RHS of the first of  Eq.~\eqref{projector} is one sixth the sum over all 6 permutations in $S_3$, and so projects onto the totally symmetric representation. 
The RHS of the second of  Eq.~\eqref{projector} is one sixth the signed sum over all 6 elements of the permutation group, and so projects onto the totally antisymmetric representation. The third line of  Eq.~\eqref{projector} projects onto whatever is left after these two representations have been removed, and so projects onto the ${\bf 2}_M$ representation. In this notation $\Omega=(012)$ and 
$\Omega^2=(021)$ (see  Eq.~\eqref{permb} for a definition of $\Omega$) so that 
\begin{equation}\label{Z3}
{\mathfrak S}=  \frac{1}{3}\left(1+ \Omega + \Omega^2\right) = P_{\mathbf{1}_{S}}+ P_{\mathbf{1}_{A}} = 1-P_{\mathbf{2}_{M}}~.
\end{equation}
In other words, ${\mathfrak S}$ is the projector onto the completely symmetric and completely antisymmetric subspaces, but projects out the mixed $\mathbf{2}_M$ subspace. Consequently, 
\eqref{finaragain} can be rewritten as 
\begin{equation}\label{popf}
\left( P_{\mathbf{1}_{S}}+ P_{\mathbf{1}_{A}}\right)\circ 
\left[  \frac{\ln Z^{\rm rot}}{\beta} \right] =0~.
\end{equation}
Now the condition $P_{\mathbf{1}_{A}}\circ\left[  \frac{\ln Z^{\rm rot}}{\beta} \right] =0$
follows trivially from the symmetry of $\ln Z^{\rm rot}$ under the interchange of $b_1^2$ and $b_2^2$. The nontrivial condition in   Eq.~\eqref{popf} is the assertion 
\begin{equation}\label{ninfo}
P_{\mathbf{1}_{S}}\circ\left[\frac{\ln Z^{\rm rot}}{\beta} \right]=0~, ~~~\frac{\ln Z^{\rm rot}}{\beta} \in {\bf 2}_M~.
\end{equation}
Note also that while the 
${\bf 2}_M$ representation
is two dimensional, exactly one state in each representation is symmetric 
under the interchange of $b^2_1$
and $b^2_2$ (of course $\frac{\ln Z}{\beta}$ enjoys this symmetry). 

In order to make our discussion more quantitative, we now recall that $\frac{\ln Z^{\rm rot}}{\beta}$ can be written in the form   Eq.~\eqref{Hdef}), where $H(y_0, y_1, y_2)$ admits a power series expansion in its arguments. Let us recall (see e.g. \cite{Choudhury:2018iwf} for a review)  that the space of polynomials of arbitrary degree, of the three variables $(b^2_0,b^2_1,b^2_2)$, constitute a freely generated module. We sometimes use the following terminology: we call the degree of a monomial (in the three variables above) the level of the monomial. The generators of this module are the three $S_3$ invariant monomials  $E_1$, $E_2$ and $E_3$ (see  Eq.~\eqref{invar}). The module of all  polynomials is not irreducible; it has four distinct submodules (each of which is freely generated, i.e. has no null states) and so  has four distinct primaries. The first of these (at `level zero') is  monomial unity, in the ${\bf 1}_S$ representation. There are two primaries in the ${\bf 2}_M$ representation. The first of these, at level one, is given by linear combinations  $(b_0^2-b_1^2)$,  $(b_1^2-b_2^2)$, 
$(b_2^2-b_0^2)$ (note these three functions span a two dimensional basis of functions). The second of these, at level two,  is spanned by linear combinations of three functions $(b_0^2b_2^2-b_1^2 b_2^2)$,  $(b_1^2b_0^2-b_2^2 b_0^2)$, $(b_2^2b_1^2-b_0^2 b_1^2)$, once again a two dimensional space of functions. The last primary, at level three, lies in the ${\bf 1}_A$ representation, is given by ${\mathcal V}$ defined in  Eq.~\eqref{deltadef}. 

We can verify that the primaries listed above do indeed transform in the assigned representations by checking (for instance) that 
\begin{equation}\label{projrepver}
\begin{split}
&P_{1_{S}}\circ\left(1\right)=1~, ~~~ P_{1_{A}}\circ\left(1\right)=P_{2_{M}}\circ\left(1\right)= 0~,\\
&P_{1_{S}}\circ\left(b_1^2-b_2^2\right) =P_{1_{A}}\circ\left(b_1^2-b_2^2\right)=0~,~~~P_{2_{M}}\circ\left(b_1^2-b_2^2\right) = b_1^2-b_2^2~,\\
&P_{1_{S}}\circ\left(b_1^2b_0^2-b_2^2 b_0^2\right) =P_{1_{A}}\circ\left(b_1^2b_0^2-b_2^2 b_0^2\right)=0~,~~~P_{2_{M}}\circ\left(b_1^2 b_0^2-b_2^2 b_0^2\right) = b_1^2b_0^2-b_2^2 b_0^2~,\\
&P_{1_{S}}\circ\left(P\right)=0~,~~~ P_{1_{A}}\circ\left(P\right)= P~, ~~~P_{2_{M}}\circ\left(P\right)= 0~.\\
\end{split}
\end{equation}

We will now count polynomials weighted by degree. Let 
$z(x)$ denote the `trace' ${\rm Tr} x^L$ over the space of polynomials, where $L$ denotes the level or degree of the monomials. Let $z_{\rm all}(x)$ denote the partition function over all polynomials of our three variables. Let 
$z_{{\bf 1}_S}(x)$, $z_{{\bf 1}_A}(x)$ and $z_{{\bf 2}_M}(x)$, respectively, denote the partition functions over polynomials that transform in the corresponding representations. Then it follows from the construction of the free modules above that 
\begin{equation}\label{modulepfs}
\begin{split}
&z_{{\bf 1}_S}(x)=
\frac{1}{(1-x)(1-x^2)(1-x^3)}~,\\
&z_{{\bf 1}_A}(x)=
\frac{x^3}{(1-x)(1-x^2)(1-x^3)}~,\\
&z_{{\bf 2}_M}(x)=
\frac{2x+2x^2}{(1-x)(1-x^2)(1-x^3)}~,\\
&z_{\rm all}(x) =\frac{1}{(1-x)^3}= z_{{\bf 1}_S}+z_{{\bf 1}_A}(x) +z_{{\bf 2}_M}(x)~.
\end{split}
\end{equation}

We will now use the formulae  Eq.~\eqref{modulepfs} to count both the number 
of polynomials $H(b_0^2, b_1^2, b_2^2)$
that are symmetric under
$1 \leftrightarrow 2$, as well as the number of constraints on these polynomials. 

Let $z_{\rm naive}(x)$ represent the partition function of polynomials symmetric under $1 \leftrightarrow 2$. Since every  polynomial in the ${\bf 1}_S$ representation,  exactly half the polynomials in the 
${\bf 2}_M$ representation, and none of the polynomials in the ${\bf 1}_A$ representation enjoy this symmetry, it follows that 
\begin{equation}\label{znaive}
z_{\rm naive}(x)=z_{{\bf 1}_S}(x) + \frac{1}{2}
z_{{\bf 2}_M}(x)= 
\frac{1+x+x^2}{(1-x)(1-x^2)(1-x^3)}
\end{equation}
in agreement with 
\eqref{firstcount}. 

Of course the count  Eq.~\eqref{znaive} does not account for the fact that the polynomials in the function $H$ are constrained to obey the 
equation  Eq.~\eqref{cyc}. 
In order to understand the $S_3$ representation content of 
\eqref{cyc}, we rederive this equation as follows. We use $\frac{\ln Z^{\rm rot}}{\beta}=\frac{P H(b_1^2-b_2^2)}{2 \pi^2}$ to obtain 
\begin{equation}\label{altder}
P_{{\bf 1}_S}\circ \left[\frac{\ln Z^{\rm rot}}{\beta} \right]
=\frac{1}{2\pi^2} P_{{\bf 1}_S} \circ\left[P (b_1^2-b_2^2)H \right]=
\frac{P}{2 \pi^2} P_{{\bf 1}_A}\circ\left[(b_1^2-b_2^2)H \right]~.
\end{equation}
Note that pulling out the completely antisymmetric object $P$ has turned  the projector onto the symmetric representation in the middle of  Eq.~\eqref{altder} into a projector onto the antisymmetric representation in the last of 
\eqref{altder}.
 Inserting  Eq.~\eqref{altder} into the first of  Eq.~\eqref{ninfo} 
yields the equation 
\begin{equation}\label{altcyc}
 P_{{\bf 1}_A}\circ\left[(b_1^2-b_2^2)H \right]=0~.
 \end{equation}
 But  Eq.~\eqref{altcyc} is simply a rewriting of the equation  Eq.~\eqref{cyc}, as may be verified by 
 expanding out the projector on the LHS of  Eq.~\eqref{altcyc} and using the $1\leftrightarrow 2$ symmetry of $H$. 
 
 We see that the LHS of 
  Eq.~\eqref{cyc} lies in the ${\bf 1}_A$ representation. A little thought will convince the reader that any polynomial ${\bf 1}_A$ representation can be obtained from the constraint equation (LHS of  Eq.~\eqref{cyc}) for a suitable choice of the function $H$. We thus conclude that the constraint equation  Eq.~\eqref{cyc} on the degree $L$ part of $H$ is in one to one correspondence with polynomials of degree $L+1$ in the ${\bf 1}_A$ representation. 
The partition function over such polynomials equals $\frac{z_{{\bf 1}_A}(x)}{x}$ (the division by $x$ accounts for the shift in level by unity, i.e. the `one' in  `$L+1$'), in perfect agreement with 
\eqref{constcount}. 
The partition function over polynomials $H$ subject to the constraint Eq.~\eqref{cyc} is thus given by $z_{\rm acc}(x)$
\begin{equation}\label{pfandco}
z_{\rm acc}(x)=z_{{\bf 1}_S}(x) + \frac{1}{2}
z_{{\bf 2}_M}(x) -\frac{z_{{\bf 1}_A}(x)}{x} 
= \frac{1+x}{(1-x)(1-x^2)(1-x^3)}
\end{equation}
in perfect agreement with  Eq.~\eqref{secondcount}.

\subsection{Polynomial nature of ${\mathcal I}$ in $2\kappa$ dimensions, $\kappa\geq 2$}\label{gensec}

In this section we demonstrate that ${\mathcal I}$, the local integrand of the thermal effective action, takes the form of a polynomial (more precisely a power series expansion) in a suitable set of variables. The result of this subsection will be a key input into our derivation, presented in subsequent subsections, of 
the analytic structure of the expansion \eqref{expallord} and the central equation  \eqref{omegapartzed} 
for CFTs in arbitrary even dimensions. 

The discussion of this subsection can be viewed as a generalization (also an independent rederivation) of 
\eqref{explicitf} to arbitrary even dimensions.  As in the rest of this paper, we work with CP even theories whose partition function is well described by a local thermal effective action.

Throughout this Appendix we work in the embedding space, $R^{2\kappa, 2}$, (see e.g. Appendix B of \cite{Chandorkar:2021viw} for a review). The manifold $S^{2\kappa-1} \times R$ (up to Weyl transformations) on which our CFT lives is isomorphic to the space of null rays, $N$, of this embedding spacetime (we work with the  Lorentzian continuation of the Euclidean set up in the main text). Points on this spacetime are parameterized by 
$N^P$ ($P=-1, 0, 1 \ldots 2 \kappa$) 
that obey $N^2=0$. 
$N$ dependent scaling of the null vector $N$ 
change the Weyl frame in our boundary spacetime. In this section we choose to work in the Weyl frame relevant to the computation of the partition function $\ln Z^{\rm rot}$ (see Appendix A of \cite{Advant:2026cqt}).
We can reach this Weyl gauge starting with the metric $g_{\mu\nu}$
in any Weyl frame, and then performing the Weyl rescaling that sets the metric equal to 
\begin{equation}\label{wfom}
\frac{g_{\mu\nu}}{v^2},
\end{equation}
where $v^2$ is computed in the metric 
$g_{\mu\nu}$.
\footnote{The metric used in the computation of $\ln Z^{\rm rot}$ in \cite{Advant:2026cqt} had an extra factor of $\beta^2$. As $\beta$ is a constant, we obtain the same partition function on dropping this factor, as we have done in \eqref{wfom}.}

\subsubsection{The vector $v$ in embedding-space variables and the Cayley--Hamilton relation}\label{genalg_inv}

Recall that the vector $v$ is a conformal Killing vector. Conformal Killing vectors are in one-to-one correspondence with the generators of $SO(2\kappa, 2)$, and so with dimension $(2\kappa+2) \times (2\kappa+2)$ matrices $A^P_{~Q}$ s.t. \footnote{In other words, the flow generated by $A$ preserves the embedding space inner product.} 
\begin{equation}\label{Acond}
\eta A^T = - A \eta, ~~~\implies ~~~(A \eta)^T=- A \eta
\end{equation}
where $\eta=\eta^{-1}$ is the diagonal metric on $R^{2\kappa, 2}$. We see that the matrix $ A\eta$ 
(which has both upper indices) is antisymmetric; i.e. the matrix $A$ is antisymmetric  with respect to the embedding-space metric. 

Now let us suppose that $\lambda$ is an eigenvalue of the matrix $A$. This implies that  
\begin{equation}\label{pmev}
\begin{split}
&{\rm det}\left( A-\lambda I\right)=0 \\
\implies & {\rm det}\left( A\eta-\lambda \eta\right)=0 \\
\implies & {\rm det}\left( (A\eta)^T-\lambda \eta^T\right)=0 \\
\implies & {\rm det}\left( -A\eta-\lambda \eta\right)=0 \\
\implies & {\rm det}\left( A+\lambda I \right)=0 
\end{split}
\end{equation}
so that $-\lambda$ is also an eigenvalue of $A$. 
It follows, therefore, that all eigenvalues of the matrix $B$ defined by 
\begin{equation}\label{genalg_B}
B:=-A^2~.
\end{equation}
have degeneracy two ($B$, like $A$, has index structure $B^P_{~Q}$). 
It follows that  $B$ has $\kappa+1$ distinct eigenvalues, each with multiplicity two.  We denote them by
\begin{equation}\label{genalg_b}
(c_0,c_1,\ldots,c_\kappa)~.
\end{equation}
In order to understand the physical meaning of $c_i$, 
consider the standard form vector $(v_L)_\mu$, which, in Lorentzian space, takes the form 
\begin{equation}\label{nhg}
v_L= i v_E =\beta\left( \partial_t+\omega_1 \partial_{\phi_1}+\cdots+\omega_\kappa\partial_{\phi_\kappa} \right) 
\end{equation} 
where ``$L$'' denotes Lorentzian and ``$E$'' denotes Euclidean. This CKV clearly corresponds to a block diagonal matrix $A$, whose $2\times 2$ blocks are 
\begin{equation}\label{aform}
\begin{pmatrix}
0 & \beta\\
-\beta & 0
\end{pmatrix}~,\quad\mathrm{and}\quad \begin{pmatrix}
0 & \beta\omega_i\\
-\beta\omega_i & 0
\end{pmatrix}~~\mathrm{for}~~i=1,2\ldots,\kappa~.  
\end{equation}
Computing the eigenvalues of $B=-A^2$, we identify \
\begin{equation}\label{genalg_bident}
(c_0,c_1,\ldots,c_\kappa) = \big(b_0^2, b_1^2 \ldots b_\kappa^2\big)= \big(\beta^2,\beta^2\omega_1^2,\ldots,\beta^2\omega_\kappa^2\big)~.
\end{equation}
Below we will also often use the completely symmetric polynomials  $E_r$ defined in Eq.~\eqref{invar};
or equivalently via 
\begin{equation}\label{genalg_E}
Q(t):=\prod_{i=0}^{\kappa}(t-c_i)
=t^{\kappa+1}-E_1t^\kappa+E_2t^{\kappa-1}-\cdots+(-1)^{\kappa+1}E_{\kappa+1}~.
\end{equation}
We emphasize that $E_r$ depend only on the unordered `compactification data' and are therefore $S_{\kappa +1}$ invariant. 

\subsubsection{Fixing the Weyl gauge}

It is now a simple matter to fix Weyl `gauge' as in Eq.~\eqref{wfom}. 
Recall that $N$ is a null ray and so obeys 
\begin{equation}\label{nullray}
N^2=0~.
\end{equation}
In the Weyl gauge \eqref{wfom} it follows from definitions that $v^2=1$. A little thought will convince the reader that $v^2$ is given, in embedding space notation, by  
\begin{equation}\label{vsquareembspace}
v^2=-(AN)^2=N\cdot A^2 N~=-N \cdot B N~.
\end{equation}
(the minus sign converts the Lorentzian norm of $AN$ into the Euclidean $v^2$; compare \eqref{genalg_u}). 
Thus the Weyl gauge Eq.~\eqref{wfom} is implemented by rescaling all null rays to ensure that 
\begin{equation}\label{weylgauge}
N\cdot BN=-1~~~
\qquad
\end{equation}
\eqref{weylgauge} implements our choice of Weyl gauge.

\subsubsection{Coordinates in embedding space}

Below, we will make frequent use of the position-dependent quantities
\begin{equation}\label{coords}
m_j:=-N\cdot B^jN~,
\qquad
m_0=0~,
\qquad
m_1=1~.
\end{equation}
Recall that $m_j$ depend both on the compactification data, as well as on the location on the sphere. In contrast, the 
quantities $E_r$ above,  are position independent, but depend only on the compactification data.

We will now explain that each of the $m_j$, $(j=2\ldots \infty)$ and each ${\mathrm{tr}(B^j)}$ $(j=1\ldots \infty)$ 
can be rewritten as a polynomial in the `generators' 
\begin{equation}\label{generators}
E_1,\ldots,E_{\kappa+1}~,
\qquad
m_2,\ldots,m_\kappa~.
\end{equation}

The simple argument for this result goes as follows. As every matrix obeys its characteristic equation, it follows that
\begin{equation}\label{genalg_CH}
Q(B)=B^{\kappa+1}-E_1B^\kappa+E_2B^{\kappa-1}-\cdots+(-1)^{\kappa+1}E_{\kappa+1}=0~.
\end{equation}
Multiplying Eq.~\eqref{genalg_CH} by $ B^r$ and placing the result between two $N$'s gives
\begin{equation}\label{genalg_mrec}
\begin{split}
m_{r+\kappa+1}
={}&E_1m_{r+\kappa}-E_2m_{r+\kappa-1}+\cdots
+(-1)^\kappa E_{\kappa+1}m_r~.
\end{split}
\end{equation}
Using the same equation on every $m_s$ with $s \geq \kappa +1$ term on the RHS of Eq.~\eqref{genalg_mrec}, and repeatedly iterating this process, we conclude that  every $m_j$ with $j>\kappa$ can be rewritten as a polynomial in the generators listed in \eqref{generators}. 
In a similar manner, it follows that all of the $\Tr B^j=2\sum_a c_a^j$ can be rewritten in terms of the $E_r$.  

The $m_j$ defined above, constitute a set of $\kappa-1$ coordinates on $(S^{2\kappa-1} \times R)/G_{v}$; where $G_v$ is the subgroup of $SO(2\kappa, 2)$ that leaves $v$ invariant.
When $v$ takes the standard form \eqref{nhg}, the $G_v$
is the $\kappa+1$ dimensional abelian subgroup consisting of rotations (translations) in each of the $\phi_i$ and $t$. Consequently, the $(\kappa-1)$ quantities $m_j$, together with $t$ and $\phi_i$, constitute a full set of coordinates on $S^{2\kappa-1}\times R$. Indeed, as we will see below, $\kappa-1$ $m_j$ can be written as functions of the $\kappa-1$  
direction cosines on $S^{2 \kappa-1}$. As the coordinates $m_j$ have been defined in a completely $SO(d, 2)$ invariant manner, however, they will prove more useful than the direction cosines themselves, in our analysis below.

\subsubsection{The projector}

The metric of $S^{2\kappa-1} \times R$, in our particular choice of Weyl frame, is given by the metric on $R^{2\kappa,2}$ restricted to the codimension-two submanifold defined by 
\eqref{nullray} and \eqref{weylgauge}.
Consider any point $N_0$ on this manifold. Let $N = N_0+\delta N$. Working to first order in $\delta N$, the equations \eqref{nullray} and \eqref{weylgauge} reduce, at first nontrivial order in $\delta N$, to 
\begin{equation}\label{genalg_tangent}
\delta N\cdot N_0=0~,
\qquad
\delta N\cdot BN_0=0~.
\end{equation}
\eqref{genalg_tangent}, viewed as two constraints on the vectors $\delta N$,  defines the tangent space of our submanifold around the point $N_0$. We will now construct $\Pi$,  the projector onto the $2\kappa$ dimensional tangent space described above. The simplicity (polynomial nature) of this projector will play a key role in the remaining analysis below. 

Consider a general $n$ dimensional linear vector space. 
Consider a collection of $m$ vectors $v_{(a)}^\mu$ ($a=1\ldots m$) in this space. Let $G_{ab}$ denote the $m\times m$ matrix of inner products of these vectors, and let $G^{ab}$ denote the inverse of $G_{ab}$.
Then the projector onto the orthogonal complement of the space spanned by $v_{(a)}$ is given by  
\begin{equation}\label{pmu}
\Pi^{\mu}_{~\nu}= \delta^\mu_{~\nu} - G^{ab} v_{(a)}^{~~\mu} v_{(b)\nu}~.
\end{equation}
Note, in particular, that independent rescalings of $v_{(a)}$ leave this projector unchanged. 

We can now apply the general formula \eqref{pmu} to obtain the projector onto the tangent space of a general point $N$ on the submanifold described by \eqref{nullray} and \eqref{weylgauge}. In this case $m=2$, and the two vectors $v_{(a)}$ are $N$ and $B N$ respectively.  The matrix of inner products of these two vectors is given by  
\begin{equation}\label{genalg_normmat}
G_{ab}= \begin{pmatrix}
N^2 & N\cdot BN\\
BN\cdot N & (BN)^2
\end{pmatrix}
=
\begin{pmatrix}
0&-1\\
-1&-m_2
\end{pmatrix}~.
\end{equation}
A key simplification is that the determinant of this matrix is $-1$. For this reason,  the inverse  of this matrix contains no denominator depending on the position on the sphere, and
is given by 
\begin{equation}\label{gabup}
G^{ab} = \begin{pmatrix}
m_2&-1\\
-1&0
\end{pmatrix}~.
\end{equation}
It follows from \eqref{pmu} that projector onto our tangent space  is given by 
\begin{equation}\label{proj}
\begin{split}
\Pi^Q_{~P}&= \delta^Q_{~P}-m_2 N^Q N_P+N^Q(BN)_P+(BN)^Q N_P\\
\implies\Pi V &
=V-m_2N(N\cdot V)+N(BN\cdot V)+BN(N\cdot V)~.\\
\end{split}
\end{equation}

\subsubsection{Covariant derivatives preserve the polynomial nature}\label{genalg_poly}

Whenever a manifold can be viewed as a submanifold of a flat space, the covariant derivative (of any tensor on the submanifold) is obtained by taking the ordinary derivative of that tensor 
(along the relevant tangent derivatives of the manifold), and projecting the result of this computation back onto the tangent space of the manifold.

In the current context, our boundary spacetime consists of the codimension two submanifold of $R^{2\kappa, 2}$ defined by \eqref{nullray} and \eqref{weylgauge}, and the projector onto tangent space is given by Eq.~\eqref{proj}. 
Applying the prescription spelled out in the previous paragraph, we conclude that 
\begin{equation}\label{genalg_covder}
\bar\nabla_UV=\Pi D_UV~,
\end{equation}
where $D_U$ is the ordinary flat space derivative dotted with the tangent vector $U$ (which we assume to lie in the tangent space of our manifold). 

Now recall that $A$ (the generator dual to $v$) and $B=-A^2$ are both constant matrices in embedding-space. Because the ordinary derivatives of these constant matrices vanishes, it follows that 
\begin{equation}\label{basicder}
D_UA=0~,
\qquad
D_UB=0~. 
\end{equation}
Moreover, the covariant derivative of the vector field $N^A$ is given by
\begin{equation}\label{cdn}
(D_U N)^P= \Pi^P_{~Q} U^M \partial_M N^Q=
\Pi^P_{~Q} U^Q=U^P, ~~~~\implies D_U N= U~.
\end{equation}

It follows that differentiating a polynomial expression in $A,B,N$ gives another polynomial expression before projection. The key equation \eqref{proj} then shows us that the projection itself introduces no denominator.

Since the $m_j$ are themselves polynomials of $N$ and $B$, 
the discussion above implies that covariant derivatives of 
$m_j$ are themselves polynomials in $m_j$ and $E_r$. 
We can also directly verify this conclusion as follows. 
It follows immediately from the definition of the $m_j$ 
in \eqref{coords} that 
\begin{equation}\label{genalg_gradm}
\bar{\nabla}_P m_j = \Pi_P^{~Q} \partial_Q m_j = 
-2 \Pi_P^{~Q} (B^j)_{QR} N^R= -2(B^jN)_P-2(m_2m_j-m_{j+1})N_P+2m_j (BN)_P~.
\end{equation}
The RHS is polynomial, as expected. Repeated derivatives of \eqref{genalg_gradm}, followed by repeated use of \eqref{proj}, therefore remain polynomial as well.

\subsubsection{The Riemann Tensor }\label{genalg_curvature}

A similar conclusion also holds for all curvature invariants. 
As our submanifold is embedded in flat space, its Riemann tensor is given by the Gauss–Codazzi equations 
\begin{equation}\label{gausscodazzi}
R_{PQRS}
=
G^{ab}
\left(
K^{(a)}_{PR}K^{(b)}_{QS}
-
K^{(a)}_{PS}K^{(b)}_{QR}
\right)~,
\end{equation}
where $G^{ab}$ is given in \eqref{gabup}. Here $K^{(a)}_{\mu\nu}$ are the extrinsic curvatures defined by 
\begin{equation}\label{genalg_Mdef}
K^{(a)}_{PQ}:=\Pi_P^{~P'}\Pi_Q^{~Q'} D_{P'} n^{(a)}_{Q'}~.
\end{equation}
As around \eqref{genalg_normmat} we choose our two normal vectors $n^{(a)}$ to be $N^P$ and $(B N)^P$. The two extrinsic curvatures are thus
\begin{equation}\label{extexp}
    \begin{split}
        &K^{(1)}_{PQ}=\Pi_P^{~P'}\Pi_Q^{~Q'} D_{P'} N_{Q'}=\Pi_P^{P'}\Pi_{P'Q}=\Pi_{PQ}~,\\
        &K^{(2)}_{PQ}=\Pi_P^{~P'}\Pi_Q^{~Q'} D_{P'}(BN)_{Q'}=\Pi_P^{~P'}\Pi_Q^{~Q'}B_{P'Q'}:=M_{PQ}~,
    \end{split}
\end{equation}
where we have used $D_PN_Q=\eta_{PQ}$.
Putting \eqref{extexp} into \eqref{gausscodazzi}, the Riemann tensor is
\begin{equation}\label{riemannten}
R_{PQRS}
=m_2\left(\Pi_{PR}\Pi_{QS}-\Pi_{PS}\Pi_{QR}\right)
-M_{PR}\Pi_{QS}-\Pi_{PR}M_{QS}
+M_{PS}\Pi_{QR}+\Pi_{PS}M_{QR}~.
\end{equation}
The following identities can be obtained using \eqref{extexp} and \eqref{proj}
\begin{equation}\label{mrel}
    \begin{split}
        &M_P^{~\,P}=M_{PQ}\eta^{QP}=\Tr{\Pi B}=\Tr{B}-m_2~,\\
        &M_{PQ}M^{PQ}=\Tr{[\Pi B\Pi B]}=\Tr{B^2}+m_2^2-2m_3~.
    \end{split}
\end{equation}
Hence, these scalars are also polynomials. Using \eqref{riemannten} and \eqref{mrel}, we get
\begin{equation}\label{genalg_Rscalar}
\begin{split}
     &R_{PQ}=\left(d\,m_2-\operatorname{Tr}B\right)\Pi_{PQ}
-(d-2)M_{PQ}~,\\
& R=(d-1)\left[(d+2)m_2-2\operatorname{Tr}B\right]~.
\end{split}
\end{equation}

\subsubsection{Sample curvature invariants}

The other curvature invariants are
\begin{equation}
    \begin{split}
        & R_{PQ} R^{PQ}
~~~~~~~={}d\left(d\,m_2-\operatorname{Tr}B\right)^2
-2(d-2)\left(d\,m_2-\operatorname{Tr}B\right)
\left(\operatorname{Tr}B-m_2\right)
+(d-2)^2
\left[\operatorname{Tr}(B^2)+m_2^2-2m_3\right]~,\\
& R_{PQRS} R^{PQRS}
={}2d(d-1)m_2^2
-8(d-1)m_2(\Tr B-m_2)+4(\Tr B-m_2)^2
+4(d-2)\left(\Tr B^2+m_2^2-2m_3\right)~.
    \end{split}
\end{equation}
One can easily see that all these scalars are polynomials. Derivatives of curvature are also polynomial as the derivatives act only on $m_i$s since $B$ is spacetime independent. From \eqref{genalg_gradm},
\begin{equation}\label{genalg_gradinner}
(\bar\nabla m_j)\cdot(\bar\nabla m_k)
=4\left[-m_{j+k}+m_jm_{k+1}+m_km_{j+1}-m_2m_jm_k\right]~.
\end{equation}
Since $\Tr B$ is independent of position, the second of \eqref{genalg_Rscalar} gives, for instance,
\begin{equation}\label{genalg_gradR2}
(\bar\nabla R)^2
=4(d-1)^2(d+2)^2
\left[-m_4+2m_2m_3-m_2^3\right]~.
\end{equation}
When $m_4$ is not part of the set \eqref{generators}, it is eliminated using \eqref{genalg_mrec}.  The same procedure applies to any finite number of covariant derivatives.

In the metric induced on the codimension two manifold \eqref{genalg_tangent}, the unit thermal vector  is
\begin{equation}\label{genalg_u}
u:=AN~,
\qquad
u^2=-N A^2 N=N B N=-1~,
\end{equation}
($u^2=-1$ makes sense because we are in Lorentzian space; this continues to $u^2=1$ in Euclidean space).

The projection orthogonal to $u$ therefore also introduces no denominator.  Upon completely contracting a finite local scalar, every term reduces to products of traces and contractions involving two $N$'s,
\begin{equation}\label{genalg_loops}
\Tr A^{2j}=\Tr (-B)^j~,
\qquad
N\cdot A^{2j}N=(-1)^{j+1} m_j~.
\end{equation}
Odd traces and $N\cdot A^{2j+1}N$ vanish because $A$ is antisymmetric.  Combining these statements with \eqref{genalg_mrec} and the fact that $\Tr{B^j}=2\sum_ic_i^j$ is a polynomial in the symmetric invariants $E_r$, we conclude that at every fixed order in the local thermal effective action
\begin{equation}\label{poly}
{\mathcal I}={\mathcal I}(E_1,\ldots,E_{\kappa+1};m_2,\ldots,m_\kappa)
\end{equation}
is a polynomial.  Its coefficients are theory-dependent Wilson coefficients and do not depend on the compactification data.  In $d=4$ there is only one position-dependent variable $m_2$. In the next subsection we will demonstrate that $m_2$ is related to the  variable $z$ (defined in \eqref{hatzed} and used extensively in the main text) via 
\begin{equation}\label{zrel}
z=E_1-m_2~.
\end{equation}
Thus the fact that ${\mathcal I}$, in the four dimensional case,  is polynomial, both in the variables $z$ and  in the variables $b_i^2$ (see \eqref{explicitf}) 
follows directly from the general discussion of this subsection. 

\subsection{The integral over $S^{2 \kappa -1}$ for $\kappa \geq 2$}
\label{genalg_integral}

In the previous subsection we have demonstrated that the integrand, ${\mathcal I}$, of the thermal effective action is a polynomial in the variables $m_i$ and $E_r$ (see e.g. \eqref{genalg_poly}). In this subsection we study the integral over such ${\mathcal I}$ over the rotating sphere and, in particular, demonstrate that its analytic structure takes the form presented in the first of \eqref{expallord}. The final formulae of this subsection will also form the starting point for our derivation of \eqref{omegapartzed} in the next subsection. 

Below, we  will find it convenient to switch between several distinct coordinates for $S^{2 \kappa -1}$. We first pause to define and introduce each of these coordinate systems. 

\subsubsection{Direction cosines}

One familiar set of coordinates on $S^{2 \kappa-1}$ consists of 
the $\kappa$ angles $\phi_i$ (dual to the $\kappa$ angular momenta $J_i$, together with the $\kappa$
`direction cosines' $l_i$ which are constrained to sum to unity. 
More concretely, these coordinates are defined as follows. Let us view our unit sphere as embedded in $C^{ \kappa}$ (and centered around its origin). 
Then $S^{2\kappa-1}\subset {\mathbb C}^{\kappa}$ is given by the locus of points 
\begin{equation}\label{genalg_spherecoord}
z_i=\sqrt{\ell_i}\,e^{i\phi_i}~,
\qquad
\ell_i\geq0~,
\qquad
\sum_{i=1}^{\kappa}\ell_i=1~.
\end{equation}
In these coordinates, the round sphere metric times $R$ is given by 
\begin{equation}\label{rscoord}
ds^2= d\tau^2 + \sum_{m=1}^\kappa \left(\frac{d\ell_m^2}{4\ell_m}+ \ell_m d \phi_m^2\right)~. 
\end{equation}
In the coordinates \eqref{rscoord}, the norm of the vector $v$ (see \eqref{vedef}) 
is given by 
\begin{equation}\label{genalg_v2ell}
v^2=\beta^2-\sum_{i=1}^\kappa \ell_i \beta^2 \omega_i^2
= \beta^2\left(\sum_{i=1}^\kappa \ell_i\right) -\sum_{i=1}^\kappa \ell_i \beta^2 \omega_i^2=
\sum_{i=1}^{\kappa}(c_0-c_i)\ell_i~.
\end{equation}
In these coordinates, the thermal effective action is given in terms of the density 
${\mathcal I}$ by 
\begin{equation}\label{genalg_lint}
\frac{\ln Z^{\rm rot}}{\beta}
=\frac{(2\pi)^\kappa}{2^{\kappa-1}}
\int_{\substack{\ell_i\geq0,\\  \sum_i\ell_i=1}}
\frac{d^{\kappa-1}\ell}{(v^2)^\kappa}
\,{\mathcal I}(E,m)~,
\end{equation}
(the factor of  $(2\pi)^\kappa$ comes from performing the integrals over the $\kappa$ angles $\phi_i$ and $2^{\kappa-1}$ in the denominator comes from $\sqrt{g}$). 

We can visualize the integration region in $\ell_i$ as living on the plane $\sum_j \ell_j=1$ embedded in the $\kappa$ dimensional space, whose axes are $\ell_i$
($i=1\ldots \kappa$). 
The boundary of the integration region is given by the intersection of this plane with the  $\kappa$ codimension one surfaces $\ell_i=0, ~~~(i=1\ldots \kappa)$. These are the surfaces on which the angles $\phi_i$ shrink in the full geometry (this shrinking happens in a smooth way, like at the origin of 2d polar coordinates).

 \subsubsection{The coordinates $\lambda_i$}

For reasons that will soon become clear, we will sometimes find it convenient to change coordinates from $\ell_i$ to $\lambda_i$ where
\begin{equation}\label{genalg_lambda}
\lambda_i:=\frac{(c_0-c_i)\ell_i}{v^2}~,
\qquad
\lambda_i\geq0~,
\qquad
\sum_{i=1}^\kappa \lambda_i=1~.
\end{equation}
This coordinate change is easily inverted to find $\ell_i$ in terms of $\lambda_i$ via  
\begin{equation}\label{genalg_lambdainv}
\ell_i=
\frac{\lambda_i/(c_0-c_i)}{\sum_j\lambda_j/(c_0-c_j)}~.
\end{equation}
Inserting \eqref{genalg_lambdainv}
allows us to express $v^2$ as a function of the $\lambda_i$,
yielding 
\begin{equation}
v^2=\left(\sum_j\frac{\lambda_j}{c_0-c_j}\right)^{-1}~.
\end{equation}
The Jacobian of the variable change from $\ell_i$ to $\lambda_i$ is easily determined from \eqref{genalg_lambdainv}. After some algebra we find 
\begin{equation}\label{lambdameasure}
\frac{d^{\kappa-1}\ell}{(v^2)^\kappa}
=
\frac{d^{\kappa-1}\lambda}
{\prod_{i=1}^{\kappa}(c_0-c_i)}~.
\end{equation}
The key point here is that the $\lambda_i$ (unlike $\ell_i$) are `flat' coordinates; all the position dependence in the denominator of \eqref{lambdameasure} has disappeared.

The integration domain for the $\lambda_i$ can be visualized in the same way as for the $\ell_i$ above. As in that case, the  $\phi_i$ circles shrink to zero on the codimension one surfaces $\lambda_i=0$.

The integration domain in the $\lambda_i$ can be restated as follows. Working in the Cartesian space $(\lambda_1, \lambda_2 \ldots \lambda_\kappa)$, consider the $\kappa$ distinct unit vectors 
$e_1=(1, 0, \ldots 0)$, $e_2=(0, 1, 0, \ldots 0)$, $\ldots e_\kappa= (0, 0, 0, \ldots 1)$. The integration domain is given by 
\begin{equation}
\label{intdomain}
\sum_j \lambda_j e_j, ~~~\lambda_i\geq 0, ~~~\sum_{j=1}^\kappa \lambda_j=1~.
\end{equation}
However \eqref{intdomain}
is the definition of a polytope built out of the $\kappa$ vectors $e_1 \ldots e_\kappa$. We conclude that the integration domain in $\lambda_i$ is a polytope built out of the vectors described above.

\subsubsection{$m_i$ as coordinates}

As we have already explained in the previous subsection, the quantities $m_i$, defined in \eqref{coords}, can also be used as a set of coordinates for $S^{2 \kappa -1}$. This can be made quantitative by rewriting the definition of $m_i$ in a `Weyl' gauge invariant form 
\begin{equation}\label{winvdef}
m_j=\frac{N\cdot B^j N}{N \cdot B N}
\end{equation}
\eqref{winvdef} is the Weyl invariant generalization of \eqref{coords} because 
\begin{itemize}
\item 
The RHS of \eqref{winvdef} is left unchanged by local scalings of $N$ and so is Weyl invariant. 
\item \eqref{winvdef} reduces to \eqref{coords} in the Weyl frame \eqref{weylgauge}.
\end{itemize}

We can now use the Weyl invariant definition \eqref{winvdef}, to compute $m_j$ directly in the `round sphere' Weyl frame adopted in this subsection. In the round sphere Weyl frame, the null vector $N$
is given, in terms of angles and direction cosines by
\begin{equation}\label{Ninterms}
N=\left(\cos t , \sin t, \sqrt{l_1} \cos \phi_1, \sqrt{l_1} \sin \phi_1 , \sqrt{l_2} \cos \phi_2, \sqrt{l_2} \sin \phi_2 \ldots \sqrt{l_\kappa} \cos \phi_\kappa, 
\sqrt{l_\kappa} \sin \phi_\kappa\right)~,
\end{equation}
(where $t$ is global Lorentzian time; note that  the first two coordinates in Eq. \eqref{Ninterms} are timelike, while the remaining are spacelike). In the same basis, the matrix $A$
takes the block diagonal form \eqref{aform}, and $B=-A^2$. A simple computation then yields  
\begin{equation}\label{genalg_mell}
m_j=
\frac{c_0^j-\sum_i c_i^j\ell_i}{v^2}~,
\end{equation}
\footnote{We can use this formula to derive \eqref{zrel}. First note that 
\eqref{hatzed} can be rewritten as $z = c_0 - \frac{(c_0-c_1)(c_0-c_2)}{v^2}$, 
with $v^2 = (c_0-c_1)l_1 + (c_0-c_2)l_2$. 
In $d=4$, and at $j=2$ \eqref{genalg_mell} reduces to 
$m_2=\frac{c_0^2 -c_1^2l_1 -c_2^2 l_2}{v^2}$. It follows that $z + m_2 = c_0 - \frac{(c_0-c_1)(c_0-c_2)}{v^2} + \frac{c_0^2 - (c_1^2 l_1 + c_2^2 l_2)}{v^2}
= c_0 + \frac{(c_1+c_2)v^2}{v^2} = c_0 + c_1 + c_2 = E_1$, as asserted in 
\eqref{zrel}.}
and hence
\begin{equation}\label{genalg_mlambda}
\begin{split}
m_j
&=\sum_{i=1}^{\kappa}\lambda_i
\frac{c_0^j-c_i^j}{c_0-c_i}\\
&=\sum_{i=1}^{\kappa}\lambda_i
\left(c_0^{j-1}+c_0^{j-2}c_i+\cdots+c_i^{j-1}\right)~.
\end{split}
\end{equation}
Note that the apparent denominator in the first line cancels. It follows that  $m_j$ is a linear function of the $\lambda_i$, with coefficients that are polynomials in the $c_a$.

\subsubsection{Integration measure in $m_i$ coordinates}

The integration measure over the $m_a$ may be worked out as follows.  
\eqref{lambdameasure} can be rewritten as 
\begin{equation}\label{intmeasone}
\frac{1}{\prod_{j=1}^\kappa (c_0-c_j)} \int d\lambda_1 \wedge d \lambda_2 \ldots d \lambda_\kappa \delta(\lambda_1 +\lambda_2 +\ldots \lambda_\kappa -1)~.
\end{equation}
Using the delta function to  integrate out (say) $\lambda_1$, we can rewrite the first of \eqref{genalg_mlambda} as 
\begin{equation} \label{mlamnew} 
m_j = \frac{c_0^j - c_1^j}{c_0 - c_1} + \sum_{i=2}^{\kappa}\lambda_i\underbrace{\left(\frac{c_0^j - c_i^j}{c_0 - c_i} - \frac{c_0^j - c_1^j}{c_0 - c_1}\right)}_{\equiv\,\tilde{A}_{ji}}~,
\end{equation}
so that \eqref{intmeasone} can be rewritten as 
\begin{equation}\label{inn}
\frac{1}{{\rm det}{\tilde A} \prod_{j=1}^\kappa (c_0-c_j) } dm_2 \wedge dm_3 \ldots d m_\kappa~,
\end{equation}
where the $(\kappa-1)\times (\kappa-1)$ matrix ${\tilde A}$ (see \eqref{mlamnew}) equals 
\begin{equation}\label{nmc}
\begin{split}
&{\tilde A}_{ji}=B_{ji}-B_{j1}, ~~~~(i, j=2\ldots \kappa)\\
&B_{ji}=\frac{c_0^j-c_i^j}{c_0-c_i}, ~~~B_{j1} 
= \frac{c_0^j-c_1^j}{c_0-c_1}~. 
\end{split}
\end{equation}
As defined, $B$ is a $(\kappa-1)\times \kappa$ 
matrix. We can enhance it to a $\kappa\times \kappa$ matrix by adding a row of ones, i.e by defining 
\begin{equation} \hat{B} = \begin{pmatrix} 1 & 1 & \cdots & 1 \\ B_{2,1} & B_{2,2} & \cdots & B_{2,\kappa} \\ \vdots & & & \vdots \\ B_{\kappa,1} & B_{\kappa,2} & \cdots & B_{\kappa,\kappa} \end{pmatrix}~.
\end{equation}
By subtracting the first column of ${\hat B}$ from each of its last $\kappa-1$ columns, we see that 
\begin{equation}
\label{detba}
\det\hat{B} = \det\begin{pmatrix}1& 0 & 0 & \cdots & 0  \\ B_{2,1} & \tilde{A}_{2,2} & \tilde{A}_{2,3} & \cdots & \tilde{A}_{2,\kappa} \\ \vdots & & & \vdots & \vdots \\ B_{\kappa,1}& \tilde{A}_{\kappa,2} & \tilde{A}_{\kappa,3} & \cdots & \tilde{A}_{\kappa,\kappa} \end{pmatrix}~.
\end{equation}
We conclude that 
\begin{equation}\label{detbaa}
{\rm det}{\hat B}
={\rm det} {{\tilde A}}~.
\end{equation}
Now the determinant of ${\hat B}$ is clearly of degree $1+2 +\ldots \kappa-1=\frac{\kappa(\kappa-1)}{2}$ in the variables $c_0, c_1 \ldots c_\kappa$. Moreover this determinant vanishes whenever $c_i=c_j$ for any $i, j \in 1\ldots \kappa$. It follows that the determinant has $\prod_{i<j}(c_i-c_j)$ as a factor. As this factor already has the required degree, it follows that ${\rm det}{\hat B}$ is, in fact, independent of $c_0$. The proportionality constant can be obtained by examining the coefficient of a particular term, and we find 
\begin{equation}\label{detb}
\det\hat{B}=
\prod_{1\leq i < j\leq \kappa}(c_j - c_i)~.
\end{equation}
Inserting \eqref{detb} and \eqref{detbaa} into \eqref{inn} we have
\begin{equation}\label{innfin}
\frac{1}{{\rm det}{\tilde A} \prod_{j=1}^\kappa (c_0-c_j) } dm_2 \wedge dm_3 \ldots d m_\kappa=\frac{(-1)^{\kappa}}{\prod_{0\leq i<j \leq \kappa} (c_j-c_i) } dm_2 \wedge dm_3 \ldots d m_\kappa~.
\end{equation}

Defining ${\cal V}$ (the arbitrary dimensional generalization of the four dimensional Vandermonde 
$\frac{2\pi^2}{P}$) by 
\begin{equation}\label{genalg_fullprod}
\mathcal{V}:=\prod_{0\leq i<j\leq\kappa}(c_j-c_i)~,
\end{equation}
(note that $\mathcal{V}$ defined in \eqref{genalg_fullprod}
is the same quantity as $\mathcal{V}$ defined in \eqref{deltadef})
and adding in the factors of $(2\pi)^\kappa$ from the angle integrals we obtain our final integration measure
\begin{equation}\label{innfinnfanc}
(-1)^{\kappa}\frac{2\pi^\kappa}{\mathcal{V}} dm_2 \wedge dm_3 \ldots d m_\kappa~.
\end{equation}

\subsubsection{The integration range in $m_i$ coordinates}

As the $m_j$ are linear functions of the $\lambda_i$, it follows that the integration domain in the $m_j$ continues to be given by a polytope. The polytope has $\kappa$ distinct `vertices' (i.e. defining vectors). The $V^{(a)}_0$th vertex $(a=1 \ldots \kappa)$ of this polytope is  given by the vector 
\begin{equation}
(m_2, m_3 \ldots m_{\kappa}) = \left(\left(V_{0}^{(a)}\right)_2, \left(V_{0}^{(a)}\right)_3, \ldots \left(V_{0}^{(a)}\right)_{\kappa}\right)~,
\end{equation}
where (see \eqref{genalg_mlambda})
\begin{equation}\label{vab}
\left(V^{(b)}_0\right)_j=\frac{c_0^j-c_b^j}{c_0-c_b}~.
\end{equation}

For use below, we define the quantities
\begin{equation}\label{sdef}
s_j(a,b):=\frac{c_a^j-c_b^j}{c_a-c_b}
=c_a^{j-1}+c_a^{j-2}c_b+\cdots+c_b^{j-1}~,
\qquad b\neq a~, 
\end{equation}
in terms of which the vertices of our polytope can be rewritten as 
\begin{equation}\label{vabfanc}
\left(V^{(b)}_0\right)_j=s_j(0,b)~.
\end{equation}
Consequently, the integration region in the variables $m_j$ is given by the polytope built out of the vertex vectors \eqref{vabfanc}. More explicitly, the integration region, ${\mathcal D}_0$, is given by 
\begin{equation}\label{genalg_mzero}
{\mathcal D}_0: ~~~~m_j=\sum_{b=1}^\kappa \lambda_b\,s_j(0,b)~,
\qquad
\sum_{b=1}^\kappa \lambda_b=1~,\qquad \lambda_b\geq 0~,
\end{equation}
with an orientation we will describe in more detail below. 

\subsubsection{The analytic structure of $\ln Z^{\rm rot}$}

It follows from \eqref{genalg_mzero} that 
\begin{equation}\label{lnzbb}
\left(\frac{\ln Z^{\rm rot}}{\beta}\right)=
\frac{(-1)^\kappa 2\pi^\kappa}{\mathcal{V}} \int_{{\mathcal D}_0} dm_2 \wedge dm_3 \ldots \wedge d m_\kappa ~{\mathcal I}\left(E,m^{(a)} \right)
\end{equation}
\eqref{lnzbb} is the generalization (to arbitrary dimension) of the four dimensional formula \eqref{lnzchin}. 

In the previous subsection we have demonstrated that the integrand in \eqref{lnzbb} is a polynomial in the $E_r$ and $m_i$. As the integration limits in \eqref{lnzbb} are all polynomials in $c_a$ (see  the second of \eqref{genalg_mlambda}), it follows that the result of the integral in \eqref{lnzbb} is a polynomial in $E_r$ and $c_a$, and so is a polynomial in $c_b$, $ b=0\ldots \kappa$.
 
We can actually say more. As we have explained above, at generic values of $c_a$, the integration region ${\mathcal D}_0$ is a $\kappa-1$ dimensional polytope constructed out of the $\kappa$ vectors $V^{(a)}$, $a= 1\ldots \kappa$. 
In the limit that $c_a=c_b$ 
(for any two particular values of $a$ and $b$), however, the vectors $V^{(a)}$ and $V^{(b)}$ tend to each other, and the 
polytope tends to a $\kappa-2$ dimensional surface. 
As the integral of a $\kappa-1$ form over a $\kappa-2$
dimensional surface always vanishes, it follows that 
$$\int_{{\mathcal D}_0} dm_2 \wedge dm_3 \ldots \wedge d m_\kappa ~{\mathcal I}\left(E,m^{(a)} \right)$$
vanishes (and, in fact, generically develops a simple zero) in the limit $c^a \to c^b$. Since this holds for every pair $(a, b)$, we conclude that 
\begin{equation}\label{anstruc}
\int_{{\mathcal D}_0} dm_2 \wedge dm_3 \ldots \wedge d m_\kappa ~{\mathcal I}\left(E,m^{(a)} \right) 
= \prod_{1\leq a < b \leq \kappa} (c_a-c_b) \times Q
\end{equation}
where $Q$ is a polynomial (more generally, admits a power series expansion) in $c_a$. Plugging \eqref{anstruc} into \eqref{lnzbb}, and using the definition of $\cal V$, we conclude that 
\begin{equation}
\label{eq:2}
\left(\frac{\ln Z^{\rm rot}}{\beta}\right)\propto 
\frac{1}{\prod_{b=1}^\kappa (c_0-c_b) } \times Q
\end{equation}

demonstrating that $\ln Z^{\rm rot}$ has the analytic structure expressed in the second of \eqref{expallord}. Our proof applies in every even dimension.

\subsection{Derivation of our central equation in every even dimension}

In this subsection 
we demonstrate that \eqref{omegapartzed} holds for any CP invariant CFT$_{2 \kappa}$ for $\kappa \geq 2$ (whose partition function is well defined by a thermal effective action), generalizing the $d=4$
($\kappa=2$) derivation presented in the main text. The starting point for our derivation is \eqref{lnzbb}, derived in the previous subsection.

\subsubsection{Action of permutations}

We now study the action of the $\mathbb{Z}_{\kappa+1}$ generator $\Omega$ (defined in \eqref{permb}) on 
$\frac{\ln Z^{\rm rot}}{\beta}$. Recall from \eqref{omegadelta} that $\Omega^a\circ {\cal V} = (-1)^{\kappa a} {\cal V}$. Recall also that ${\mathcal I}$ is function of $E_r$ and $m_i$, both of which are $S_{\kappa+1}$ (hence $\mathbb{Z}_{\kappa+1}$) invariant. We conclude that 
$\Omega^a\circ {\mathcal I}= {\mathcal I}$.  Finally, $\Omega^a$ maps the integration range ${\mathcal D}_0$ to ${\mathcal D}_a$, where ${\mathcal D}_a$ is the polytope built out of the $\kappa$ distinct vectors $V_a^{(b)}$, $b=0, \ldots \kappa, b\neq a$, where
\begin{equation}\label{vabgen}
\left(V_a^{(b)}\right)_j=\frac{c_a^j-c_b^j}{c_a-c_b}
= s_j(a, b)~.
\end{equation}
Note that $V_a^{(b)}=V_b^{(a)}$. Below we will sometimes also use the notation
\begin{equation}\label{vabpab}
p_{ab}=V_a^{(b)}=V_b^{(a)}~.
\end{equation}

In order to properly define the domain of integration described above, we need to specify both the polytope as well as its orientation. 
There is a natural way of associating an orientation of the polytope with an ordering of its defining vectors; given the ordered vector set $(v^1, v^2 \ldots v^{n+1})$,  we declare that the $n$ tuples of vectors $\big( (v^{2}-v^{1}), (v^3-v^1), \ldots (v^{n+1}-v^1) \big)$ are positively/ negatively oriented, depending on whether the $n \times n$ determinant of ordered column vectors is positive or negative. We can, more carefully, specify the integration range ${\mathcal D}_0$
by specifying that the vectors \eqref{vabgen} that form the edges of the polytope are ordered as $(1, 2, 3, \ldots \kappa)$. The permutation operator $\Omega^a$ maps this choice of ordering to the polytope specified by the vectors \eqref{vabgen} ordered as 
\begin{equation}\label{vorder}
\left(V_a^{(a+1)}, V_a^{(a+2)}, \ldots  V^{\kappa}_a, V^{(0)}_a, \ldots, V^{(a-1)}_a \right)~.
\end{equation}
However, we choose to define the polytope ${\mathcal D}_a$
with $V^{(a)}$ ordered in increasing order. Moving to this ordering from that used in \eqref{vorder} requires us to first pull $V^{(0)}$ to the left of all vectors, then pull $V^{(1)}$ to the left of all vectors and so on. The net sign induced by this process equals $(-1)^{ (\kappa -a)a}$. 
Including the sign from the action of $\Omega^a$ on ${\mathcal V}$, we get $(-1)^{a^2}=(-1)^a$. With this `increasing' definition of 
${\mathcal D}_a$, we 
conclude that 
\begin{equation}\label{genalg_Za}
\Omega^a \circ \left(\frac{\ln Z^{\rm rot}}{\beta}\right)
=\frac{2\pi^\kappa(-1)^{\kappa -a}}{\mathcal{V}}
 \int_{{\mathcal D}_a} dm_2 \wedge dm_3 \ldots \wedge d m_\kappa ~{\mathcal I}\left(E,m^{(a)} \right)
\end{equation}
where the factor of $(-1)^\kappa$ is carried over from \eqref{lnzbb}.
In what follows, it will be useful to rewrite \eqref{genalg_Za} in the slightly more compact form 
\begin{equation}\label{genalg_Zaform}
\left(\frac{\ln Z^{\rm rot}}{\beta}\right)_a
=
\frac{2\pi^\kappa}{\mathcal{V}}
(-1)^{\kappa-a}
\int_{{\cal D}_a}\Upsilon~,
\end{equation}
where the $\kappa-1$ form $\Upsilon$ is defined by 
\begin{equation}\label{upsilon}
\Upsilon=
dm_2 \wedge dm_3 \ldots dm_{\kappa} 
~{\mathcal I}~.
\end{equation}

\subsubsection{d=4}\label{genalg_real}

In the simple special case of four dimensions, we have already seen in \eqref{lnzchin} that the contour for the original integral, the contour for the action of $\Omega$, and the contour for the action of $\Omega^2$, respectively, are given by the three oriented intervals
\begin{equation}\label{oldcontfour}
[c_1,c_2]~,
\qquad
[c_2,c_0]~,
\qquad
[c_0,c_1]~.
\end{equation}
in the variable
$z=E_1-m_2$. As explained around \eqref{lnzchin}, the integral then vanishes as a consequence of the fact that the  intervals
in \eqref{oldcontfour} double back on themselves. As a sanity check, we first rederive this result using \eqref{genalg_Za},  specialized to $d=4$ (and the `increasing' definition of ${\mathcal D}_a$). In this case $\kappa=2$, so the $\kappa-1$ component vectors $V_a^{(b)}$ are simply numbers denoting the value $m_2$, and equal
\begin{equation}\label{vabfd}
V_a^{(b)}=c_a+c_b
\end{equation}
\eqref{genalg_Za} thus instructs us to integrate $m_2$  over the intervals 
\begin{equation}\label{intervalsinteg}
[c_1+c_0, c_2+c_0 ] ~, ~~~
[c_1+c_0, c_1+c_2 ]~, ~~~
[c_2+c_0, c_2+c_1 ]~, 
\end{equation}
but to weight three integrals above with the relative signs $+$, $-$ and $+$ 
respectively. Using  $E_1=c_0+c_1+c_2$, and absorbing the middle minus sign into a reversal of the contour,  we see this exactly agrees with the integral of  $z$ over the contour \eqref{oldcontfour}, as anticipated. 

\subsubsection{$d=6$}

Let us now study the case $d=6$. 
For the purposes of visualization, we assume in this subsubsection that $c_a$ take real and distinct values; however the fact that ${\mathcal I}$ is 
a polynomial in $c_a$ allows for a straightforward analytic continuation, so all our final results also apply at complex values of $c_a$.

In this case the vectors $V_a^{(b)}=V_b^{(a)}=p_{ab}$ are two component vectors $(m_2, m_3)$. The four regions ${\mathcal D}_0$, ${\mathcal D}_1$, ${\mathcal D}_2$ and ${\mathcal D}_3$ 
are each triangles on the $m_2, m_3$ plane. The vertices of the triangle corresponding to ${\mathcal D}_a$ are the three points $p_{ab}$, $b=0,\ldots, 3$, $b \neq a$, where 
\begin{equation}\label{pabsix}
p_{ab}
=
\left(
 c_a+c_b,
 c_a^2+c_a c_b+c_b^2
\right)~.
\end{equation}
Each of these four triangles shares exactly one vertex with every other triangle (the common vertex between ${\mathcal D}_a$ and ${\mathcal D}_b$ is $p_{ab}$).
According to \eqref{genalg_Za}, the ${\mathfrak S}\circ \frac{\ln Z}{\beta}$ is obtained by computing the integral of the given two form over the surface
\begin{equation}\label{sixdregion}
{\cal D}_0-{\cal D}_1+{\cal D}_2-{\cal D}_3~.
\end{equation} 

As the 12 components of the 6 vectors $p_{ab}$ are all determined by  four numbers $c_a$; as a consequence the six vectors $p_{ab}$ are far from generic. It will nonetheless prove  illuminating to first  study the  integration domain \eqref{sixdregion}
for six generic vectors $p_{ab}$, and only subsequently specialize to $p_{ab}$ given by \eqref{pabsix}.

When $p_{ab}$ are generic, the integration domain \eqref{sixdregion} can be visualized as  
in Fig. \ref{fig:genalg-octahedron} below. ${\mathcal D}_0$, 
${\mathcal D}_1$,
${\mathcal D}_2$,
and ${\mathcal D}_3$ are the four shaded triangles in Fig. \ref{fig:genalg-octahedron}; note that every triangle shares exactly one common vertex with every other triangle. While Fig. \ref{fig:genalg-octahedron}
has been drawn in a 3 dimensional manner, to aid visualization, 
the reader should mentally  visualize the points $p_{02}$ and $p_{13}$ as being squashed inwards (towards the center of the figure) so that the entire octahedron of Fig. \ref{fig:genalg-octahedron} is flattened onto the  single plane (the plane of $p_{12}, p_{23}, p_{03}, p_{01}$). 
 If we visualize shaded triangles as triangular sheets of paper, the dark blue triangles  carry one orientation (upper flap of the flat paper origami) and the two light blue triangles represent the opposite orientation (lower flap of the flat paper origami). 
The resultant flat origami (integration domain)  lies on the $(m_2, m_3)$ plane, and  covers parts of this plane. While some portions of this plane are covered by only the upper flap (positive orientation), other portions are covered by only the lower flap (negative orientation) and still other portions are covered by both flaps. While the integral receives no contributions from the last region (as the integrals over oppositely oriented surfaces cancel), it  receives contributions from the first and second regions, and so does, in general, evaluate to a nonzero result.  

Continuing to work with generic $p_{ab}$, let us now study an augmented  integration domain. Our augmented domain is defined to include also the four triangular holes (which can be visualized as white paper triangles) in Fig. \ref{fig:genalg-octahedron} (in addition to the four blue shaded triangles already studied above). Orientations are chosen so that the normal to every triangle is outward pointing. 
This augmented integration domain is a flat origami whose upper and lower flaps each consist of two blue and two white triangles, and can be thought of as a `closed kite shaped envelope'.
When we paste this new flat origami onto the $(m_2, m_3)$ plane, every covered point is covered by both an upper and a lower flap. As a consequence, the integral over the augmented integration domain simply vanishes for arbitrary choices of the six vectors  $p_{ab}$.

Now the augmented integration domain effectively coincides with the actual domain \eqref{sixdregion}
 when each of the four holes (white triangles) in Fig. \ref{fig:genalg-octahedron} has zero area, i.e.    when the  three vectors ($p_{ab}$, $p_{ac}$, $p_{bc}$)
are collinear, for each of the four possible choices of the triplet $(a, b, c)$. This collinearity condition is met only at finely tuned choices for  $p_{ab}$.
 As emphasized  above, however,
$p_{ab}$ given by \eqref{pabsix} are very far from generic; and the collinearity conditions above are, in fact, always obeyed. 
One way of seeing this is to note (from \eqref{pabsix}) that  the  components ($m_2, m_3$) of each of the three vectors $p_{ab}$, $p_{bc}$, 
and $p_{ac}$ obey the equation  
\begin{equation}\label{genalg_line}
m_3=(c_a+c_b+c_c)m_2
-(c_a c_b+c_a c_c+c_b c_c)~.
\end{equation} \footnote{\eqref{genalg_line} is the assertion that each of these three vectors lies  on the line  $y=mx+c$ where  here $y \rightarrow m_3$, $x\rightarrow m_2$, $m\rightarrow c_a+c_b+c_c$ and $c \rightarrow -(c_a c_b+c_a c_c+c_b c_c)$. }
for each of the four choices of the triplet $(a, b, c)$. It follows that in the case under study, the original integration domain \eqref{sixdregion} effectively coincides with the augmented integration domain, and the integration over \eqref{sixdregion} actually vanishes, so  that  
\begin{equation}\label{mfoo}
{\mathfrak S}\circ \left( \frac{\ln Z^{\rm rot}}{\beta} \right)=0
\end{equation}
as claimed. 

\begin{figure}[!ht]
\centering
\includegraphics[width=0.92\linewidth]{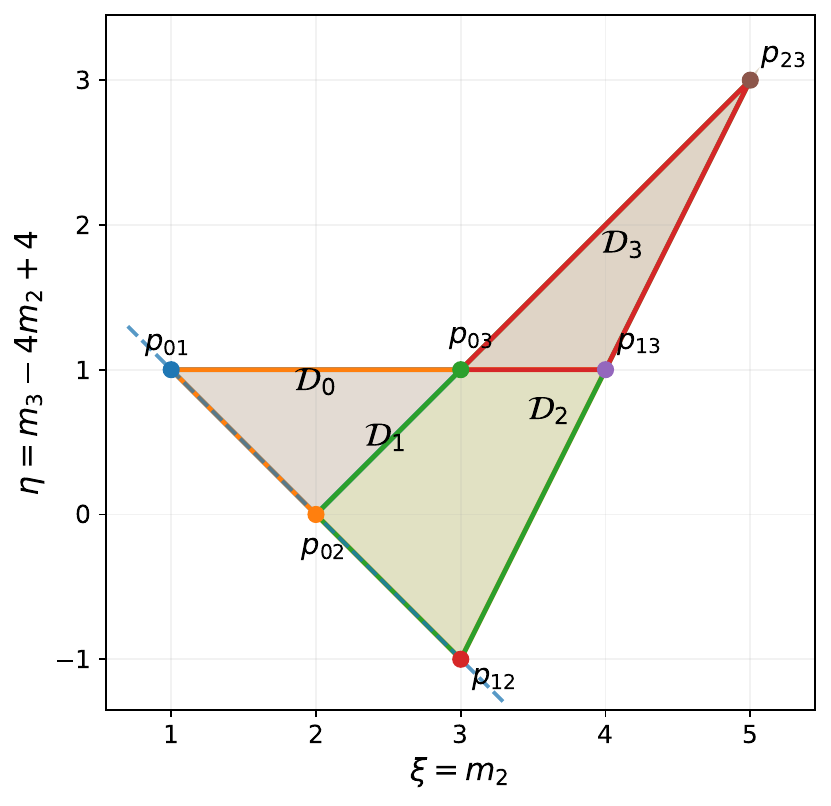}
\caption{The four integration regions in $d=6$ for real $(c_0,c_1,c_2,c_3)=(0,1,2,3)$, shown in the coordinates $\xi=m_2$, $\eta=m_3-4m_2+4$.  The three corners $p_{01},p_{02},p_{12}$ lie on one straight line.  The same statement holds for every choice of three labels, and the corresponding oriented boundary segments cancel.}
\label{fig:genalg-real6}
\end{figure}

Before generalizing to higher dimensions,  it is interesting to pause to  visualize the integration region \eqref{sixdregion}
in the physically relevant collinear case. 
Upon mentally squashing the octahedron of Fig. \ref{fig:genalg-octahedron} down to plane, in this case,  we obtain a flattened origami of the form depicted in 
Fig. \ref{fig:genalg-real6}. Fig. \ref{fig:genalg-real6} has been plotted for the illustrative choice $(c_0,c_1,c_2,c_3)=(0,1,2,3)$.  In order to conveniently display the triangles (without changing their incidence relations), the figure uses the equivalent coordinates $\zeta=m_2$ and $\eta=m_3-4m_2+4$.

We have explained above that the collinearity condition on $p_{ab}$ ensures that the (originally open) integration domain \eqref{sixdregion} actually effectively coincides with the closed augmented integration domain. This point can directly be understood from Fig. \ref{fig:genalg-real6} as follows. The integration region \eqref{sixdregion} has four boundaries that lie along the four straight lines in 
Fig. \ref{fig:genalg-real6}. However these boundaries actually double back on each other, and so can (for the purposes of computing an integral) equally well be sewn  together to form a closed surface (this closed surface is the augmented domain). For example, the part of the oriented boundary involving the labels $0,1,2$ lies on the leftmost slanting line in Fig. \ref{fig:genalg-real6}, and equals
\begin{equation}\label{fourdint}
[p_{01},p_{02}]-[p_{01},p_{12}]+[p_{02},p_{12}]~.
\end{equation}
A glance at Fig.~\ref{fig:genalg-real6} will convince the reader that the three segments in \eqref{fourdint} lie on the same line, and the line segment (second term in \eqref{fourdint}) is the sum of the two shorter ones but with opposite orientation. Hence the integration domain does not really have a boundary. The same argument applies to each choice of three labels (each of the four straight lines in Fig.~\ref{fig:genalg-real6} so the oriented boundary of
\begin{equation}\label{genalg_C6}
{\cal D}_0-{\cal D}_1+{\cal D}_2-{\cal D}_3
\end{equation}
effectively vanishes, and the integration domain is effectively closed. 

The vanishing of the integral over the domain depicted in Fig.~\ref{fig:genalg-real6} can also be verified very explicitly as follows. In Fig.~\ref{fig:genalg-real6}, the triangles $(p_{01},p_{02}, p_{03})$ and $(p_{20},p_{21}, p_{23})$ are positively oriented (their normal points out of the paper) and constitute the upper flap of the flattened origami. Together, these two triangles cover the shaded region once. On the other hand, the triangles 
$(p_{10},p_{12}, p_{13})$ and $(p_{30},p_{31}, p_{32})$ are effectively negatively oriented (because of the minus signs in \eqref{sixdregion}); together they also cover the shaded region in Fig.~\ref{fig:genalg-real6}. We thus see that every point on the integration domain in Fig.~\ref{fig:genalg-real6} occurs twice; once with positive and once with negative orientation. Consequently the integral vanishes. 

Finally, the vanishing of the integral over the surface of Fig.~\ref{fig:genalg-real6} can also be argued in the following equivalent fashion. As the real $(m_2,m_3)$ plane is contractible, the integrand two-form $\Upsilon$ (that appears in  \eqref{genalg_Za}) can be written as $\Upsilon=d\alpha$ (recall that every top form is always closed, and the contractibility of the integration region guarantees that this closed form is also exact). 
Stokes' theorem then gives
\begin{equation}\label{genalg_stokes6}
\int_{{\cal D}_0}\Upsilon-
\int_{{\cal D}_1}\Upsilon+
\int_{{\cal D}_2}\Upsilon-
\int_{{\cal D}_3}\Upsilon
=0~,
\end{equation}
where we have used that the boundary of the integration region is effectively trivial.

\subsubsection{General even dimensions}

We now turn to the generalization of the discussion above to every even dimension. In each such dimension, we have 
\begin{equation}\label{intrelgen}
\mathfrak{S}\circ \left(\frac{\ln Z^{\rm rot}}{\beta}\right)
=
\frac{2\pi^\kappa}{(\kappa+1)\,\mathcal{V}}
(-1)^{\kappa}
\int_{{\cal D}}\Upsilon~,
\end{equation}
where the integration domain ${\mathcal D}$ is given by 
\begin{equation}\label{md}
{\mathcal D}={\mathcal D}_0-{\mathcal D}_1+{\mathcal D}_2-\ldots +(-1)^\kappa {\mathcal D}_\kappa~.
\end{equation}
Here ${\mathcal D}_a$ is the $\kappa-1$ dimensional simplex 
constructed by the $\kappa$ vectors $p_{ab}, ~~~(b=0 \ldots \kappa, ~b\neq a)$, with the `increasing' orientation. Note that this simplex has $\kappa$ distinct corners (these are the vectors $p_{ab}$, with $b \neq a$). These simplices are line intervals when $\kappa=2$ (see \eqref{fourdint}), triangles when 
$\kappa=3$ (see Fig. \ref{fig:genalg-octahedron}) and tetrahedra when $\kappa=4$. They are simply called  $(\kappa-1)$ simplices) for $\kappa \geq 5$.

\eqref{md} tells us that the integration domain is the union of $\kappa+1$ distinct simplices. Each simplex shares exactly one common corner (or vertex) with every other simplex. As in our $d=6$ discussion above, we first study the integration domain for arbitrary values of the $\kappa-1$ component vectors $p_{ab}$ (the components being $(m_2,\dots ,m_\kappa)$), and only later examine the implications of the fact that these $\frac{(\kappa+1)\kappa}{2}$  $(\kappa-1)$ component vectors  are actually determined by the $\kappa+1$ numbers $c_a$ via Eq.~\eqref{vabgen}. 

In our $d=6$ discussion above,  we have explained above (see around Fig. \ref{fig:genalg-octahedron}) that the integration domain for generic $p_{ab}$ can be visualized as consisting of four of the eight flat surfaces of a 3d solid octahedron. The generalization of this statement is the following.  Topologically speaking, the integration domain in $d=2 \kappa$, $\kappa\geq 3$,  consists of $\kappa+1$ of the $2(\kappa+1)$ flat boundaries of $\kappa$ dimensional solid surface called a $\Delta(2, \kappa+1)$, which we now describe in detail. 

\begin{figure}[t]
  \centering
\includegraphics[width=\columnwidth]{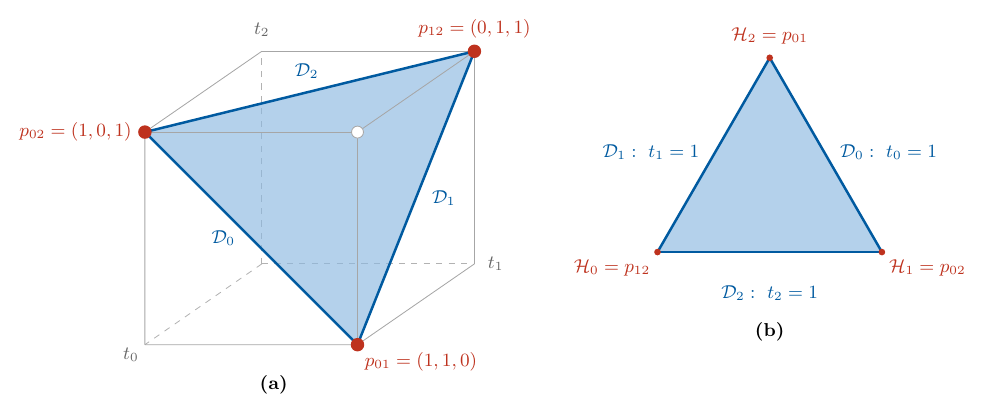}
  \caption{The polytope $\Delta(2,\kappa+1)$ for $\kappa=2$. Panel (a) shows
  it as the triangle cut out of the unit cube by the plane $\sum_a t_a=2$;
  panel (b) shows the same triangle flat. The blue edges are the faces
  $t_a=1$, which map to the integration regions $\mathcal D_a$. The red
  vertices are the faces $t_a=0$, which map to the boundary regions
  $\mathcal H_a$ and here degenerate to points. For $\kappa=2$ the three
  $\mathcal D_a$ form a closed contour.}
  \label{fig:polytope-kappa2}
\end{figure}

Consider a Cartesian $R^{\kappa+1}$ space. 
$\Delta(2,\kappa+1)$ is the intersection of the plane $\sum_a t_a=2$ with the unit cube of this 
Cartesian space. In equations
\begin{equation}\label{polytope}
\Delta(2,\kappa+1):=
\left\{
(t_0,\ldots,t_\kappa)\,\Big|\,
0\leq t_a\leq1~,
\sum_{a=0}^{\kappa}t_a=2
\right\}~.
\end{equation}
This shape is easily visualized in the case $\kappa=2$; see Fig. \ref{fig:polytope-kappa2} where it consists of the interior of a triangle. The boundaries of this triangle are the three integration domains ${\mathcal D}_0$, ${\mathcal D}_1$ and ${\mathcal D}_2$, together with the three red points in this figure. The fact that the red dots are of one lower dimension than the integration domains is a non generic feature that does not persist to $\kappa \geq 3$. At these values of $\kappa$, this  $\kappa$ dimensional `polytope' (see Eq.~\eqref{polytope}) has two different kinds of flat $\kappa-1$ dimensional  boundaries; first the $\kappa+1$ `shaded' surfaces $t_a=1, ~~(a=0\ldots \kappa)$, and second the $\kappa+1$ `white' surfaces $t_a=0, ~~~(a=0\ldots \kappa)$. Topologically speaking, we identify the  $a^{th}$
shaded surface $t^a=1$  with our integration surfaces ${\mathcal D}_a$, and the $a^{th}$ `white' surface $t_a=0$ with `holes' between these integration surfaces. 

The identification spelled out above can be justified as follows.  Notice that the surface element $t^a=1$ is given by the equation $\sum t^b=1$, $b \neq a$, $t^b>0$,  and so defines  a $(\kappa-1)$ simplex. Moreover, the $a^{th}$ simplex touches the $b^{th}$ simplex at exactly one point $t_a=t_b=1$. This is exactly the connectivity of the 
${\mathcal D}_a$ above, justifying our identification. Note that it follows from this discussion that the corners $t_a=t_b=1$, $b=0\ldots \kappa$, $b\neq a$ should be topologically identified with $p_{ab})$.

The full boundary of $\Delta(2, \kappa+1)$ is a closed oriented surface; topologically speaking, our integration domain consists of a part of this surface. We refer to the remainder of this surface as holes. The holes on this surface are made up of the $(\kappa+1)$ patches $t^a=0$.  \footnote{Topologically speaking, these are  the general dimensional analogues of the four
`holes' or `white shaded surfaces' in Fig. \ref{fig:genalg-octahedron}.} It follows immediately from the definition 
\eqref{polytope}, that each of these `holes' is, itself, a $\Delta(2, \kappa)$. \footnote{In the special case $\kappa=2$, this surface consists of the intersection of the line 
$t_1+t_2=2$ with the unit square in $R^2$. Clearly this intersection is given by a single point $t_1=t_2=1$, explaining the degenerate nature of the holes when $\kappa=2$.}
Notice that the hole $t^a=0$ shares a boundary with each of the shaded surfaces $t^b=1$ with $b\neq a$. Clearly, the  common boundary between these surfaces is itself a $\kappa-2$ dimensional simplex constructed out of the $\kappa-1$ vectors 
\begin{equation}\label{vecho}
p_{cb}, ~~c=0\ldots \kappa, ~c\neq b,~ c\neq a.
\end{equation}
In addition, the hole $t^a=0$ also shares a boundary with the hole $t^b=0$, for all $b \neq a$. It follows from definitions
that this common boundary is, in turn, a $\Delta(2, \kappa-1)$. 

We reemphasize that the discussion of the previous paragraphs describes our 
integration domain only in a topological sense. The $R^{\kappa+1}$ space in which \eqref{polytope} is embedded is purely fictional (it has nothing to do with the physical space parameterized by  $m_i$). The precise statement is that there exists a continuous map from the surface of the polytope \eqref{polytope} into the $R^{\kappa-1}$ parameterized by 
$m_2 \ldots m_\kappa$, such that the shaded surfaces $t_a=1$ map to ${\mathcal D}_a$, the $a^{th}$ component of the integration domain ${\mathcal D}_a$. Under the same map, the `white' surfaces $t_a=0$ map to holes in the integration domain. 

In the physical problem of interest, our integration surfaces 
${\mathcal D}_a$ are given and fixed. On the other hand the holes ${\mathcal H}_a$ are, of course, not specified; we may take them to be any surfaces that complete the integration contour ${\mathcal D}$ into a closed oriented contour. The only thing that is fixed about these surfaces is their $\kappa-2$ dimensional boundaries with the $\kappa$ integration regions ${\mathcal D}_b$ $b\neq a$. This $\kappa-2$ dimensional boundary is itself a simplex 
generated by the $\kappa-1$ vectors $p_{bc}$, $c\neq a, ~c\neq b$ (as mentioned, we have one such boundary for each ${\mathcal D}_b$, $b\neq a$).

We denote this $\kappa-1$ dimensional oriented region  of $R^{\kappa-1}$ (parameterized by $m_i$) as ${\mathcal  H}_a$. \footnote{Orientations of the surfaces of the polytope \eqref{polytope} are chosen to make them all `outward pointing', and the orientation of the domain ${\mathcal H}_a$ then follows from the continuous map that takes us from the surface of \eqref{polytope} to $R^{\kappa-1}$.}

In the discussion above we have characterized the `shaded
integration surfaces' and the `white holes' in every $\kappa$; however our characterization was somewhat abstract. As a brief aside, in the table below we provide an explicit listing of the shape of the solid surfaces ($\kappa-1$ simplices) and holes 
($\Delta(2, \kappa)$) in low dimensions:
\begin{center}\label{tablelisting}
\begin{tabular}{c|c|c|c}
\toprule
$d$ & polytope & $t_a=1$ faces & $t_a=0$ faces\\
\midrule
$4$ & $\Delta(2,3)$ & intervals & points\\
$6$ & $\Delta(2,4)$, octahedron & triangles & triangles\\
$8$ & $\Delta(2,5)$, rectified 5-cell & tetrahedra & octahedra\\
$10$ & $\Delta(2,6)$ & $4$-simplices & rectified 5-cells\\
\bottomrule
\end{tabular}
\end{center}

As an aside for the interested reader, we pause to describe the case $d=8$ in a little more detail (the reader who finds herself uninterested in this exercise in visualization can safely skip to the next paragraph). \begin{figure}[!ht]
\centering
\includegraphics[width=0.96\linewidth]{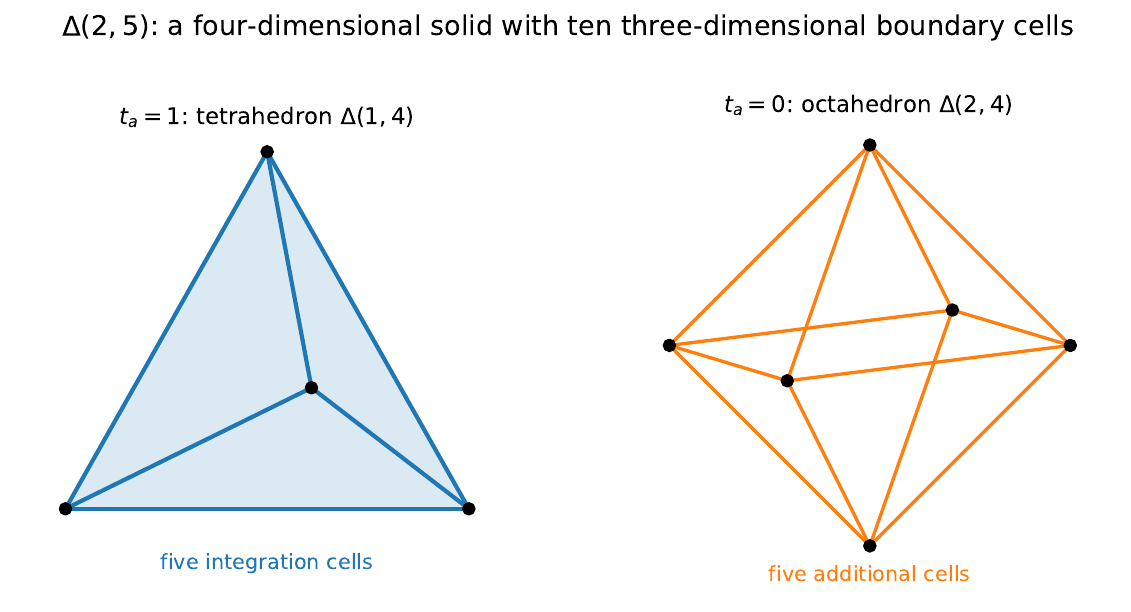}
\caption{The boundary cells of $\Delta(2,5)$ in $d=8$.  There are five tetrahedral faces $t_a=1$, one for each term in the cyclic sum, and five octahedral faces $t_a=0$.}
\label{fig:genalg-d8}
\end{figure}
In this case the polytope is $\Delta(2,5)$, a four-dimensional solid, usually called the rectified 5-cell (see the table above). As $\Delta(2,5)$ is four dimensional it is hard to visualize. However its boundary components are three dimensional, and so can be visualized. The boundary consists of five tetrahedra $t_a=1$ (these constitute our integration domain) and five octahedra $t_a=0$ (these constitute our holes).  Thus the five terms in the cyclic sum are represented by tetrahedral integration regions, while the five additional faces are octahedra.  Figure~\ref{fig:genalg-d8} displays the two kinds of three-dimensional boundary cells.

The remaining discussion proceeds as in $d=6$. As the integrand on the RHS \eqref{intrelgen} is (order by order) a polynomial (hence analytic) in the $\kappa-1$ $m_i$, its integral over any closed $\kappa-1$ dimensional surface vanishes as a consequence of Stokes's theorem. Applying this to the map of the boundary of the polytope $\Delta(2, \kappa+1)$, we conclude that the RHS of \eqref{intrelgen} vanishes when the integration domain ${\mathcal D}$ is generalized to ${\mathcal D} +{\mathcal H}$ where 
\begin{equation}\label{deftilded}
{\mathcal H}=\sum_{a=0}^\kappa {\mathcal H}_a~.
\end{equation}
In equations 
\begin{equation}\label{whatvan}
\int_{{\mathcal D} +{\mathcal H}} \Upsilon=0~.
\end{equation}

As in $d=6$, this is the most we can say when $p_{ab}$ are arbitrary vectors. As in $d=6$, however, it is very important that $p_{ab}$  are not actually arbitrary, but are instead given by Eq.~\eqref{vabgen}. This has the following crucial consequence. Recall that we are free to choose the shapes of the holes ${\mathcal H}_a$ subject only to the following constraints; each of $\kappa$ boundaries of the region ${\mathcal  H}_a$ with the integration domain ${\mathcal D}_b$ is given by Eq.~\eqref{vecho}, (for the $\kappa$ values of $b=0, \ldots \kappa, ~~b \neq a$). In addition, the boundary of the hole ${\mathcal H}_a$ with the hole ${\mathcal H}_b$ (for each of the $\kappa$ possible value of 
$b$ with $b\neq a$) must, of course, match with the boundary of ${\mathcal H}_b$ with ${\mathcal H}_a$.

In the subsequent paragraph, we will demonstrate that each of the vectors in Eq.~\eqref{vecho} actually lie in a particular $\kappa -2$ dimensional hyperplane of $R^{\kappa -1}$; the same hyperplane for each of the values of $b$ listed above (i.e. for each $b\neq a$)\footnote{The  plane in question depends on the `which hole' label $a$, but not on the `which boundary' label $b$. } \footnote{This remarkable fact is the analogue of the fact that all boundaries of a particular hole lie on the same straight line (and in fact form a closed loop) in $d=6$, as explained in detail in the previous subsubsection.} Since all boundaries of the hole ${\mathcal H}_a$ with the fixed integration regions ${\mathcal D}_b$ lie in the same $\kappa-2$ dimensional hyperplane, it follows that region ${\mathcal H}_a$ can be chosen to lie entirely in the same $\kappa-2$ hyperplane. \footnote{So that the boundary between ${\mathcal H}_a$ and ${\mathcal H}_b$ lies in the generically $\kappa-3$ dimensional intersection of the relevant subspaces.} With this choice, the  integral over this surface vanishes. Since this is true for every value of $a$, the integral over ${\mathcal H}$ vanishes. Consequently \eqref{whatvan} becomes 
\begin{equation}\label{whatvann}
\int_{{\cal D}}\Upsilon=0
\end{equation}
and it follows from \eqref{intrelgen} that 
$${\mathfrak S} \circ \left( \frac{\ln Z^{\rm rot}}{\beta} \right)=0$$
as we set out to prove.

As promised above, we now demonstrate that the vectors 
$p_{bd}$ with $b \neq a$ and $d \neq a$ all lie in a given 
$\kappa-2$ plane. 
Define 
\begin{equation}\label{qadef}
q_a(t):=\prod_{m\neq a}(t-c_m)
=t^\kappa-e_1^{(a)}t^{\kappa-1}+e_2^{(a)}t^{\kappa-2}-\cdots+(-1)^\kappa e_\kappa^{(a)}~,
\end{equation}
(notice the product above runs over all $m \neq a$). Now consider some $d\neq a$ and another $b\neq a$.  It follows 
immediately from the definition \eqref{qadef} that 
$q_a(c_d)=q_a(c_b)=0$. Consequently $q_a(c_d)-q_a(c_b)=0$. Dividing this by $c_d-c_b$, using \eqref{qadef} and the 
definition \eqref{sdef}, we conclude that 
\begin{equation}\label{hyperplane}
m_\kappa-e_1^{(a)}m_{\kappa-1}
+e_2^{(a)}m_{\kappa-2}-\cdots
+(-1)^{\kappa-1}e_{\kappa-1}^{(a)}=0~,
\end{equation}
where $m_j$ are the components of the vector $p_{db}$. Recall that \eqref{hyperplane} holds for the components of any $p_{db}$, for any choice of  $d, b \neq a$. Since \eqref{hyperplane} defines a codimension one hyperplane in the $R^{\kappa-1}$ parameterized by the $m_j$, it follows that the simplices that make up all the boundaries of 
the $a^{th}$ hole lie in this hyperplane, as we needed to prove. 

\subsection{Evaluation of cyclic sum of the Kerr-AdS$_{2\kappa+1}$ black hole partition function }
In this small section, we will derive the cyclic identity for the Kerr-AdS$_{2\kappa+1}$ black hole partition function. We start with the expression $\frac{\ln Z_{\rm BH}}{\beta}$ as written in \eqref{eq:general-thermal-action}. We apply $\Omega^a$ ($\Omega$ or $\Omega_{\rm bh}$ as defined in \eqref{zthreebh}) on it and it gives us the following
\begin{equation}
\begin{split}
\Omega^a\circ\left[\frac{\ln Z_{\rm BH}}{\beta}\right]&=\frac{\pi^{\kappa-1}}
{8(\kappa-1)!G_{2\kappa+1}}
\left(1-q_a^2\right)
\prod_{\substack{i=0\\i\neq a}}^{\kappa}
\frac{1+q_i^2}{q_a^2-q_i^2}\\&=\frac{\pi^{\kappa-1}}
{8(\kappa-1)!G_{2\kappa+1}}
\left[\prod_{j=0}^\kappa\left(1+q^2_j\right)\right]\frac{1-q_a^2}{1+q_a^2}
\prod_{\substack{i=0\\i\neq a}}^{\kappa}
\frac{1}{q_a^2-q_i^2}~,\qquad\mathrm{for}~~a=0,1,\cdots,\kappa~.    
\end{split}    
\end{equation}
Summing over $a$ from $0$ to $\kappa$, we obtain
\begin{equation}\label{eq:orbit_sum_of_BH}
\Big(1+\mathrm{\Omega}+\cdots+\mathrm{\Omega}^\kappa\Big)\circ\left[\frac{\ln Z_{\rm BH}}{\beta}\right]=\frac{\pi^{\kappa-1}}
{8(\kappa-1)!G_{2\kappa+1}}
\left[\prod_{j=0}^\kappa\left(1+q^2_j\right)\right]\sum_{a=0}^{\kappa}\frac{1-q_a^2}{1+q_a^2}
\prod_{\substack{i=0\\i\neq a}}^{\kappa}
\frac{1}{q_a^2-q_i^2}~.   
\end{equation}
We will evaluate the $a$ sum by using contour integration method. To do that, we will define the following function $f(z)$ such that
\begin{equation}
 f(z):=\frac{1-z}{1+z}\prod_{i=0}^{\kappa}
\frac{1}{z-q_i^2}~.   
\end{equation}
We can see that $f(z)$ has simple poles at $z=q_a^2$ for $a=0,1,\cdots,\kappa$ and also at $z=-1$. For the large value of $|z|$, the behavior of $f(z)\sim-\frac{1}{z^{\kappa+1}}$. So $f(z)$ falls off at least  as fast as $\frac{1}{z^2}$ for $\kappa\ge 1$. Hence, we conclude that the contour integration of $f(z)$ on a circle (with center at the origin of the complex $z$ plane) vanishes as its radius tends to infinity. Using the residue theorem, we can write
\begin{equation}\label{eq:resdue_theorem}
\begin{split}
\sum_{a=0}^{\kappa}\mathrm{Res}_{z=q_a^2}f(z)+\mathrm{Res}_{z=-1}f(z)=0~.    
\end{split}    
\end{equation}
The residues of $f(z)$ at $z=q_a^2$ are given by
\begin{equation}
\mathrm{Res}_{z=q_a^2}f(z)=\frac{1-q_a^2}{1+q_a^2}\prod_{\substack{i=0\\i\neq a}}^{\kappa}\frac{1}{q_a^2-q_i^2}~,    
\end{equation}
and the residue of $f(z)$ at $z=-1$ is given by
\begin{equation}
\mathrm{Res}_{z=-1}f(z)=2\prod_{j=0}^{\kappa}\frac{1}{\left(-1-q_j^2\right)}=-2(-1)^\kappa \prod_{j=0}^{\kappa}\frac{1}{\left(1+q_j^2\right)}~.    
\end{equation}
Substituting the residues into Eq.~\eqref{eq:resdue_theorem}, we get
\begin{equation}
\sum_{a=0}^\kappa \frac{1-q_a^2}{1+q_a^2}\prod_{\substack{i=0\\i\neq a}}^{\kappa}\frac{1}{q_a^2-q_i^2}=2(-1)^\kappa \prod_{j=0}^{\kappa}\frac{1}{\left(1+q_j^2\right)}~.   
\end{equation}
Substituting the above identity in Eq.~\eqref{eq:orbit_sum_of_BH}, we have obtained the Eq.~\eqref{bhident}
\begin{equation}\label{bhident}
\Big(1+\mathrm{\Omega}+\cdots+\mathrm{\Omega}^\kappa\Big)\circ\left[\frac{\ln Z_{\rm BH}}{\beta}\right]=(-1)^\kappa\frac{\pi^{\kappa-1}}
{4(\kappa-1)!G_{2\kappa+1}}=(-1)^\kappa \kappa!a_{2\kappa}    
\end{equation}
where the $a$ anomaly $a_{2\kappa}$ in $d=2\kappa$ dimensional holographic CFT is $a_{2\kappa}=\frac{\pi^{\kappa-1}}
{4\kappa!(\kappa-1)!G_{2\kappa+1}}$.

\subsection{Spectral equivalence between $S^3\times \mathbb{R}$ and $AdS_3\times S^1$}

\begin{figure}[H]
    \centering
    \includegraphics[width=0.7\linewidth]{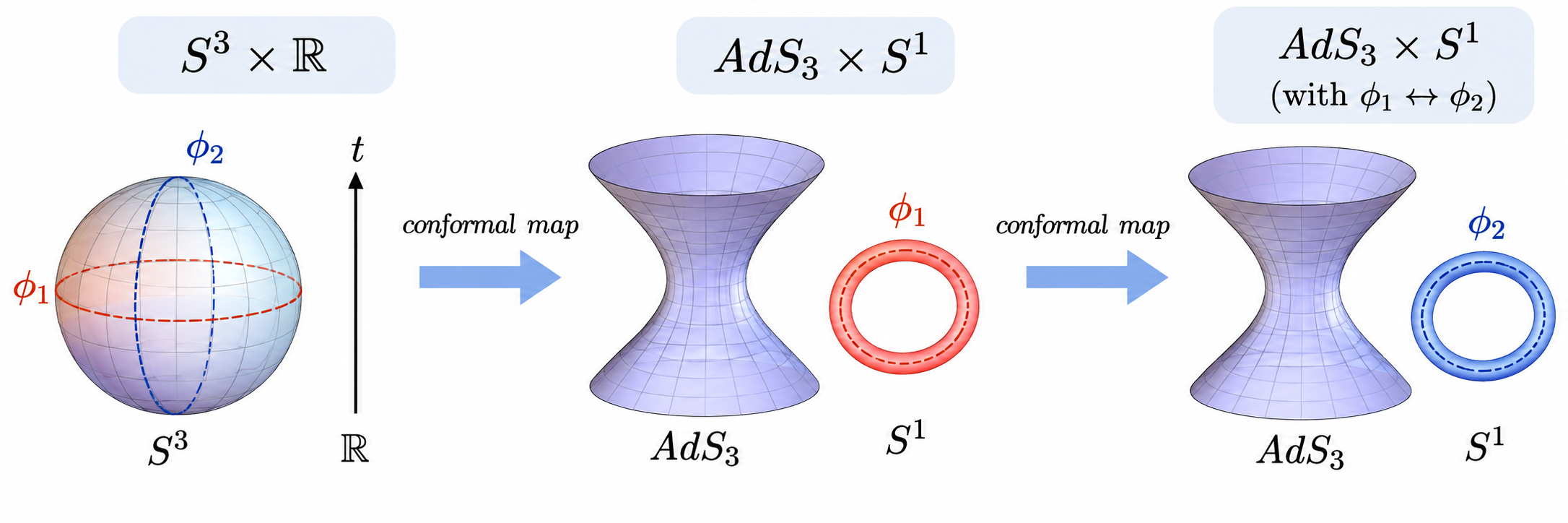}
    \caption{The map from spheres to hyperbolic spaces}
    \label{sads}
\end{figure}

We consider a conformally coupled scalar on $S^3\times \mathbb{R}$ which satisfies the following equation,
\begin{equation}\label{scalarEOM}
   \left (\Box - \xi R_{S^3}\right) \Phi(x) = 0
\end{equation}
The metric of $S^3\times \mathbb{R}$ in the Hopf coordinates is given by,
\begin{equation}
ds_{S^3 \times R}^2=
-dt^2 +d\theta^2
+\sin^2\theta\,d\phi_1^2
+\cos^2\theta\,d\phi_2^2,
\end{equation}
where the scalar modes can be written as
\begin{equation}
\Phi=e^{-iEt} Y_{n,m,p}\left(\theta,\phi_1,\phi_2\right),~~~
n,m\in\mathbb Z.
\end{equation}
where the scalar harmonics $Y_{n,m,p}\left(\theta,\phi_1,\phi_2\right)$ on $S^3$ can be written as
\begin{equation}
Y_{n,m,p}\left(\theta,\phi_1,\phi_2\right)
\propto
e^{in\phi_1+im\phi_2}
(\sin\theta)^{|n|}
(\cos\theta)^{|m|}
P_p^{(|n|,|m|)}(\cos 2\theta),~~ p=0,1,2,\ldots,
\end{equation}
where, $P_p^{(|n|,|m|)}(z)$ is the Jacobi polynomial of degree $p$.

Scalar harmonics $Y_{n,m,p}$ of $S^3$ satisfy
\begin{equation}\label{S3spectrum}
-\nabla_{S^3}^2Y_l=
l(l+2)Y_l ~~~\qquad
l=|n|+|m|+2p.
\end{equation}
Using  Eq.~\eqref{S3spectrum} and including the contribution of the conformal coupling $\xi R_{S^3}=\frac{1}{6}\times 6=1$ to Eq.~\eqref{scalarEOM}, 
we obtain 
\begin{equation}
E^2=l(l+2)+1=(l+1)^2.
\end{equation}
Finally, we obtain the energy as
\begin{equation}\label{energyS3}
E_{S^3\times \mathbb{R}}=
|n|+|m|+2p+1.
\end{equation}

Next, we perform the following coordinate change,
\begin{equation}\label{cotrans}
l=\sin^2\theta=\frac{1}{\cosh^2\rho},
\end{equation}
and rewrite the metric of
$S^3\times \mathbb{R}$ as $AdS_3\times S^1$ up to some Weyl factor. The metric takes the form,
\begin{equation}
ds_{S^3\times R}^2
= W^2
\left[
d\phi_1^2
+d\rho^2
-\cosh^2\rho\,d t^2
+\sinh^2\rho\,d\phi_2^2
\right] = \frac{1}{\cosh^2\rho}ds^2_{AdS_3\times S^1}; ~~~W^2=\frac{1}{\cosh^2\rho}
\end{equation}
where, $ds^2_{AdS_3\times S^1}$ is the metric of $AdS_3\times S^1$, with $S^1$ parametrized by $\phi_1$. Moreover, we impose smoothness of the scalar field where the $\phi_1$ circle shrinks, i.e.,
\begin{equation}
\theta=0,~~~
l=\sin^2\theta\rightarrow 0,
\end{equation}
and locally, the metric takes the form
\begin{equation}
ds^2 \approx
dr^2+r^2d\phi_1^2+\cdots,
\qquad
r\approx\sin\theta=\sqrt{l}.
\end{equation}
For a mode with angular momentum $n$ $$\Phi\sim e^{in\phi_1}$$ Hence, the smoothness at the origin implies 
\begin{equation}
\Phi_{S^3}
\sim
r^{|n|}
\sim
l^{|n|/2}.
\end{equation}
Upon using  Eq.~\eqref{cotrans},
this gives
\begin{equation}
\Phi_{S^3}
\sim
(\cosh\rho)^{-|n|}.
\end{equation}

Since the scalar has Weyl weight one, we obtain
\begin{equation}\label{bdy}
\Phi_{AdS_3\times S^1}=
W \,\Phi_{S^3} \sim
(\cosh\rho)^{-(|n|+1)},
\qquad
\rho\rightarrow\infty.
\end{equation}

Thus, the smoothness boundary condition at the shrinking $\phi_i$ corresponds to the normalizable boundary condition on $AdS_3$.

For a conformally coupled scalar on $AdS_3\times S^1$, if we expand it along the non-contractible circle,
\begin{equation}
\Phi_{AdS_3\times S^1}=
e^{in\phi_1}\varphi_n,
\end{equation}
the Kaluza--Klein momentum contributes $n^2$, and conformal coupling contributes $-1$.Therefore the
effective $AdS_3$ mass is
\begin{equation}
m_n^2=n^2-1.
\end{equation}

Using the AdS-CFT map, we obtain
\begin{equation}\label{deltan}
m_n^2
=\Delta_n(\Delta_n-2). \implies \Delta_n = 1 \pm |n|
\end{equation}

Next, using the falloff condition in  Eq.~\eqref{bdy} we obtain,
\begin{equation}
\Delta_n=|n|+1.
\end{equation}

 A scalar of dimension $\Delta_n$ in global $AdS_3$, with angular momentum $m$ and radial quantum number $p$, has energy
\begin{equation}
E_{\mathrm{AdS}_3}
=\Delta_n+|m|+2p.
\end{equation}
Upon using  Eq.~\eqref{deltan}, we finally obtain the expression for the energy as
\begin{equation}
E_{AdS_3\times S^1}=|n|+|m|+2p+1.
\end{equation}
which exactly agrees with $E_{S^3\times \mathbb{R}}$ in  Eq.~\eqref{energyS3}.

\end{document}